\documentclass[%
 reprint,
 amsmath,amssymb,
prb,
]{revtex4-1}
\usepackage{xparse}

\usepackage{appendix}
\usepackage{graphicx}
\usepackage{dcolumn}
\usepackage{bm}

\renewcommand{\d}[2]{\ensuremath{\frac{\text{d} #1}{\text{d} #2}}}

\newcommand{\ket}[1]{\ensuremath{\left| #1 \right>}}
\newcommand{\bra}[1]{\ensuremath{\left< #1 \right|}}
\newcommand{\braket}[2]{\ensuremath{\left< #1 \ \vphantom{#2} \right| 
\left. #2 \vphantom{#1} \right>}}

\newcommand{\Tr}{\text{Tr}}
\usepackage{bbold}
\begin{document}

\title{Loschmidt Amplitude and Work Distribution in  Quenches of the Sine-Gordon Model}

\author{Colin Rylands} 
\email{rylands@physics.rutgers.edu}
\author{Natan Andrei}
\affiliation{Department of Physics, Rutgers University,
Piscataway, New Jersey 08854.
}

\date{\today}

\begin{abstract}

   The Sine-Gordon - equivalently, the massive Thirring - Hamiltonian is ubiquitous in low-dimensional physics, with applications that range from cold atom and strongly correlated systems to quantum impurities. We study here its non-equilibrium dynamics  using the quantum quench protocol - following the system  as it evolves under the Sine-Gordon Hamiltonian  from initial  Mott type  states  with large potential barriers.  By means of the Bethe Ansatz we calculate exactly the Loschmidt amplitude, the fidelity and work distribution characterizing these quenches for different values of the interaction strength. Some universal features are noted as well as an interesting duality relating  quenches in different parameter regimes of the model.

\end{abstract}

\maketitle

\section{Introduction}
The study of non equilibrium physics has exploded in recent years in large part thanks to advances in the experimental realization of quantum systems which exhibit long coherence times when far from equilibrium\cite{CA, PolRev}. Questions of the  thermalization of isolated quantum systems or the existence and classification of dynamical phase transitions can now be pondered from both experimental and theoretical viewpoints. A standard protocol to study non equilibrium systems is the quantum quench. The protocol consists of  preparing the  system in a state $\ket{\Phi_i}$, often an eigenstate of some Hamiltonian $\bar{H}$, and then  evolving it in time with a different Hamiltonian, $H$.  Typically the post quench Hamiltonian differs from $\bar{H}$ by changing some parameter, introducing new interaction terms, applying an external field or some combination thereof. The most commonly studied protocol is the sudden quench in which the change in parameter is deemed to happen instantaneously ({\it i.e.} over a time scale much shorter than any set by $H$). Non sudden quenches, wherein the parameter is changed over a finite period of time are sometimes considered\cite{Smacchia} however the sudden quench is the most tractable theoretically\cite{AndNonEq, MitraRev, EssCau} and is readily achievable in experiment\cite{CA}.

A fundamental quantity of interest in a quantum quench is the Loschmidt amplitude,
\begin{eqnarray}\label{Echo}
\mathcal{G}(t)=\bra{\Phi_i}e^{-iHt}\ket{\Phi_i}
\end{eqnarray}
which is the overlap between the initial state and its post quench evolution. This  allows one to calculate the Loschmidt echo, $\mathcal{L}(t)=\mathcal{G}^*(t)\mathcal{G}(t)$ which is of interest in many fields from quantum information and quantum chaos\cite{delCamp} to nuclear physics\cite{LosChaos} and is the central quantity in studies of dynamical quantum phase transitions (DQPT)\cite{HeylPolKeh} which occur when the amplitude vanishes as a function of $t$. Our main source of interest in the amplitude is its relation to the work done during the quench process.

\begin{figure}
 \includegraphics[trim={35mm 140mm 45mm 42mm},clip,width=.4\textwidth]{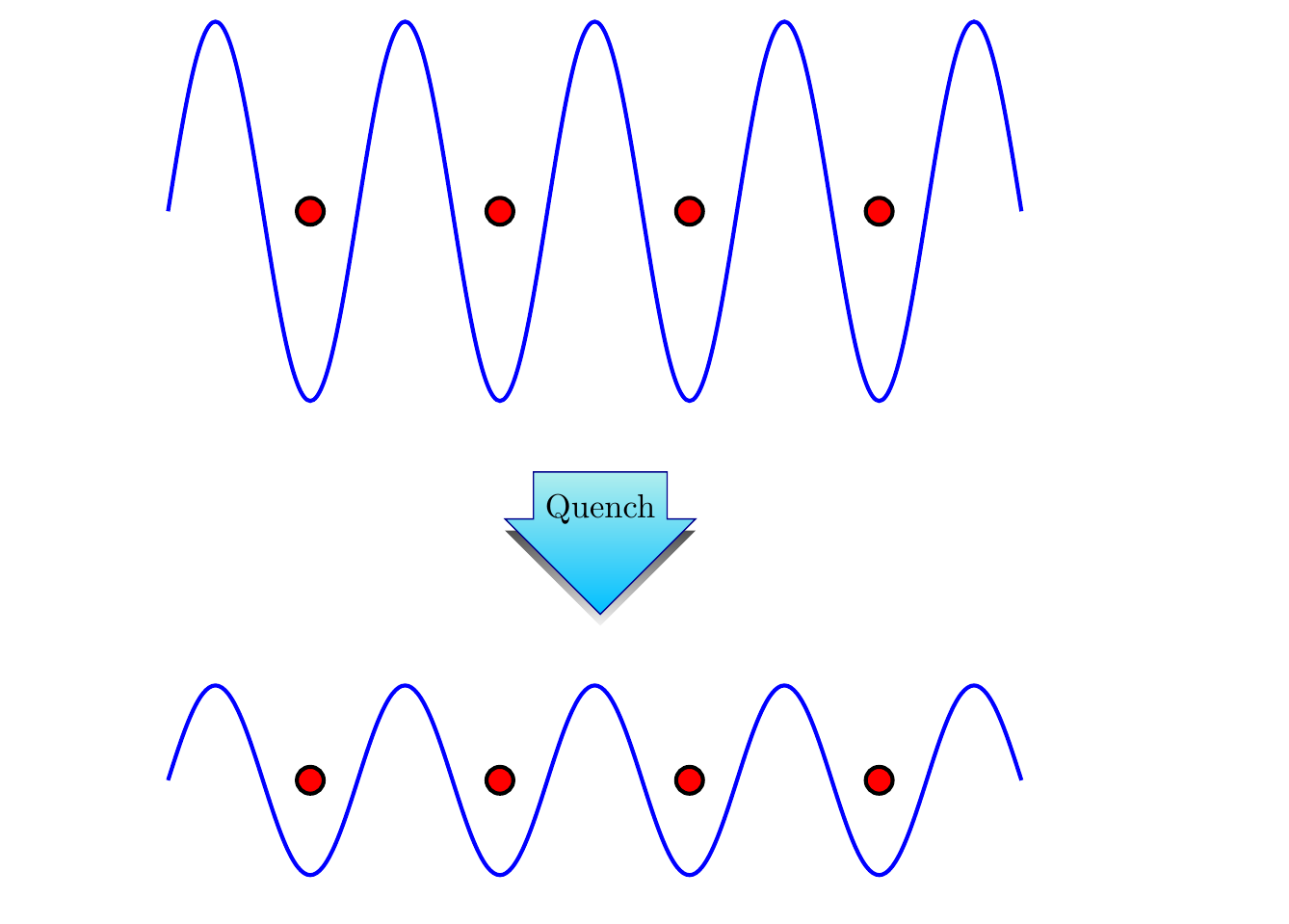}
\caption{A typical quantum quench protocol occurring in cold atom systems. Atoms are initially held in a deep optical lattice. The lattice depth is then suddenly lowered and the system allowed to evolve. The Sine-Gordon/massive Thirring model is an effective description of this process. }
\end{figure}

The concepts of work and entropy  in the context of  far from equilibrium, isolated quantum systems \cite{FlucRMP, Goold} are closely related to the Loschmidt amplitude. Work, in a quantum setting, is defined as the difference between two projective energy measurements, pre and post quench. It is therefore described by a probability distribution, $P(W)$ defined by\cite{TakLutHan, Silva08}

\begin{eqnarray}\label{Pdef}
P(W)=\sum_n \delta\left(W-(E_n-\epsilon_i)\right)|\braket{\Psi_n}{\Phi_i}|^2
\end{eqnarray}
with $\ket{\Psi_n}$  the eigenstates of $H$. 
It is not too difficult to see that up to a phase this is the Fourier transform of the Loschmidt amplitude, 
\begin{eqnarray}\label{p}
P(W)=\int_{-\infty}^\infty\frac{\mathrm{d}t}{2\pi} e^{iWt+i\epsilon_it}\mathcal{G}(t).
\end{eqnarray}
$\epsilon_i $ being the energy of the initial state as measured with respect to $\bar{H}$. $P(W)$ has the form of the spectral function familiar from many body physics\cite{Mahan} and accordingly shares many of the same generic features. 
It is defined for $W\geq \delta E$ with $\delta E= E_0-\epsilon_i$ being the energy difference between the initial state and the post quench ground state and possesses a delta function peak at $W=\delta E$ signifying the transition from the initial state to the ground state. This peak is weighted by the fidelity, $\mathcal{F}=|\braket{\Psi_0}{\Phi_i}|^2$ the probability for such a transition. If $H$ is gapped there exists a continuum of excited states separated from this peak into which $\ket{\Phi_i}$ can transition during the quench. For translationally invariant initial states and provided momentum is conserved by the quench the lower threshold for the  continuum is at $W=2m+\delta E$ with $m$ being the mass of the lightest quasiparticle. This signifies the emission of two quasiparticles with opposite momentum from the initial state.  At the threshold, $P(W)$  exhibits an edge singularity similar to the Anderson and Mahan effects in the X-ray edge problem\cite{Mahan}.  The exponent of the edge singularity depends upon the dimensionality of the system, its statistics and also if $\ket{\Phi_i}$ and $\ket{\Psi_0}$ lie within the same phase or not. It is independent of the quench protocol which only affects the coefficients of the singularity\cite{Smacchia}. For theories containing multiple particle species with say, mass $m_l$, similar edge singularities will occur at $W=2m_l+\delta E$ as well as when new channels open up. For example the emission of four particles with zero momentum causes an edge singularity at $W=4m+\delta E$.

Below the threshold, in the region $\delta E<W <2m+\delta E$ additional delta function peaks may appear under certain circumstances.  If bound states are supported by $H$ then delta function peaks may appear in this region at $W=m_b+\delta E$, $m_b$ being the masses of such states. The position and number of these peaks depending upon the symmetry of the initial state\cite{Palmai,Palmai2,GritDemPol}. In the absence of bound states a delta function peak may still appear below the threshold at $W=m+\delta E$ depending on the symmetry of the initial state. This is related to the existence of a kinematic pole in the boundary matrix describing the initial state and can be further traced back the system being quenched across a critical point\cite{GamSil1,Palmai2}. If the quasiparticles have finite lifetime one can expect broadening of these features to occur  (however this will not be the case in the model we consider which is integrable.) 

For a global quench the work done is extensive and therefore in the thermodynamic limit the distribution becomes sharply peaked about the average, $\left<W\right>$, with fluctuations vanishing as $1/\sqrt{L}$, $L$ being the system size\cite{Smacchia, SotGamSil}. The large deviations of $P(W)$, however retain interesting features and in particular the threshold edge singularity exhibits universal behaviour similar to those of critical systems\cite{GamSil2}. 
Lastly we note that in studying the work distribution one can  identify the work done due to reversible and irreversible processes in the quench. For example the transition to the ground state is a reversible process as it could equally be achieved adiabatically. The rest of the distribution describes the irreversible work done during the quench which can be related to the entropy production and the spread of entanglement in the system\cite{Goold}. 

Much intuition about quantum work has been obtained from calculations of  non-interacting models or using heuristic arguments\cite{Silva08,GamSil1,GamSil2,Smacchia,SotGamSil,Palmai2}. While there have been some notable calculations concerning work statistics in interacting models the list is short\cite{Palmai, Chiara1,Chiara2}. Here we will study these questions in the context of the  Sine-Gordon model, a full fledged field theory with many applications throughout low-dimensional physics. In high energy physics it is the archetypal integrable quantum field theory exhibiting many fundamental properties of the field such as scattering and renormalization.  In condensed matter it serves as the low energy description of many one dimensional  systems including the Hubbard model, spin chains and disordered systems as well as serving an important role in understanding quantum impurity systems like the Kondo model, while in statistical physics it provides a description of the  Kosterlitz-Thouless transition \cite{TG, gogolin2004bosonization}. Most pertinent to our present discussion is its use in the field of cold atom experiments where it can be used to describe   a pair of tunnel coupled quasi-condensates\cite{GritDemPol,GritDemPol2} or the superfluid-Mott transition of a Bose Einstein condensate (BEC) loaded into an optical lattice\cite{FishFish,Coldatom}.

We will calculate the Loschmidt amplitude for some  quenches in the Sine-Gordon model using its fermionic equivalent, the  massive Thirring model. There exists a wealth of information on the quench dynamics of the Sine-Gordon model \cite{GritDemPol,GritDemPol2, Bertini, Moca, Kormos, Cubero1,Horvath} mostly focusing on the time evolution of observables. Here we shall study the work distribution describing these  quenches and also examine the existence of associated DQPTs.

\section{The Sine-Gordon/Massive Thirring model}

The Sine-Gordon model is described by the Hamiltonian,
\begin{eqnarray}\label{Hbos}\nonumber 
H=\frac{1}{2}\int\mathrm{d}x\,\Big\{\Pi^2(x)+\left[\partial\phi(x)\right]^2-M^2\cos{\left[\beta\phi(x)\right]}\Big\}
\end{eqnarray}
where $\Pi(x)$ and $\phi(x)$ are conjugate bosonic fields related through bosonisation to the current and density of the fermion fields in the massive Thirring (MTM) model,

\begin{eqnarray}\nonumber 
H&=&-i\int \left[\psi_+^\dag\partial_x\psi_+-\psi_-^\dag\partial_x\psi_-\right]
\\\label{Hmtm}&&+m_0\int \left[\psi^\dag_+\psi_-+\psi_-^\dag\psi_+\right]
+4g\,\int\psi^\dag_+\psi_-^\dag\psi_-\psi_+.
\end{eqnarray}
Here $\psi_\pm^\dag(x)$ are left and right moving fermions which are coupled together via a bare mass term $m_0$ and  local density-density interactions of strength  $4g$. The bosonic parameters $M$ and $\beta$ are related to the fermionic  $m_0$ and $g$ through\cite{sGMTM},
\begin{eqnarray}
\frac{\beta^2}{4\pi}=1-\frac{2}{\pi}\arctan(g)
\end{eqnarray}

The model is exactly solvable and its spectrum, thermodynamics and scattering properties well understood \cite{ZamZam,BT, Thacker,Fowler,Korepin,Korepinmtm}. It has two distinct regimes corresponding to  repulsive ($g<0$, $4\pi<\beta^2<8\pi$) or attractive ($g>0$,  $0<\beta^2<4\pi$) interactions respectively. The value $\beta^2>8\pi$ may also be considered however in this region the bosonic theory is gapless and does not map to the massive Thirring model. In the repulsive regime the spectrum consists of two types of particles known as solitons and anti solitons of mass $m$, the bare mass, $m_0$ being renormalized to $m$ in  the presence of the interactions. In the attractive regime bound states of solitons and anti solitons known as breathers appear, their number and mass depending upon the value of $\beta$ and $m$. In this work we will focus on the fermionic MTM  formulation of the model to perform the calculations and in the end make the relation to the bosonic parameters.

Two quenches of particular interest are those when the prequench Hamiltonian $\bar{H}$ has a very large mass, $M^2\to\infty$, while   the quenching Hamiltonian $H$ is obtained by suddenly reducing the mass parameter to a finite value which also allows the interaction to come into full effect.
The initial state $|\Phi_i\rangle$ will be either the ground  state of $\bar{H}$ (see FIG. 1)  or the maximally excited state, with the bosons initially located on top of the potential wave (we shall also consider a combination of both). We note that  the maximally excited state can be viewed as the ground state of an initial Hamiltonian with $M^2 \to - \infty$). 
Such quench scenarios  can be achieved in a cold atom system by having a very deep optical lattice in the superfluid-Mott description or a  phase difference of $0$ or $\pi$ in the coupled quasi-condensates case.

We shall compute the Loschmidt amplitudes  associated with these quenches and  study the related work distributions, $P(W)$ paying particular attention to the region around the threshold.  
To do this we make use of the light cone lattice regularization of the time evolution operator\cite{Lightcone}. This will allow us derive a single non linear integral equation (NLIE) describing the amplitude directly in real time without having to analytically continue. The NLIE is valid for both the  repulsive and attractive regimes as well as for cases when the quench is within or across a phase. We carry out this in section III below. It is shown that there exists a non-equilibrium duality relating quenches in the repulsive regime to the attractive.  Furthermore we show that for the  considered quenches no DQPT may occur since the NLIE that describes these quenches in the presence of interactions does not admit solutions that satisfy the required conditions, see below.  In the subsequent section we examine the non interacting limit and discuss some general features present in the Loschmidt amplitude and work distribution.  The work distribution is found in the region around the threshold and many of the features described above are seen. In section IV we study the repulsive case and determine $P(W)$ in the vicinity of the edge threshold. In the penultimate section we discuss the attractive case and see how the work distribution is affected by bound states. In the final section we make a connection between the NLIE  governing the amplitude and observable quantities providing a firm link between the definition of a DQPT and its manifestation in the dynamics of a system.

\section{Loschmidt amplitude as a Lattice problem}
Our objective is to calculate the Loschmidt amplitude, $\mathcal{G}(t)$ which we do in terms of a single non linear integral equation (NLIE). Our strategy will be to utilize the light cone lattice regularization\cite{Lightcone} of the time evolution operator $e^{-iHt}$ . Within this approach  $\mathcal{G}(t)$ is seen to be the same as the partition function of classical six-vertex model with boundaries. In the thermodynamic limit the partition function is given by the largest eigenvalue of the transfer matrix of the model which, for particular types of initial state can be calculated using Bethe Ansatz.  This strategy has recently be used to great effect in calculating the Loschmidt echo in the XXZ Heisenberg model\cite{poz1,Poz2, poz3}. Unlike that instance however we will obtain the NLIE directly in real time without the need to Wick rotate back.
\begin{figure}
\includegraphics[trim={0mm 28mm 0mm 0mm},clip,width=.3\textwidth]{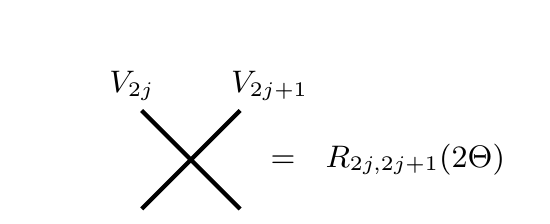}
\includegraphics[trim={20mm 20mm 15mm 5mm},clip,width=.35\textwidth]{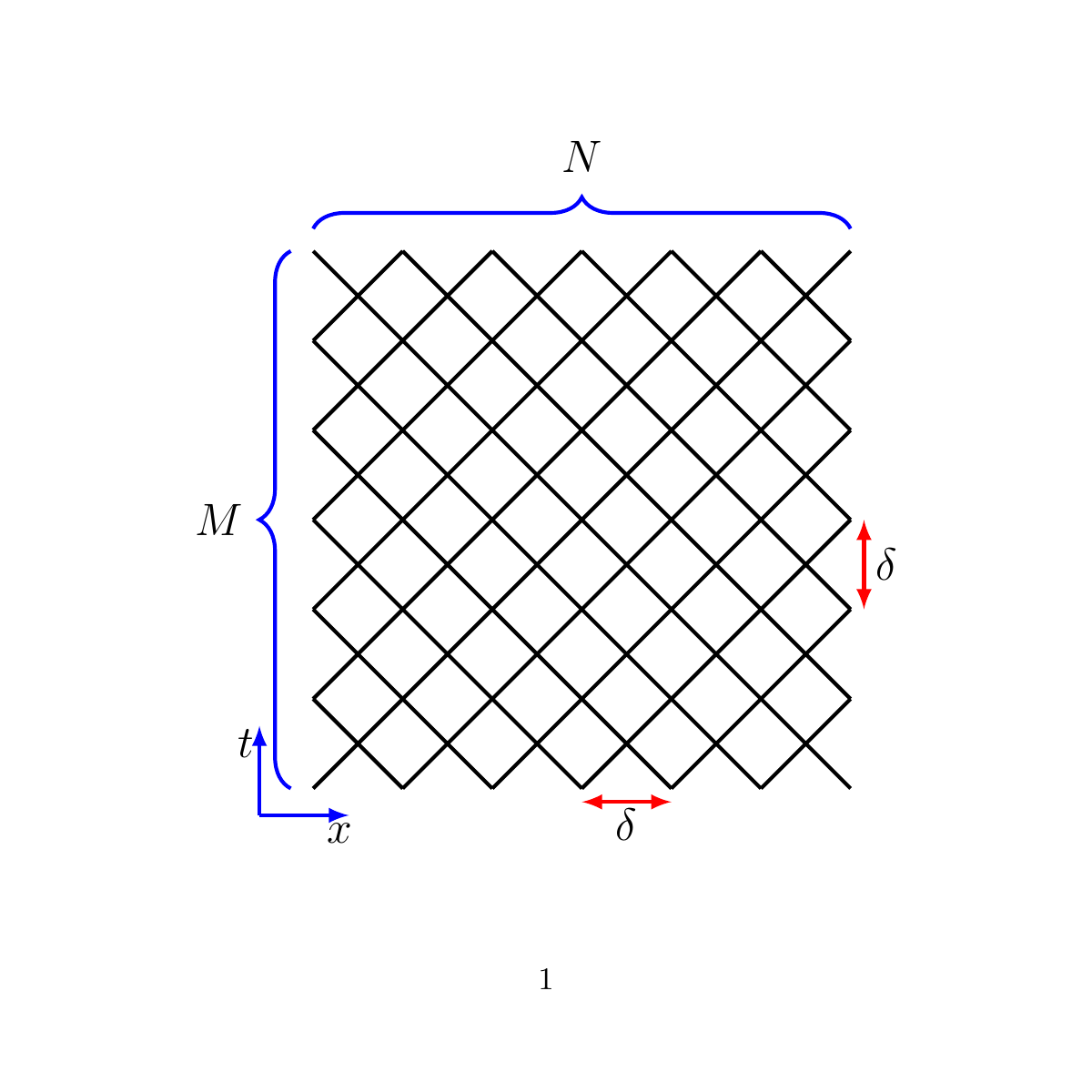}
\caption{The light cone lattice consists of  left and right oriented diagonal lines representing the world lines of bare relativistic particles. The vertical direction is time and the horizontal is space with  $\delta$ being the lattice spacing in both directions. The number of intersections in the spacelike direction is $N$ while the number in the timelike direction is $M$. To each intersection we associate a matrix of transition amplitudes $R_{k,k+1}(2\Theta)$. We assume periodic boundary conditions in the spatial direction.}
\end{figure}

\subsection{Light cone lattice}
The light cone lattice method is an approach to regularizing integrable quantum field theories in which the integrability of the continuum model is preserved. The  solution of the massive Thirring model through this method  will form the starting point of our calculation of the Loschmidt amplitude. The formulation is thoroughly explained in several works\cite{Lightcone, Lightcone2, Lightcone3, Vernier} and we give only the main points below. In this approach, one starts from a discretized 2D Minkowski  spacetime formed by a regular diagonal lattice of right-oriented and left-oriented straight lines, see FIG. 2. These represent the world-lines of bare right and left movers with each intersection of lines being a discretized point in spacetime. 
Every row of the lattice consists of $N$ such intersections labelled by $j=1,\dots,N$ and to each link emanating from these points we associate a vector space $V_k=\mathbb{C}^2$ with $V_{2j}$ corresponding to left movers and $V_{2j-1}$ to right movers emanating from the $j^{\text{th}}$ intersection. The fermion creation and annihilation operators then correspond, via Jordan-Wigner transformation to $\sigma^\pm_{k}$ which act on these vector spaces. Furthermore to each vertex we associate a matrix $R_{kk+1}(2\Theta)$ acting on $V_k\otimes V_{k+1}$, defined by
\begin{eqnarray}
R_{ij}(u)\!&=&\!\!\begin{pmatrix}
\sinh(u+\eta)&0&0&0\\
0&\sinh{(u)}&\sinh{(\eta)}&0\\
0&\sinh{(\eta)}&\sinh{(u)}&0\\
0&0&0&\sinh(u+\eta)
\end{pmatrix}
\end{eqnarray}
representing the microscopic transition amplitudes for the fields. The parameter $\eta=i\gamma$ encodes the interactions in the model according to
\begin{eqnarray}
\frac{\gamma}{\pi}=1-\frac{\beta^2}{8\pi}
\end{eqnarray}
while $\Theta$ is a rapidity cutoff for the bare particles. It is related  to $m$ the physical mass by 
\begin{eqnarray}
m=\frac{4N}{L}e^{-\frac{\pi}{\gamma}\Theta}
\end{eqnarray}
with $N$ being the total particle number and $L$ the system size. The bare mass is given by $m_0=\frac{4N}{L}\sin\gamma e^{-2\Theta} $ with the two mass parameters coinciding when the  interaction is turned off at $\beta^2=4\pi$ or $\gamma=\pi/2$. The thermodynamic limit takes $N\to\infty$ with $\delta=L/N$, the lattice spacing held fixed while the continuum limit requires taking $\delta\to 0$ and $\Theta\to\infty $ such that $m$ is held fixed. Upon taking these limits the continuum theory is recovered. 

This light cone lattice MTM thus takes the form of a 2d statistical model, namely an inhomogeneous 6-vertex model\cite{Baxter}. From this point of view the time evolution operator is the vertical transfer matrix on this lattice which is given by 
\cite{Lightcone},
\begin{eqnarray}\label{Hlatt}
e^{-iH\delta}&=&\left[\frac{\sinh{(2\Theta+\eta)}}{\sinh{(2\Theta-\eta)}}\right]^{N}\tau(-\Theta)\tau^{-1}(\Theta)\\\nonumber
\tau(u)&=&\Tr_{k}\left[R_{k1}(u-\Theta)R_{k2}(u+\Theta)\dots\right.\\\label{tau}&&\left.\dots R_{k2N-1}(u-\Theta)R_{k2N}(u+\Theta)\right]
\end{eqnarray} 
with $\Tr_k$ indicating taking the trace over the $k$ space. One can interpret $\tau(\pm\Theta)$ as evolving by one step along the world lines of the right or left moving particles, i.e.  in the $x\mp t$ directions. The combination above  therefore gives evolution in the timelike direction by one lattice unit $\delta$. Moreover, by taking the number of rows in the lattice to be $M$ we have that the total transfer matrix is   
\begin{eqnarray}\label{U}
e^{-iHt}&=&\lim_{X\to\infty}\left[\frac{\sinh{(2\Theta+\eta)}}{\sinh{(2\Theta-\eta)}}\right]^{NM}\left[\tau(-\Theta)\tau^{-1}(\Theta)\right]^M
\end{eqnarray}
where $t=M\delta$ and the limit $X\to\infty$ stands for the the thermodynamic and continuum limits discussed before in addition to $M\to\infty$ which is taken while  holding $t$ constant.

Before utilizing, \eqref{U} in our calculation of $\mathcal{G}(t)$ we close this section by commenting that within this regularization the wavefunctions of the system take the form
\begin{eqnarray}\label{wf}
\ket{\Psi}=\sum_{\{j\}}a_{\{j\}}\prod_{l=1}^\mathcal{N} \sigma^+_{j_l}\ket{\Downarrow}
\end{eqnarray}
with $\mathcal{N}\leq N$, $a_{\{j\}}$ being some coefficients, $\{j\}=\{j_1,\dots, j_{\mathcal{N}}\}$ the positions of the flipped spins, corresponding to the creation of a bare particle and $\ket{\Downarrow}=\otimes^{2N}\ket{\downarrow}$ is  the regularized, drained Fermi sea. In the continuum limit  with $ \sigma_{2j-1}^+/\sqrt{\delta}\to \psi^\dag_{+}(j\delta )$ and $
\sigma_{2j}^+/\sqrt{\delta}\to \psi^\dag_{-}(j\delta )$ these become the standard MTM wavefunctions\cite{Thacker}.
\begin{figure}
 \includegraphics[trim={20mm 20mm 15mm 5mm},clip,width=.49\textwidth]{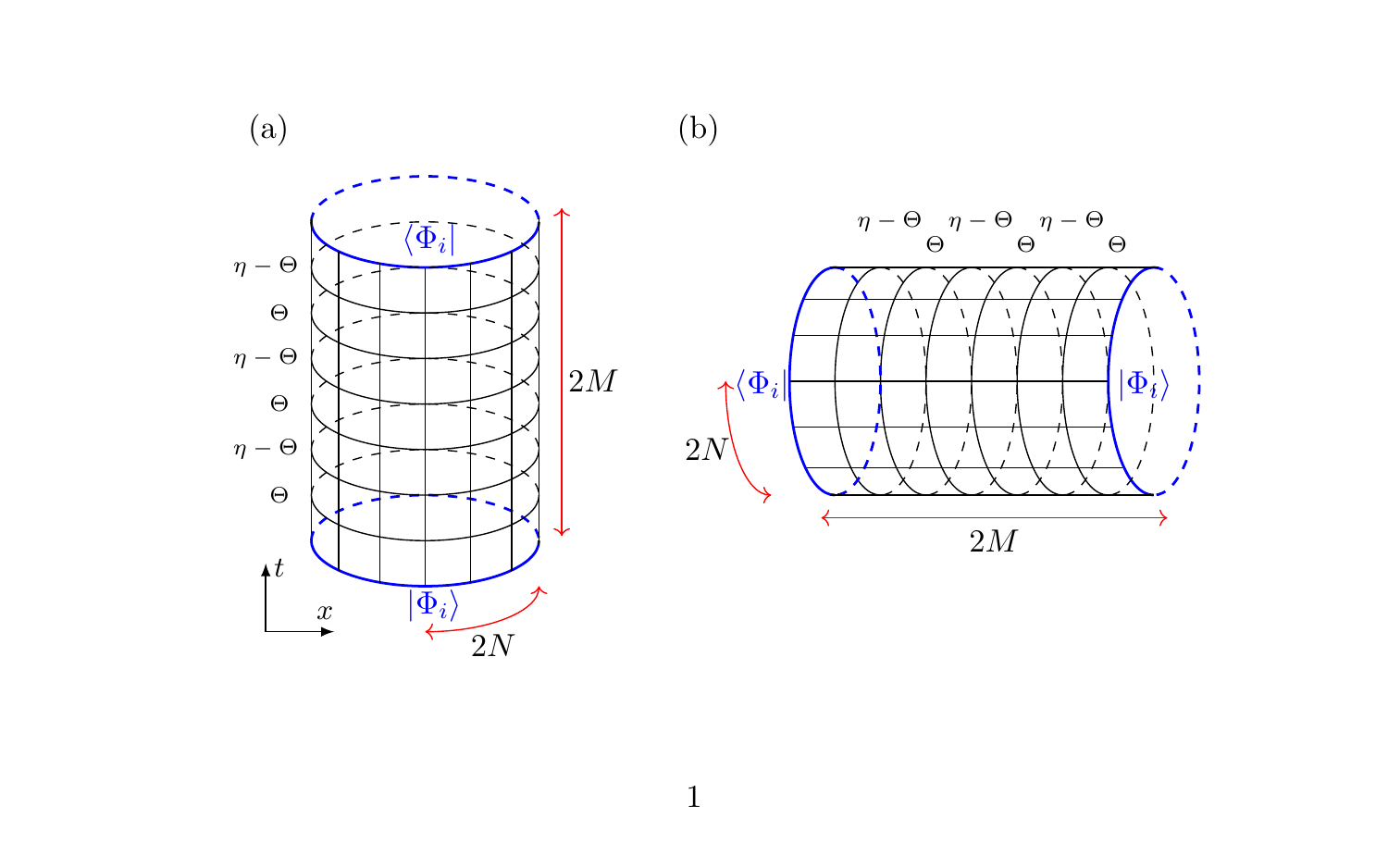}
\caption{(a)The Loschmidt amplitude as given in \eqref{Gg} takes the form of the partition function of a 2d classical lattice with the initial state appearing as boundaries along the top and bottom. Each horizontal line represents  $\tau(\Theta)$ or $\tau(\eta-\Theta)$. The trace present in $\tau(u)$, \eqref{tau} represents periodic boundary  conditions imposed in the spacelike direction. (b) Under a rotation the amplitude is then the partition function of a different 2d classical lattice model with $N$ and $M$ exchanged and the initial conditions becoming spatial boundaries. Each horizontal line here represents $T(\Theta)$ or $T(-\Theta)$ given by \eqref{T}. Although the partition functions calculated using the viewpoint in (a) or (b) are the same the transfer matrices in either case are not the same owing to the fact that the model is Lorentz invariant rather than Euclidean invariant.}
\end{figure}
\subsection{Regularized Loschmidt amplitude}
Having defined the time evolution operator in \eqref{Hlatt} the Loschmidt amplitude is given by 
\begin{eqnarray}\nonumber
\mathcal{G}(t)&=&\lim_{X\to\infty}\left[\frac{\sinh{(2\Theta+\eta)}}{\sinh{(2\Theta-\eta)}}\right]^{NM}\bra{\Phi_i}\left[\tau(-\Theta)\tau^{-1}(\Theta)\right]^M\ket{\Phi_i}\\\label{Gg}&=&\lim_{X\to\infty}\left[\frac{\bra{\Phi_i}\left[\tau(-\Theta)\tau(\Theta-\eta)\right]^M\ket{\Phi_i}}{\sinh^{2NM}{(2\Theta-\eta)\sinh^{2NM}{(\eta)}}}\right]
\end{eqnarray}
where in going to the second line we have used the fact that $\tau^{-1}(\Theta)\propto \tau(\Theta-\eta)$\cite{ODBA} and we have implicitly regularized the initial state in the light cone formalism \'{a} la \eqref{wf}. When written in this fashion calculating the Loschmidt amplitude becomes a classical 2d lattice problem. Indeed $\tau(-\Theta)\tau(\Theta-\eta)$
is the vertical transfer transfer matrix of an inhomogeneous 6-vertex model on a $2N\times 2M $ lattice with $\pm\Theta$ associated to each vertical line, $\Theta $ and $\eta-\Theta$ associated to each horizontal line and periodic boundary conditions in the horizontal direction. Note that this 6-vertex model is not the same as the one mentioned in relation to the light cone lattice, it is an auxiliary system on to which we have mapped our calculation. The Loschmidt amplitude is therefore the expectation value of this transfer matrix in the initial state, see FIG 3(a). By rotating the whole lattice we see in FIG. 3(b) that this is equivalent to computing the partition function on a similar but more convenient 6-vertex model. The initial state becomes boundary conditions in the horizontal direction, the periodic boundary conditions becoming a trace and parameters associated to vertical lines being exchanged with those associated to horizontal lines.
This gives us 
\begin{eqnarray}\nonumber
\mathcal{G}(t)&=&\lim_{ X\to\infty}\left[\frac{1}{\sinh{(2\Theta-\eta)\sinh{(\eta)}}}\right]^{2NM}\\\label{Gh}&&\times\Tr\left[\bra{\Phi_i}\otimes^N_{j=1}\left[T_{2j-1}(-\Theta)\otimes T_{2j}(\Theta)\right]\ket{\Phi_i}\right].
\end{eqnarray}
where we have introduced the  so called quantum transfer matrix which acts in the original horizontal direction\cite{Klumper}
 \begin{eqnarray}\nonumber
  T_j(u)&=&R_{j1}(u-\Theta)R_{j2}(u-(-\Theta+\eta))\dots\\\label{T}
  &&\dots R_{j2M-1}(u-\Theta) R_{j2M}(u-(-\Theta+\eta))
 \end{eqnarray}
and the trace is over the original horizontal spaces, $2j-1,~2j$.

At this point the light cone formalism has provided us with a quantity which although regularized is no easier to calculate. Inspecting \eqref{Gh} we see that the operator is a tensor product of operators acting on the horizontal vector spaces but a generic initial state will cause mixing between them all. 
For a particular type of initial state however the problem simplifies considerably. 
We consider here initial states which can be written as the product of two site states, 
 \begin{eqnarray}
 \ket{\Phi_i}=\otimes^{N}\left[\frac{1}{\sqrt{\braket{v}{v}}}\ket{v}\right],\\\label{v}
 \ket{v}=c_1\ket{\uparrow\downarrow}+c_2\ket{\downarrow\uparrow}.
 \end{eqnarray}
Included among this category are the states $c_1=\pm c_2$ which are of particular interest. They correspond to the ground state and highest excited state of $\bar{H}$ when the initial mass is very large, $m_i\to\infty$. In bosonic language this means taking the coefficient of the $\cos{\left[\beta\phi(x)\right]}$ term in \eqref{Hbos} to be very large with the states corresponding to those which minimize or maximize this particular term. 

In the context of cold atom experiments there are numerous ways to realize the model and  these states. In some cases the Sine-Gordon model is the actual model of the system, for example when describing a pair of coupled condensates with the initial states corresponding to an initial phase difference of $0$ or $\pi$ \cite{GritDemPol}. In other cases the Sine-Gordon model provides an effective low energy description, for example  when used as the low energy description of the superfluid-Mott transition  with the initial states corresponding to states in a very deep optical lattice with either positive of negative on-site interaction\cite{FishFish}. 

It should be noted however that the use of effective low energy Hamiltonians, while very useful in equilibrium, needs to be carefully examined when applied out of equilibrium since  quench process involves all eigenstates of the post quench Hamiltonian which may result in the Sine-Gordon description being invalid. \footnote{A relevant figure of merit is the average energy density $\epsilon_\text{av}(t)=\bra{\Phi_i(t)}\bar{H}\ket{\Phi_i(t)}/L$ which should not be too large in order to invoke the Sine-Gordon description.}
Having  firm  results, such as those presented herein, allows one to confront them with corresponding experimental data to determine to what extent low energy Hamiltonians  are valid for nonequilibrium physics.

\begin{figure}
 \includegraphics[trim={20mm 20mm 15mm 5mm},clip,width=.45\textwidth]{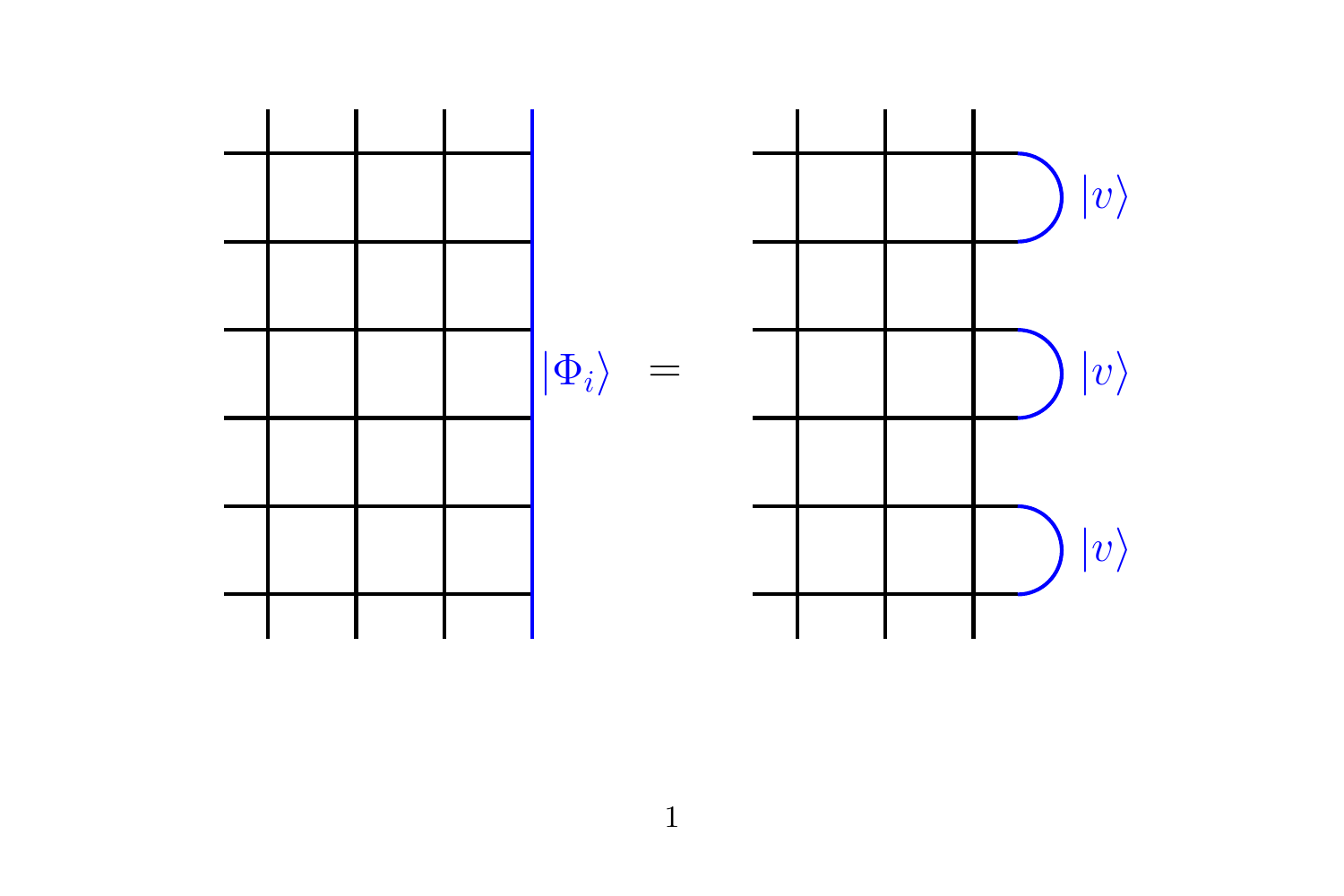}
\caption{By choosing the initial state appropriately the Loschmidt amplitude simplifies considerably. If $\ket{\Phi_i}=\ket{v}^N$ with $v$ given in \eqref{v} only two horizontal lines are coupled. The amplitude then reduces to the form \eqref{Ghorizontal}.}
\end{figure}

In what follows we will concentrate mostly on these two initial states along with the more general case of $c_1=e^{2\xi}c_2$ with $\xi$ a purely imaginary parameter that interpolates between the two. With this type of initial state no mixing occurs between the different parts of the tensor product in \eqref{Gh}, see FIG. 4, and the amplitude reduces to
\begin{eqnarray}\nonumber
 \mathcal{G}(t)=\lim_{X\to\infty}\left[\frac{1}{\sinh{(2\Theta-\eta)\sinh{(\eta)}}}\right]^{2NM}\\\label{Ghorizontal}
 \times\Tr\left[\frac{\bra{v}T(-\Theta)\otimes T(\Theta)
 \ket{v}}{\braket{v}{v}}\right]^N
 \end{eqnarray}
  Denoting the eigenvalues of the operator $\bra{v}T(-\Theta)\otimes T(\Theta)
 \ket{v}$ by $\Lambda_j, \,j=1\dots 2^{2M}$ we have that
 \begin{eqnarray}\nonumber
 \mathcal{G}(t)&=&\lim_{X\to\infty}\left[\frac{1}{\sinh{(2\Theta-\eta)\sinh{(\eta)}}}\right]^{2NM}\sum_j^{2^{2M}}\left[\frac{\Lambda_j}{\braket{v}{v}}\right]^N\\\label{Glam}
 &=&\lim_{X\to\infty}\left[\frac{1}{\sinh{(2\Theta-\eta)\sinh{(\eta)}}}\right]^{2NM}\left[\frac{\Lambda_{\text{max}}}{\braket{v}{v}}\right]^N
 \end{eqnarray}
In going to the second line it was  assumed that there exists a unique maximal eigenvalue, $\Lambda_{\text{max}}$ with a non vanishing gap to the rest of the spectrum. Accordingly in the thermodynamic limit $N\to\infty$, only this maximal eigenvalue contributes\cite{Baxter}. If the next eigenvalue crosses it at some time then a DQPT takes place\cite{AndSirk, poz3}. We discuss this point later.
 

 \subsection{Boundary quantum inverse problem}
 Our task now is to find the maximal eigenvalue of the operator  $\bra{v}T(u)\otimes T(-u)
 \ket{v}$ evaluated at $u=-\Theta$. 
 It was shown in [\!\!\citenum{poz1}] however that this is equivalent to finding the ground state of the XXZ model with open boundary conditions. This quantity is well studied and we get that $\Lambda_{\text{max}}=\Lambda(-\Theta)$ where the function $\Lambda(u)$ is given by \cite{Poz2} \begin{widetext}
\begin{eqnarray}\nonumber
 \Lambda(u)=\frac{\sinh{(2u+\eta)}}{\sinh{(2u)}}\sinh^2{(u+(\xi-\eta/2))}\left[\sinh{(u+\Theta)}\sinh{(u-\Theta+\eta)}\right]^{2M}\\\label{Lambda}
 \times\prod_{k}^{M}\frac{\sinh{(u-\lambda_k-\eta)}}{\sinh{(u-\lambda_k)}}\frac{\sinh{(u+\lambda_k-\eta)}}{\sinh{(u+\lambda_k)}}~~~  ~~\\\nonumber+\frac{\sinh{(2u-\eta)}}{\sinh{(2u)}}\sinh^2{(u-(\xi-\eta/2))}\left[\sinh{(u-\Theta)}\sinh{(u+\Theta-\eta)}\right]^{2M}\\
 \times \prod_{k}^{M}\frac{\sinh{(u-\lambda_k+\eta)}}{\sinh{(u-\lambda_k)}}\frac{\sinh{(u+\lambda_k+\eta)}}{\sinh{(u+\lambda_k)}}~~  ~~~
\end{eqnarray}
The newly introduced parameters $\lambda_k$ are known as the Bethe roots. In order to give the maximal eigenvalue we take them to be  the real, non negative solutions of  the Bethe ansatz equations
\begin{eqnarray}\nonumber
\frac{\sinh{(2\lambda_j-\eta)}}{\sinh{(2\lambda_j+\eta)}}\frac{\sinh^2{(\lambda_j-(\xi-\eta/2))}}{\sinh^2{(\lambda_j+(\xi-\eta/2))}}\left[\frac{\sinh{(\lambda_j-\Theta)}\sinh{(\lambda_j+\Theta-\eta)}}{\sinh{(\lambda_j+\Theta)}\sinh{(\lambda_j-\Theta+\eta)}}\right]^{2M}\\\label{BAE}
 =-\prod_{k}^{M}\frac{\sinh{(\lambda_j-\lambda_k-\eta)}}{\sinh{(\lambda_j-\lambda_k+\eta)}}\frac{\sinh{(\lambda_j+\lambda_k-\eta)}}{\sinh{(\lambda_j+\lambda_k+\eta)}}
\end{eqnarray}
\end{widetext}
The initial state is encoded by the relations
 \begin{eqnarray}
  c_1&=&\;\; \;\sinh{(\xi-\Theta-\eta/2)}\\
  c_2&=&-\sinh{(\xi+\Theta+\eta/2)}
 \end{eqnarray}
which gives $c_1=e^{2\xi}c_2$ in the continuum limit with $\xi=i\pi/2$ being the ground state or $\xi=0$ the maximally excited state of $\bar{H}$. 
Note that the Bethe equations \eqref{BAE} allow for a solution at $\lambda=0$ which must be excluded as it results in vanishing wavefunction and furthermore that if $\lambda_j$ is a solution then $-\lambda_j$ is also. The other eigenvalues, $\lambda_j$ can also be calculated and checked that they are gapped from $\Lambda_\text{max}$. 

\subsection{ An auxiliary function $\mathfrak{a}(u)$  and its  non linear integral equation (NLIE)}

The exact Loschmidt amplitude is  obtained from $\Lambda_{max}$ given by  equation \eqref{Lambda} subject to the Bethe Ansatz equations \eqref{BAE}. However to take the thermodynamic and continuum limits we need to bring these expressions to a more convenient form.
We define the auxiliary function $\mathfrak{a}(u)$ as
\begin{eqnarray}\nonumber
 \mathfrak{a}(u)=K_\xi(u)\left[\frac{\sinh{(u-\Theta)}\sinh{(u+\Theta-\eta)}}{\sinh{(u+\Theta)}\sinh{(u-\Theta+\eta)}}\right]^{2M}\\\label{a}\times\prod_{k}^{M}\frac{\sinh{(u-\lambda_k+\eta)}}{\sinh{(u-\lambda_k-\eta)}}\frac{\sinh{(u+\lambda_k+\eta)}}{\sinh{(u+\lambda_k-\eta)}}
\end{eqnarray}
where the boundary part, $K_\xi(u)$ containing the initial state information is given by 
\begin{eqnarray}\label{k}
 K_\xi(u)=\frac{\sinh{(2u-\eta)}}{\sinh{(2u+\eta)}}\frac{\sinh^2{(u-(\xi-\eta/2))}}{\sinh^2{(u+(\xi-\eta/2))}}.
\end{eqnarray}
We shall express the Loschmidt amplitude in terms of this auxiliary function and obtain an NLIE for it. 

The Bethe equations are  succinctly written in terms of this as  $\mathfrak{a}(\lambda_j)=-1$, 
so that the function $1+\mathfrak{a}(u)$ has zeros only at $u=\pm\lambda_j$ and $u=0$ as well as a pole of order $2M$ at $u=-\Theta$. This allows us to write that for an analytic function $f(x)$\cite{poz1},
\begin{eqnarray}\nonumber
 \sum_j^{2M}\left[f(\lambda_j)+f(-\lambda_j)\right]&=&-\oint_C\frac{\mathrm{d}\mu}{2\pi i} f'(\mu)\log{\left[\frac{1+\mathfrak{a}(\mu)}{1+K_\xi(\mu)}\right]}\\\label{int}&&\quad\quad\quad+2Mf(-\Theta)
\end{eqnarray}
where the contour is taken to encircle, counterclockwise the Bethe roots which lie on the real axis as well as their negative counterparts. A typical choice is for $C=C_+\cup C_-$ with $C_-$ running from $-\infty-i\zeta$ to $\infty-i\zeta$ and $C_+$ going back from $\infty+i\zeta$ to $-\infty+i\zeta$ and $2\zeta<\text{min}(\gamma,\pi-\gamma)$. Note that the singularity at $\mu=0$ as well as any singularities coming from $K(u)$ are cancelled by the $\log[1+K(u)]$ factor in the integrand. Using this we can write \eqref{a} as\begin{widetext}
\begin{eqnarray}\nonumber
 \log{\mathfrak{a}(u)}
 &=&\log{K_\xi(u)}+2M\log\left[\frac{\sinh{(u-\Theta)}\sinh{(u+\Theta-\eta)}}{\sinh{(u+\Theta)}\sinh{(u-\Theta+\eta)}}\right]\\\label{NLIE}&&-\oint_C\frac{\mathrm{d}\mu}{2\pi} \frac{\sin{2\gamma}}{\sinh{(u-\mu+i\gamma)\sinh{(u-\mu-i\gamma)}}}\log{\left[\frac{1+\mathfrak{a}(\mu)}{1+K_\xi(\mu)}\right]}
\end{eqnarray}
After some manipulation (see Appendix A) we get a NLIE for $\mathfrak{a}(u)$\footnote{ We could have added a chemical potential term $\int\mu\left[\psi^\dag_+\psi_++\psi^\dag_-\psi_-\right]$ to the MTM Hamiltonian and maintain integrability. Such a term introduces twisted boundary conditions to the lattice problem and comes at the cost on introducing a second coupled NLIE\cite{DDV1}}. Upon taking the thermodynamic and continuum limits this is 
\begin{eqnarray}\nonumber
\log{\mathfrak{a}(u)}
=-2mt\sinh{\left(\frac{\pi}{\gamma}u\right)}+\log{\mathbb{K}_\xi(u)}
+\int_{-\infty}^{\infty}\mathrm{d}\mu\,G(u-\mu-i\zeta) \log{\left[1+\mathfrak{a}(\mu+i\zeta)\right]}\\\label{NLIEa}-\int_{-\infty}^{\infty} \mathrm{d}\mu\,G(u-\mu+i\zeta) \log{\left[1+\mathfrak{a}^{-1}(\mu-i\zeta)\right]}~.~~~~~
\end{eqnarray}\end{widetext}
where $\log{\mathbb{K}_\xi(u)}=J*\log\left[\frac{\sinh{(u-i\gamma)}}{\sinh{(u+i\gamma)}}K_\xi(u)\right] $ and,
\begin{eqnarray}\label{J}
J(x)&=&\int_{-\infty}^\infty\frac{\mathrm{d}\omega}{2\pi}e^{i\omega x}\frac{\sinh{(\pi\omega/2)}}{2\cosh{(\gamma\omega/2)}\sinh{\left[(\pi-\gamma)\omega/2\right]}}\\\label{Gw}
G(x)&=&\int_{-\infty}^\infty\frac{\mathrm{d}\omega}{2\pi}e^{i\omega x}\frac{\sinh{\left[(\pi-2\gamma)\omega/2\right]}}{2\cosh{(\gamma\omega/2)}\sinh{\left[(\pi-\gamma)\omega/2\right]}}.
\end{eqnarray}
and $*$ denotes the convolution $f*g(x)=\int f(x-y)g(y)\mathrm{d}y$. The function $G(x)$ is actually the derivative of the soliton-soliton phase shift,
\begin{eqnarray}
G(x)=\frac{1}{2\pi i}\d{}{x}\log{S_{++}(\frac{\pi}{\gamma}x)}\,.
\end{eqnarray}   
where an explicit expression for $S_{++}$ is known and presented in Appendix C. In addition the first term in \eqref{NLIEa} is related by a shift of the rapidity $u\to u+i\gamma/2$ to the soliton energy $m\cosh{\left(\frac{\pi}{\gamma}u\right)}$, while the second term, $\log{\mathbb{K}_\xi(u)}$ is related by the same shift to the renormalized boundary phase shift\cite{BdrySaleur}.  In either the repulsive or attractive regimes it has the form\cite{Ghoshal, BdrySaleur}
\begin{eqnarray}
\mathbb{K}_\xi\left(x\right)=-\frac{\left[S_0\left(\frac{\pi}{\gamma}x\right)S_1\left(\frac{\pi}{\gamma}x\right)\right]^2}{S_{++}(2\frac{\pi}{\gamma}x)}
\end{eqnarray}
with $S_{++}$ again being the soliton-soliton phase shift. The other two factors $S_0$ and $S_1$ are presented in full in Appendix C. The first term, $S_0$ along with $S_{++}$  is independent of $\xi$ and hence the initial state while the $S_1$ term does depend upon $\xi$. Within the repulsive regime $\mathbb{K}_\xi(u)$ does not contain any poles in the so called ``physical strip" $0\leq \text{Im}(x)\leq \gamma/2$ apart from a possible pole at $x=i\gamma/2$ depending on the value of $\xi$. We are most interested in the behaviour of the boundary phase shift in the vicinity of this point, where   $\mathbb{K}_\xi$ takes the form \cite{Bajnok},
\begin{eqnarray}\label{gfunc}
\mathbb{K}_\xi(x)\sim\gamma\frac{g^2_{\xi}}{2\pi}\frac{1}{x-i\frac{\gamma}{2}}
\end{eqnarray}
with the coefficient, $g_\xi$ charachterizing  the initial condition. In the free case considered below we have $g_0=2$ while $g_{i\frac{\pi}{2}}=0$ which we will see ultimately determines the low $W$ behaviour of $P(W)$.  Using the explicit expressions presented in Appendix C we see that also in the interacting case $g_{i\frac{\pi}{2}}=0 $ while $g_0\neq 0$. In the more complicated attractive case the appearance of bound states in the spectrum is reflected in additional poles appearing in the physical strip. We will see below that these correspond to the delta function peaks in $P(W)$ below the edge threshold.

The form of the NLIE depends upon the value of $\zeta$ which is chosen. The restriction on allowed values of $\zeta$ changes when crossing from the repulsive $\gamma<\pi/2$ regime to the attractive $\gamma>\pi/2$  which reflects the change in the spectrum and appearance of bound states.

\subsection{ The Loschmidt Amplitude}

Turning finally to the Loschmidt amplitude we can express it neatly in terms of   $\mathfrak{a}(u)$ using the same integral technique (see Appendix B). Taking the continuum and thermodynamic limits we get,
\begin{eqnarray}\nonumber
\log{\mathcal{G}(t)}
&=&-iE_0t+\log{\mathcal{F}}\\\nonumber&&+i\frac{mL}{4\gamma}\int_{-\infty}^\infty\mathrm{d}\mu\, e^{\frac{\pi}{\gamma}(\mu-i\zeta)}\log{\left[1+\mathfrak{a}^{-1}(\mu-i\zeta)\right]}\\ \label{Gaa}
&&-i\frac{mL}{4\gamma}\int_{-\infty}^\infty\mathrm{d}\mu\,e^{\frac{\pi}{\gamma}(\mu+i\zeta)}\log{\left[1+\mathfrak{a}(\mu+i\zeta)\right]}
\end{eqnarray}
where $E_0$, see \eqref{gsenergy} is the ground state energy of $H$ and  
\begin{eqnarray}\nonumber
\log{\mathcal{F}}
&=&-i\frac{mL}{4\gamma}\int_{-\infty}^\infty\mathrm{d}\mu\, e^{\frac{\pi}{\gamma}(\mu-i\zeta)}\log{\left[1+K_\xi^{-1}(\mu-i\zeta)\right]}\\
&&+i\frac{mL}{4\gamma}\int_{-\infty}^\infty\mathrm{d}\mu\,e^{\frac{\pi}{\gamma}(\mu+i\zeta)}\log{\left[1+K_\xi(\mu+i\zeta)\right]}.
\end{eqnarray}
This time independent term is the same as the boundary contribution to the ground state of the MTM on a finite interval\cite{LMSS} and we will see below that it appears in the work distribution as the weight of the delta function peak $\delta(W-\delta E)$. Therefore we can identify this with the fidelity,  $\mathcal{F}= |\braket{\Psi_0}{\Phi_i}|^2$.  In both the NLIE and Loschmidt amplitude it can be convenient to perform calculations prior to taking the thermodynamic and continuum limits. These cutoff expressions appear in Appendices A and B. 

An interesting feature of the amplitude and its associated NLIE is that they exhibit a non-equilibrium duality. Under the replacement $\gamma\to\pi-\gamma$ along with $\xi\to i\pi/2-\xi$ we find that they are modified by $t\to-t$ which means
\begin{eqnarray}
\mathcal{G}(t)|_{\xi,\gamma}=\mathcal{G}(-t)|_{i\frac{\pi}{2}-\xi,\pi-\gamma}.
\end{eqnarray}
This duality relates the non-equilibrium behavior of the MTM in the repulsive regime to that of the attractive regime. Similar dualities have been discovered in out of equilibrium impurity systems \cite{schill, SalFen}. In the impurity models the mapping refers only to the impurity with the bulk spectrum being invariant. In this case however the duality maps the model between two different parameter regimes with very different spectra. 
\subsection{Dynamical Quantum Phase Transition}
The similarity between the Loschmidt amplitude and the partition function   at non zero temperature\cite{CardCal} served as inspiration for  the notion of a dynamical quantum phase transition. Analogous to thermal phase transitions, DQPTs are said to occur at $t=t_c$ if $\mathcal{G}(t_c)=0$ which results in non analyticity of $\log{\mathcal{G}(t)}$, sometimes called the dynamical free energy. Unlike thermal phase transitions infinitely many of these critical points can occur periodically over the range $0<t<\infty$. It is not obvious from this thermal analogy however how a DQPT manifests itself in the dynamics the system or if in fact by crossing $t_c$ the system can be thought of being in one phase or another.  After all, time dependent observables  are not merely derivatives of this dynamical free energy as they are in the thermal case. Moreover, the amplitude itself constitutes a projection of the initial state onto only a single state $\ket{\Phi_i}$ and the quench process involves excitations over the whole spectrum of the post quench Hamiltonian. In several cases however it has been shown that order parameters exhibit non analytic behaviour at critical times in the thermodynamic limit,  a phenomenon which was later observed experimentally\cite{dqptexp}. This has been explained via another analogy with quantum phase transitions which despite being ground state phenomenon, impact upon finite temperature  properties\cite{Heylreview}. More recently, the robustness of this relationship between DQPTs and non analytic behavior of the order parameter has been called into question. It has been shown that this relationship does not necessarily carry over to non-integrable models\cite{KarrSchur} or more general non equilibrium scenarios such as a double quench\cite{KenSchurKarra} and so further study is required. 

The existence of a DQPT depends on both the initial state and Hamiltonian.
Inspecting the Loschmidt amplitude \eqref{Gaa} we can see that a DQPT can occur for our particular choice of initial states if either of the terms in the logarithms vanish. For this to be the case we require that for some real $t$ either
\begin{eqnarray}
1+\mathfrak{a}(\mu+i\zeta)=0~\text{or}~
1+\mathfrak{a}^{-1}(\mu-i\zeta)=0
\end{eqnarray}
for $\mu$ real. By construction however $\mathfrak{a}(\lambda)=-1$ only if  $\lambda\in \mathbb{R} $ and moreover only if $\lambda=\pm\lambda_k,0$, with $\lambda_k$ a Bethe parameter.  Accordingly no dynamical quantum phase transitions take place in these particular quenches of the massive Thirring model in either regime.  This by no means precludes a DQPT in quenches of the MTM. We note that using the same transfer matrix method  in the XXZ it was shown that DQPTs could  occur\cite{poz3}. These resulted from the fact that a level crossing occurred between  $\Lambda_{\text{max}}$ and the next eigenvalue. The level crossings ultimately stem from the fact that in the XXZ case the transfer matrix must be Wick rotated to give the Loschmidt amplitude. This is not the case for the MTM as the light cone lattice formulation provides us with the amplitude directly in real time with out the need to Wick rotate.  We reiterate that a DQOT depends on the initial state so even within this formalism, taking initial states requiring $\xi\in \mathbb{R}$ or those encoded by off diagonal boundary conditions may result in level crossings of $\Lambda_{\text{max}}$ leading to a DQPT. 



\section{Non-interacting limit}
As a non trivial check on this result we examine the Loschmidt amplitude at  free fermion point, $\gamma=\pi/2$. Other than a check, this will also provide us with some insight and perspective with which to appreciate the interacting case. At the free point interactions are of course absent and so $G(x)=0$ while $J(x)=1$. The  NLIE therefore reduces to 
\begin{eqnarray}\label{anon}
\log{\mathfrak{a}(u)}=-2mt\sinh{(2u)}+\log{\left[ -\frac{\sinh^2{(u-(\xi-i\pi/4))}}{\sinh^2{(u+(\xi-i\pi/4))}}\right]}.
\end{eqnarray}
Inserting this into \eqref{Ga} along with the choice $\zeta=\pi/4$ gives us 
\begin{eqnarray}\nonumber
\log{\mathcal{G}(t)}=-iE_0t+\log\mathcal{F}+L\int_{-\infty}^\infty\frac{\mathrm{d}\mu}{\pi}\, m_0\cosh{(2\mu)}\\\label{nonintg}\times
\log{\Bigg\{1+\left[\frac{\cosh{(\mu-\xi)}}{\sinh{(\mu+\xi)}}\right]^2e^{-2im_0t\cosh{(2\mu)}}\Bigg\}}
\end{eqnarray}
where the log of the fidelity is given by 
\begin{eqnarray}\nonumber
\log\mathcal{F}&=&-L\int_{-\infty}^\infty\frac{\mathrm{d}\mu}{\pi}\, m_0\cosh{(2\mu)}\\&&\times\log{\Bigg\{1+\left[\frac{\cosh{(\mu-\xi)}}{\sinh{(\mu+\xi)}}\right]^2\Bigg\}}.
\end{eqnarray}
These agree with results obtained using more standard techniques (see Appendix D)\cite{Silva08}.

We can see explicitly here   the boundary phase shift for $\xi$ and $i\frac{\pi}{2}-\xi$ are related by inversion, $K_\xi(u)= K_{i\frac{\pi}{2}-\xi}^{-1}(u)$ and consequently the Loschmidt amplitude for these cases are complex conjugate to each other or rather related by time reversal,
\begin{eqnarray}\label{symmetry}
\mathcal{G}(t)|_{\xi}=\mathcal{G}(-t)|_{i\frac{\pi}{2}-\xi}\,.
\end{eqnarray} 
this is a special case of the duality mentioned in the previous section. The non interacting point being the self dual theory. A  consequence of this is that the Loschmidt echo, $\mathcal{L}(t)$ is the same for both quenches. This ultimately results from the fact that the two initial states are related by a particle-hole transformation which is preserved by the quench. 

It may also be checked explicitly that the term inside the logarithm of \eqref{nonintg} does not vanish and so no DQPT takes place. We should reiterate that other choices of initial states can result in a DQPT, for example,  if the initial state were the ground state of the Hamiltonian with finite initial mass $m_i$\cite{VajBal}. To see this in our method one must replace the constant $\xi$ which encodes the initial states considered previously with one which is rapidity dependent. The correct choice for an initial mass of $m_i$ is shown in Appendix D to be \begin{eqnarray}\label{finitem}
\xi(\mu)=\frac{1}{2}\sinh^{-1}\left[-\frac{m_0}{m_i}\sinh{(2\mu)}\right].
\end{eqnarray}
Inserting this into \eqref{nonintg} it can be seen that the DQPT condition is be satisfied. Note that this choice of $\xi$ does not fall within the class of initial states for which the NLIE was derived. Nevertheless the simple replacement above reproduces the result using other methods.

We turn our attention now to the work probability distribution in the non interacting model. As discussed in the  introduction the edge behavior of the distribution is of most interest, however the moments of the distribution can also provide some insight to the quench process. The moments of the work distribution are given by logarithmic derivatives of the amplitude, explicitly the $n^\text{th}$ moment is defined as 
\begin{eqnarray}
\chi_n=i^n\frac{\mathrm{d}^n}{\mathrm{d}t^n}\left[\log{\mathcal{G}(t)}e^{i\epsilon_it}\right]|_{t=0}.
\end{eqnarray}
The first of these corresponds to  the average work done  during the quench, $\left<W\right>$. Differentiating \eqref{nonintg} and using 
  $\mathfrak{a}(u)|_{t=0}=K_\xi(\mu)$ one finds
 \begin{eqnarray}\label{Work}
\frac{\left<W\right>}{L}=\frac{\delta E}{L}+\int_{-\infty}^\infty\frac{\mathrm{d}\mu}{\pi}\, \left[m_0\cosh{(2\mu)}\frac{\cosh{(\mu-\xi)}}{\sinh{(\mu+\xi)}}\right]^2
\end{eqnarray}
The first term is a constant and is the amount of work done per unit length if the quench were done adiabatically instead of suddenly. Subtracting this from the average work  defines the irreversible work \cite{Goold}  done during the quench, $\left<W\right>_\text{irr}=\left<W\right>-\delta E$  which is given by the integral term. For the two particular cases of $\xi=0,i\pi/2$ this integral diverges. The reason for this divergence is the fact that these initial states are very high in energy with respect to $H$. Their overlap with the highest excited states of $H$ is not suppressed by some scale inherent in the initial states and are only cutoff by $\delta$ the lattice spacing. In contrast if the initial is the ground state of the finite initial mass Hamiltonian, then this quantity is finite. This  can be seen by replacing the constant $\xi$ in \eqref{Work} with \eqref{finitem}.

 In the remainder of this section and in subsequent sections we will concentrate on the edge behavior of $P(W)$ near the threshold. The probability distribution, for the $\xi=i\frac{\pi}{2}$ state is \begin{widetext}
\begin{eqnarray}
P(W)&=&\int_{-\infty}^\infty\frac{\mathrm{d}t}{\pi}e^{iWt+i\epsilon_i t}\mathcal{G}(t)\\\nonumber
&=&\mathcal{F}\int_{-\infty}^\infty\frac{\mathrm{d}t}{\pi}e^{iWt-i\delta E t}\left[1+L\int_{-\infty}^\infty\frac{\mathrm{d}\mu}{\pi}\, m_0\cosh{(2\mu)}\log{\Bigg\{1+\tanh^2{(\mu)}e^{-2im_0t\cosh{(2\mu)}}\Bigg\}}+\dots\right]
\end{eqnarray}\end{widetext}
where we have expanded the exponential. It is then standard to expand the logarithm also and perform the $t$ integral \cite{Silva08, Smacchia}. The advantage of doing this is that only the lowest order terms in this expansion give a contribution to $P(W)$ in the region $W<4m_0+\delta E$ with the subsequent terms contributing in the regions $W<6m_0+ \delta E$ etc. From here on we measure the work done from $\delta E$ so $W-\delta E\to W$. Keeping only the lowest term in the expansion we have

\begin{eqnarray}\label{Pgsnon}
P(W)
&=&\mathcal{F}\delta(W)+m_0\frac{\mathcal{F}L}{2\pi}\theta(W-2m_0)W\sqrt{\frac{W-2m_0}{(W+2m_0)^3}}+\dots
\end{eqnarray}
 which gives the behaviour near the threshold. Here we see the appearance of a number features alluded to in the introduction. Firstly, there is the  delta function at $W=0$  coming from the transition to the ground state which is weighted by the fidelity. Secondly, there is an edge singularity at $W=2m_0$ with exponent $1/2$. Keeping further terms in the expansion of the exponential and logarithm gives access to further edges at $W=2nm_0$.

 For the other initial state in which $\xi=0$  we get at the free point  that 
\begin{widetext}
\begin{eqnarray}
P(W)
&=&\mathcal{F}\int_{-\infty}^\infty\frac{\mathrm{d}t}{2\pi}e^{iWt-i\delta E t}\left[1+L\int_{-\infty}^\infty\frac{\mathrm{d}\mu}{\pi}\, m_0\cosh{(2\mu)}\log{\Bigg\{1+\coth^2{(\mu)}e^{-2im_0t\cosh{(2\mu)}}\Bigg\}}+\dots\right]
\end{eqnarray}
Now we would like to proceed as before and expand the logarithm, the problem with this however is that the boundary term diverges like $\coth^2(\mu)\sim 1/\mu^2$ about $\mu=0$. To deal with this we use the regularization trick of adding and subtracting a term with the same divergence\cite{Bajnok}. The $\mu$ integral then becomes 
\begin{eqnarray}\nonumber
L\int_{-\infty}^\infty\frac{\mathrm{d}\mu}{\pi}\, m_0\cosh{(2\mu)}\log{\Bigg\{\frac{1+\coth^2{(\mu)}e^{-2im_0t\cosh{(2\mu)}}}{1+4\text{csch}^2{(2\mu)}e^{-2im_0t}}\Bigg\}}\\+L\int_{-\infty}^\infty\frac{\mathrm{d}\mu}{\pi}\, m_0\cosh{(2\mu)}\log{\Bigg\{1+4\text{csch}^2{(2\mu)}e^{-2im_0t}\Bigg\}}
\end{eqnarray}
The second integral can be computed using the formula $2\pi (a-b)=\int_{-\infty}^\infty\log{\frac{a^2+x^2}{b^2+x^2}}$ using $a^2=4e^{-2im_0t}$ and $b^2=0$ and the first integral can be expanded so that we get\cite{KorPoz} 
\begin{eqnarray}\nonumber
P(W)
&=&\mathcal{F}\int_{-\infty}^\infty\frac{\mathrm{d}t}{2\pi}e^{iWt-i\delta E t}\left[1+L\int_{-\infty}^\infty\frac{\mathrm{d}\mu}{\pi}\, m_0\cosh{(2\mu)}\Bigg\{\coth^2{(\mu)}e^{-2im_0t\cosh{(2\mu)}}\right.\\\nonumber
&&\left.-4\text{csch}^2{(2\mu)}e^{-2im_0t}\Bigg\}+2m_0Le^{-im_0t}+\frac{1}{2}\left(2m_0Le^{-im_0t}\right)^2+\dots\right]
\end{eqnarray}
\end{widetext}
where we have included an extra term coming from the expansion of the exponential. Taking the Fourier transform we get\cite{GamSil1,Smacchia}
\begin{eqnarray}\nonumber
P(W)&=&\mathcal{F}\delta(W)+2m_0\mathcal{F}L\delta(W-m_0)\\\nonumber
&&+m_0\frac{\mathcal{F}L}{2\pi}\theta(W-2m_0)W\sqrt{\frac{W+2m_0}{(W-2m_0)^3}}\\&&+2\mathcal{F}^2m^2_0L^2\delta(W-2m_0)+\dots
\end{eqnarray}

This exhibits further properties which were discussed in the introduction. In addition to the delta function coming from the ground state transition there are  additional delta functions at $W=m_0$ as well as at the edge whose exponent is now $-3/2$\footnote{ We note that the  -3/2 exponent would lead to a divergence when integrated over $W$. The divergence  is however compensated by  the vanishing of the fidelity
(in the infinite volume limit) so that the integral over the work distribution yields 1.}. These new features result from the pole appearing in the boundary phase shift $K_0(\mu)$ describing the initial state. 
To understand them more physically recall that the Hilbert space of the free fermion model splits into two sectors with even and odd particle number. The previous initial state is contained only in the even sector whereas the maximal eigenstate has overlap with both sectors. The later can therefore transition to a single particle excited state with zero momentum resulting in the additional delta function \cite{Palmai2}. 
\begin{figure}\label{gs1}
 \includegraphics[width=.45\textwidth]{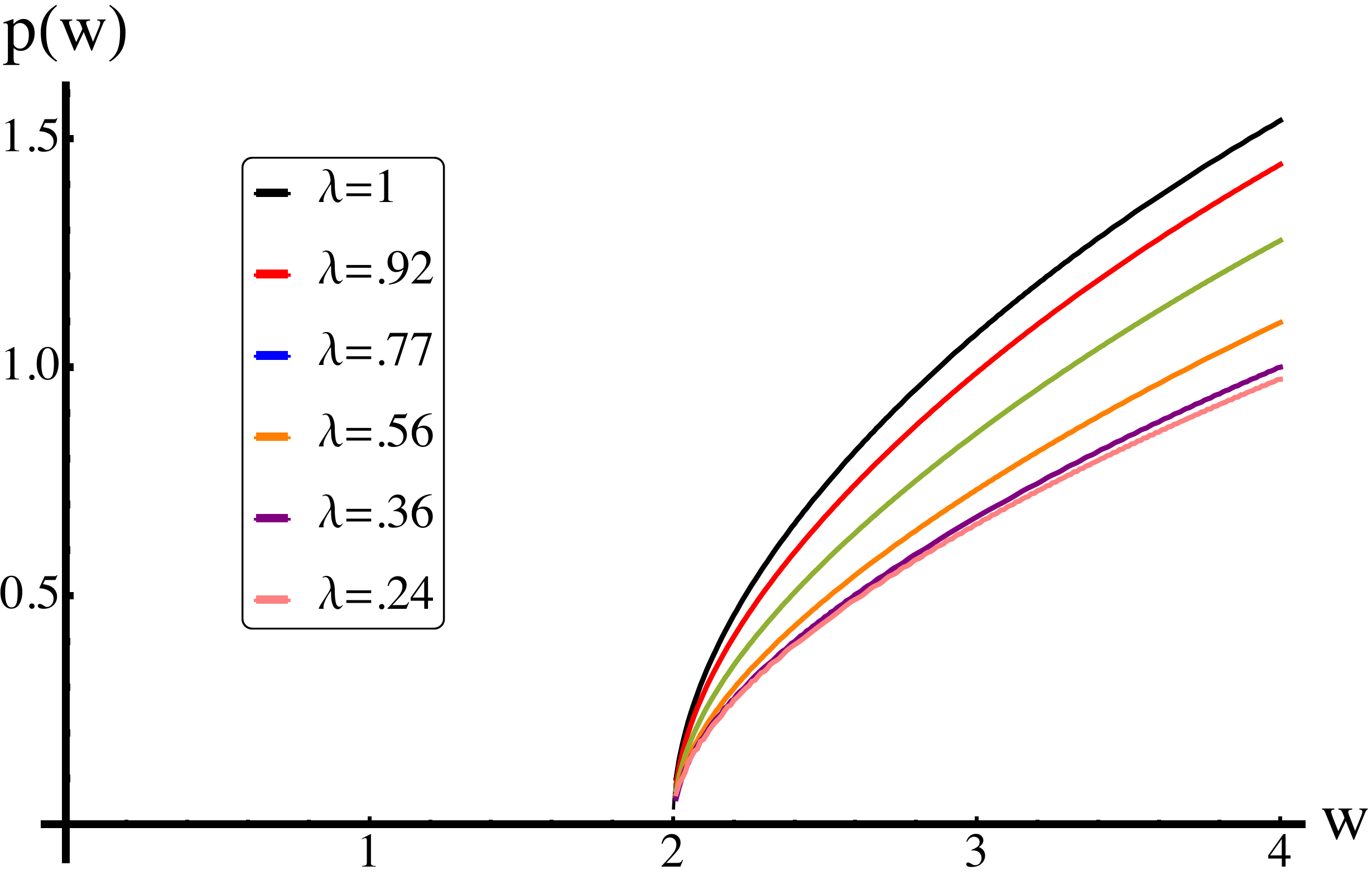}
\caption{Plot of rescaled work distribution function at different values of $\lambda=8\pi/\beta^2-1=\gamma/(\pi-\gamma)$ in the region of the threshold for $\xi=i\pi/2$. The distribution has been rescaled by the fidelity and system size, $p(w)=4\pi P(W)/(mL\mathcal{F})$ and is plotted as a function of work measured from $\delta E$ in units of $m$, $w=(W-\delta E)/m$. }
\end{figure}



\section{Repulsive regime: $\gamma<\pi/2$}
In this section we examine the repulsive case, $\gamma<\pi/2$, wherein the spectrum consists of solitons and anti-solitons but no bound states. Within this regime we choose the  contour to have $\zeta=\gamma/2-\epsilon$, with $\epsilon=0^+$ and also will find it convenient  work with a new auxiliary function
\begin{eqnarray}\nonumber
\mathfrak{y}(u)&=&\log{\mathfrak{a}(u+i\gamma/2-i\epsilon)}+2mit\cosh{\left(\frac{\pi}{\gamma}u\right)}\\&&-\log{\mathcal{K}_\xi(u)}\\\nonumber\bar{\mathfrak{y}}(u)&=&-\log{\mathfrak{a}(u-i\gamma/2+i\epsilon)}-2mit\cosh{\left(\frac{\pi}{\gamma}u\right)}\\&&+\log{\bar{\mathcal{K}}_\xi(u)}
\end{eqnarray} 
where $\mathcal{K}_\xi(u)={\mathbb{K}_\xi(u+i\gamma/2)}$ and $\bar{\mathcal{K}}_\xi(u)={\mathbb{K}^{-1}_\xi(u-i\gamma/2)}$ is the shifted boundary phase shift. In terms of this the NLIE and amplitude are\begin{widetext}
\begin{eqnarray}\nonumber
\mathfrak{y}(u)&=&\int_{-\infty}^\infty\mathrm{d}\mu\,G(u-\mu,\gamma) \log{\left[1+\mathcal{K}_\xi(\mu)e^{\mathfrak{y}(\mu)-2imt\cosh{\left(\frac{\pi}{\gamma}\mu\right)}}\right]}\\\label{NLIEY}&&-\int_{-\infty}^\infty \mathrm{d}\mu\,G(u-\mu+i\gamma-i\epsilon,\gamma) \log{\left[1+\bar{\mathcal{K}}_\xi(\mu)e^{\bar{\mathfrak{y}}(\mu)-2imt\cosh{\left(\frac{\pi}{\gamma}\mu\right)}}\right]}\\
\nonumber
\log{\mathcal{G}(t)}&=&-iE_0t+\log\mathcal{F}+\frac{mL}{4\gamma}\int_{-\infty}^\infty\mathrm{d}\mu \,\cosh{\left(\frac{\pi}{\gamma}\mu\right)}\Bigg\{\log{\left[1+\mathcal{K}_\xi(\mu)e^{\mathfrak{y}(\mu)-2imt\cosh{\left(\frac{\pi}{\gamma}u\right)}}\right]}\\\label{Gy}
&&+\log{\left[1+\bar{\mathcal{K}}_\xi(\mu)e^{\bar{\mathfrak{y}}(\mu)-2imt\cosh{\left(\frac{\pi}{\gamma}\mu\right)}}\right]}\Bigg\}
\end{eqnarray}
\end{widetext}The form of the Loschmidt echo remains similar to that of the free case but there are some important new elements. Aside from the renormalization of the mass $m_0\to m$, rapidity $2\mu\to\pi\mu/\gamma$ and boundary phase shift $K_\xi\to\mathcal{K}_\xi$ we have $\mathfrak{y}(\mu)$ which encodes the interactions in the model.

As with the non-interacting  case  we look to expand the logarithm and exponential functions in \eqref{NLIEY} and \eqref{Gy}. We then employ an approach to the  NLIE which will provide us with an exact answer expression $P(W)$ within the region of interest $0<W<4m$ \cite{Palmai}. This method, which has been used with success in studies of  the Sinh-Gordon model, entails treating the NLIE in an iterative manner. The key is that we are interested in the Fourier transform of the Loschmidt amplitude and each successive iteration of the NLIE will give us the exact answer for this below a certain value of $W$.

Expanding both the logarithm and $\exp{(\mathfrak{y})}$ for the case of $\xi=i\pi/2$ gives
\begin{widetext}
 \begin{eqnarray}\nonumber
\mathfrak{y}(u)&=&\sum_{n=1,l=0}^\infty\frac{(-n)^l}{nl!}\left[\int_{-\infty}^\infty\mathrm{d}\mu\,G(u-\mu,\gamma) \mathcal{K}_{i\frac{\pi}{2}}^n(\mu)\mathfrak{y}^l(\mu)e^{-2imnt\cosh{\left(\frac{\pi}{\gamma}\mu\right)}}\right.\\\label{NLIEyn}&&\left.-\int_{-\infty}^\infty \mathrm{d}\mu\,G(u-\mu+i\gamma-i\epsilon,\gamma) \bar{\mathcal{K}}_{i\frac{\pi}{2}}^n(\mu)\bar{\mathfrak{y}}^l(\mu)e^{-2imnt\cosh{\left(\frac{\pi}{\gamma}\mu\right)}}\right]\\
\nonumber
\log{\mathcal{G}(t)}&=&-iE_0t+\log\mathcal{F}+\frac{mL}{4\gamma}\sum_{n=l,l=0}^\infty\frac{(-n)^l}{nl!}\left[\int_{-\infty}^\infty\mathrm{d}\mu \,e^{\frac{\pi}{\gamma}\mu}\Bigg\{\mathcal{K}_{i\frac{\pi}{2}}^n(\mu)\mathfrak{y}^l(\mu)e^{-2imnt\cosh{\left(\frac{\pi}{\gamma}\mu\right)}}\right.\\\label{Gyn}
&&\left.+\bar{\mathcal{K}}_{i\frac{\pi}{2}}^n(\mu)\bar{\mathfrak{y}}^l(\mu)e^{-2imnt\cosh{\left(\frac{\pi}{\gamma}\mu\right)}}\Bigg\}\right].
\end{eqnarray}
\end{widetext}
The iterative procedure is started  from the non-interacting value,   $\mathfrak{y}_{[0]}=0$. Why this is a good starting value will be explained further below.  Inserting this into \eqref{Gyn} only the $l=0$ term survives and each term in the sum over $n$ comes with a factor of 
 \begin{eqnarray}
 e^{-2imnt\cosh{\frac{\pi}{\gamma}\mu}}.
 \end{eqnarray}
 Thus when  Fourier transformed the $n^{\text{th}}$ term gives a contribution to $P(W)$ for $W>2mn$ as in the non interacting expression. At the next step $\mathfrak{y}_{[1]}(\mu)$ is given by the $l=0$ term in \eqref{NLIEyn},   
 \begin{eqnarray}\nonumber
 \mathfrak{y}_{[1]}(u)&=&\sum_{n=1}^\infty\int_{-\infty}^\infty\frac{\mathrm{d}\mu}{n}e^{-2imnt\cosh{\left(\frac{\pi}{\gamma}\mu\right)}}\left[\,G(u-\mu,\gamma) \mathcal{K}_{i\frac{\pi}{2}}^n(\mu)\right.\\\label{NLIEy1}&&\left.- G(u-\mu+i\gamma-i\epsilon,\gamma) \bar{\mathcal{K}}_{i\frac{\pi}{2}}^n(\mu)\right]
 \end{eqnarray}
 Inserting this into the amplitude we see that only the $n=1,l=0$ term in the amplitude has a factor of $e^{-2imt\cosh{\frac{\pi}{\gamma}\mu}}$ and so only this term contributes in the region $0<W<4m$. Further iterations cannot alter this. Therefore 
 only the $n=1,l=0$ term needs to be retained and the distribution is 
\begin{eqnarray}\nonumber
P(W)
&=&\mathcal{F}\delta\left(W\right)+\frac{\mathcal{F}L}{4\pi}\theta\left(W -2m\right)\frac{W}{\sqrt{W^2/4m^2-1}}\\&&
\times\text{Re}\Big\{\mathcal{K}_{i\frac{\pi}{2}}\left(\frac{\gamma}{\pi}\text{arcosh}\left[\frac{W}{2m}\right]\right)\Big\}+\dots
\end{eqnarray}
where $\delta E$ had been absorbed  into the work,  $ W-\delta E\to W$ and the dots contain terms that have no contribution at $W<4m$. Iterating further gives the exact expression for a larger window. After the $(n-2)^\text{th}$ iteration, i.e, using $\mathfrak{y}_{[n-2]}$  an exact expression for $P(W)$ with $W<2mn$ is obtained. 

The behavior of the work distribution resembles closely \eqref{Pgsnon} and indeed after expanding $\mathcal{K}$ about the edge singularity $W\sim 2m$ we get that the edge exponent is also $1/2$. The negligible effect of interactions on $P(W)$ near the threshold is not entirely unexpected. In this region the quench process is governed by transitions to excited states containing only two quasi-particles of opposite momentum. In the thermodynamic limit the interaction effects between them can be neglected, only for a macroscopic number of excitations will the effect of the interactions be felt.
With this in mind the procedure of iterating from the non interacting value is the same as determining $P(W)$ by starting at $W=0$ and then perturbing away from this.

In order to compare the threshold behaviour for several values of the interaction we rescale the distribution by the system size, mass  and fidelity $p(W)=4\pi P(W)/(mL\mathcal{F})$.  The fidelity can  be measured experimentally as the spectral weight of the peak at $W=0$ and the mass determined by the distance to the threshold.  Furthermore we plot it as a function of $w=W/m$ in order to bring the threshold to the same point. In FIG. 5 we plot the rescaled distribution, $p(w)$ near the threshold for several values of $\lambda=8\pi/\beta^2-1$ we see that effect of the interactions is to suppress the work in this region.

For the other initial state the presence of the pole, i.e. $g_0\neq0$ indicates we should regularize the divergence in \eqref{NLIEy} and \eqref{Gy} prior to performing the expansion. Following the same procedure as before the amplitude is modified to \begin{widetext}
\begin{eqnarray}\nonumber
\log{\mathcal{G}(t)}&=&-iE_0t+\log\mathcal{F}+\frac{mL}{4\gamma}\int_{-\infty}^\infty\mathrm{d}\mu \,e^{\frac{\pi}{\gamma}\mu}\Bigg\{\log{\left[\frac{1+\mathcal{K}_0(\mu)e^{\mathfrak{y}(\mu)-2imt\cosh{\left(\frac{\pi}{\gamma}\mu\right)}}}{1+\frac{g_0^4}{4}\text{csch}^2\left(\frac{\pi}{\gamma}\mu\right)e^{\mathfrak{y}(0)-2imt}}\right]}\\
&&+\log{\left[\frac{1+\bar{\mathcal{K}}_0(\mu)e^{\mathfrak{y}(-\mu)-2imt\cosh{\left(\frac{\pi}{\gamma}\mu\right)}}}{1+\frac{g_0^4}{4}\text{csch}^2\left(\frac{\pi}{\gamma}\mu\right)e^{\mathfrak{y}(0)-2imt}}\right]}\Bigg\}+mL\frac{g_0^2}{2}e^{-imt+\sqrt{\mathfrak{y}(0)}}
\end{eqnarray}
Note that the last term receives a contribution from both the integrals, representing a soliton and anti-soliton contribution. Expanding the log terms as well as using $e^{\sqrt{\mathfrak{y}(0)}}=1+\mathfrak{y}(0)/2$ and retaining only the first terms we get 
 \begin{eqnarray}\nonumber
\log{\mathcal{G}(t)}&=&-iE_0t+\log\mathcal{F}+mL\frac{g_0^2}{2}e^{-imt}+\frac{mL}{4\gamma}\int_{-\infty}^\infty\mathrm{d}\mu \,\cosh{\left(\frac{\pi}{\gamma}\mu\right)}\Bigg\{[\mathcal{K}_0(\mu)+\mathcal{K}_0(-\mu)]e^{-2imt\cosh{\left(\frac{\pi}{\gamma}\mu\right)}}\\
&&-\frac{g_0^4}{2}\text{csch}^2\left(\frac{\pi}{\gamma}\mu\right)e^{-2imt}\Bigg\}+\dots
\end{eqnarray}\end{widetext}
here we have ignored contributions of order $e^{-2imt}$ coming from the regularization of the NLIE. The resulting work distribution is 
\begin{widetext}
\begin{eqnarray}\nonumber
P(W)
&=&\mathcal{F}\delta\left(W\right)+m\mathcal{F}L\frac{g_0^2}{2}\delta(W-m)\\\nonumber
&&+\frac{\mathcal{F}L}{4\pi}\theta\left(W -2m\right)\frac{\text{Re}\Big\{\mathcal{K}_0\left(\frac{\gamma}{\pi}\text{arcosh}\left[\frac{W}{2m}\right]\right)\Big\}}{\sqrt{W^2/4m^2-1}}\\&&+\frac{1}{2}\left[m\mathcal{F}L\frac{g_0^2}{2}\right]^2\delta(W-2m)+\dots
\end{eqnarray}
\end{widetext}
It can be checked that in this case $\mathcal{K}_\xi(x)$ diverges as $1/x$ giving an edge exponent of $-3/2$ exactly like the non-interacting case. The threshold behaviour of the rescaled distribution, $p(w)$ is plotted in FIG. 6 for several values of $\lambda$. We see that as $\lambda\to 1/2$ the distribution becomes more and more peaked. Interestingly at $\lambda=1/2$ the quantity $\mathcal{K}_{0}(\mu)+\mathcal{K}_{0}(-\mu)$ vanishes exactly and we need to go beyond this expansion to find the leading behaviour of the work distribution.


\section{Attractive Regime: $\gamma>\pi/2$}
In this section we examine the more complicated attractive regime of the MTM and consider only the $\xi=i\pi/2$ quench. The  bound states which appear in the spectrum considerably alter the quench dynamics of the system compared to the repulsive case.  We wish to apply the strategy used in the repulsive and non-interacting cases to the attractive and this time will proceed keeping $\zeta$ arbitrary. The new auxiliary function is now defined as
\begin{eqnarray}\nonumber
\mathfrak{y}(u)&=&\log{\mathfrak{a}(u+i\zeta)}+2mt\sinh{\left(\frac{\pi}{\gamma}(u+i\zeta)\right)}\\
&&-\log{\mathbb{K}_\xi(u+i\zeta)}\\\nonumber
\bar{\mathfrak{y}}(u)&=&-\log{\mathfrak{a}(u-i\zeta)}-2mt\sinh{\left(\frac{\pi}{\gamma}(u-i\zeta)\right)}\\&&+\log{\mathbb{K}_\xi(u-i\zeta)}.
\end{eqnarray} 
Rewriting the NLIE and amplitude in terms of $\mathfrak{y}(u)$ and $\bar{\mathfrak{y}}(u)$ and expanding the logarithm and exponential gives
\begin{widetext}
\begin{eqnarray}\nonumber
\mathfrak{y}(u)&=&\sum_{n=1,l=0}^\infty\frac{(-n)^l}{nl!}\int_{-\infty}^\infty\mathrm{d}\mu\,G(u-\mu,\gamma) \mathbb{K}^n_{i\frac{\pi}{2}}(\mu+i\zeta)\mathfrak{y}^l(\mu)e^{-2nmt\sinh{\left(\frac{\pi}{\gamma}(\mu+i\zeta)\right)}}\\\label{NLIEy}&&-\int_{-\infty}^\infty \mathrm{d}\mu\,G(u-\mu+2i\zeta,\gamma) \mathbb{K}^{-n}_{i\frac{\pi}{2}}(\mu-i\zeta)\bar{\mathfrak{y}}^l(\mu)e^{2nmt\sinh{\left(\frac{\pi}{\gamma}(\mu-i\zeta)\right)}}
\end{eqnarray}
\begin{eqnarray}\nonumber
\log{\mathcal{G}(t)}&=&-iE_0t+\log\mathcal{F}-i\frac{mL}{4\gamma}\sum_{n=1,l=0}^\infty\frac{(-n)^l}{nl!}\int_{-\infty}^\infty\mathrm{d}\mu\,e^{\frac{\pi}{\gamma}(\mu+i\zeta)} \mathbb{K}^n_{i\frac{\pi}{2}}(\mu+i\zeta)\mathfrak{y}^l(\mu)e^{-2mnt\sinh{\left(\frac{\pi}{\gamma}(\mu+i\zeta)\right)}}\\\label{Ga}
&&+i\frac{mL}{4\gamma}\sum_{n=1,l=0}^\infty\frac{(-n)^l}{nl!}\int_{-\infty}^\infty \mathrm{d}\mu\,e^{\frac{\pi}{\gamma}(\mu-i\zeta)} \mathbb{K}^{-n}_{i\frac{\pi}{2}}(\mu-i\zeta)\bar{\mathfrak{y}}^l(\mu)e^{2mnt\sinh{\left(\frac{\pi}{\gamma}(\mu-i\zeta)\right)}}
\end{eqnarray}
\end{widetext}
with a similar expression for $\bar{\mathfrak{y}}(u)$. 
If this were the repulsive case we would take $\zeta=\gamma/2$ and then each term in the expansion would give contributions to a certain frequency window in the Fourier transform. As was shown in the last section, keeping only $n=1,l=0$ gives us the exact answer for $P(W)$ in the region  $0<W<4m$ because of the $e^{2imt\cosh{(\pi \mu/\gamma)}}$ term.  In the attractive case  $\zeta$ cannot be taken to be this value as the functions $\mathbb{K}_\xi(u)$ and $G(u+2i\zeta)$ have poles  when Im$(u)>\pi-\gamma$. This signifies the appearance of bound states in the spectrum. In turn this leads to non-analyticites in the auxiliary function $\mathfrak{y}(u)$ and $\bar{\mathfrak{y}}(u)$.  Nevertheless, let us attempt our iterative procedure starting from the non interacting value, $\mathfrak{y}_{[0]}=0$.  The contribution to $P(W<4m)$ would nominally come from the the $n=1,l=0$ term, which is
\begin{eqnarray}\nonumber
i\frac{mL}{4\gamma}\int_{-\infty}^\infty \mathrm{d}\mu\,\left[e^{\frac{\pi}{\gamma}(\mu-i\zeta)} \mathbb{K}^{-1}_{i\frac{\pi}{2}}(\mu-i\zeta)e^{2mt\sinh{\left(\frac{\pi}{\gamma}(\mu-i\zeta)\right)}}\right.\\
-\left.e^{\frac{\pi}{\gamma}(\mu+i\zeta)} \mathbb{K}_{i\frac{\pi}{2}}(\mu+i\zeta)e^{-2mt\sinh{\left(\frac{\pi}{\gamma}(\mu+i\zeta)\right)}}\right]\,.\quad
\end{eqnarray}
The analytic structure of all the functions present here is known and  the contour can be moved up to  $\zeta=\gamma/2-\epsilon$. The cost of doing this  is to pick up the poles of $\mathbb{K}_{i\frac{\pi}{2}}(u)$ along the way. Using the expression given in Appendix C we see the poles occur at $u=i\frac{(\pi-\gamma)}{2}n$
for $n<\gamma/(\pi-\gamma)$ with $n$ even. This term thus becomes 
\begin{eqnarray}\nonumber
\sum_{\text{even}\,n}^{\lfloor \frac{\gamma}{\pi-\gamma}\rfloor} m_{b,n}L\frac{g_{b,n}}{4}e^{-im_{b,n}t}+mL\int_{-\infty}^\infty\frac{\mathrm{d}\mu }{4\gamma}\cosh{\left(\frac{\pi}{\gamma}\mu\right)}\\
\times\left[\mathbb{K}_{i\frac{\pi}{2}}(\mu+i\gamma/2)+\mathbb{K}^{-1}_{i\frac{\pi}{2}}(\mu-i\gamma/2)\right]e^{-2imt\cosh{\left(\frac{\pi}{\gamma}\mu\right)}}
\end{eqnarray}
where $g_{b,2n}$ is the residue of the pole and $m_{b,2n}$ are the masses of the even parity breathers
\begin{eqnarray}
m_{b,n}=2m\sin{\left(n\frac{\pi}{2}\left[\pi-\gamma\right]\right)}.
\end{eqnarray}
The remaining integral importantly has the same form as in the repulsive case \textit{i.e.} has a $-2imt \cosh{(\pi u/\gamma)}$ in the exponential.  This seems to work nicely but there are other terms in the expansion involving $\mathfrak{y}$ which need to be considered in this case. Lets consider the $n=1,l=1$ term.  After the first iteration it is 
\begin{eqnarray}\nonumber
i\frac{mL}{4\gamma}\int_{-\infty}^\infty\mathrm{d}\mu\,\left[e^{\frac{\pi}{\gamma}(\mu+i\zeta)} \mathbb{K}_{i\frac{\pi}{2}}(\mu+i\zeta)\mathfrak{y}_{[1]}(\mu)e^{-2mt\sinh{\left(\frac{\pi}{\gamma}(\mu+i\zeta)\right)}}\right.\\
-\left.e^{\frac{\pi}{\gamma}(\mu-i\zeta)} \mathbb{K}^{-1}_{i\frac{\pi}{2}}(\mu-i\zeta)\bar{\mathfrak{y}}_{[1]}(\mu)e^{2mt\sinh{\left(\frac{\pi}{\gamma}(\mu-i\zeta)\right)}}\right]\,\,\,\,\,\,
\end{eqnarray}
if we try to perform the same trick and move the contour up to $i\gamma/2$ so as to get the desired behaviour in the exponential we will encounter non analyticies in $\mathfrak{y}$ which are unknown. The question is will these have any effect in the region of the threshold. The first order term in the NLIE gives us that

\begin{eqnarray}\nonumber
\mathfrak{y}_{[1]}(u)=\int_{-\infty}^\infty\mathrm{d}\mu\,G(u-\mu,\gamma) \mathbb{K}_{i\frac{\pi}{2}}(\mu+i\zeta)e^{-2mt\sinh{\left(\frac{\pi}{\gamma}(\mu+i\zeta)\right)}}\\\nonumber-\int_{-\infty}^\infty \mathrm{d}\mu\,G(u-\mu+2i\zeta,\gamma) \mathbb{K}^{-1}_{i\frac{\pi}{2}}(\mu-i\zeta)e^{2mt\sinh{\left(\frac{\pi}{\gamma}(\mu-i\zeta)\right)}}\\
&&+\dots
\end{eqnarray}
with a similar term for $\bar{\mathfrak{y}}(u)$. We can move the contour up here which will pick up poles form the boundary shifts. This gives a sum of terms plus an integral with the same form as the repulsive case i.e 
\begin{eqnarray}\nonumber
\mathfrak{y}(u)_{[1]}&=&\int_{-\infty}^\infty\mathrm{d}\mu\,e^{-2imt\cosh{\left(\frac{\pi}{\gamma}\mu\right)}}\left[G(u-\mu,\gamma) \mathbb{K}_{i\frac{\pi}{2}}(\mu+i\gamma/2)\right.\\\nonumber &&\left.-G(u-\mu+2i\gamma-i\epsilon,\gamma) \mathbb{K}^{-1}_{i\frac{\pi}{2}}(\mu-i\gamma/2)\right]\\&&-
\sum_{\text{even} \, n}\alpha_n(u)e^{-im_{b,n}t}+\dots
\end{eqnarray}
 with 
\begin{eqnarray}\nonumber
\alpha_n(u)=\gamma g_{b,n}\left[G\left(u+i\frac{\pi-\gamma}{2}n+i \frac{\gamma}{2}\right)\right.\\\left.
-G\left(u-i\frac{\pi-\gamma}{2}n+i \frac{\gamma}{2}\right)\right]
\end{eqnarray}
and similar for $\bar{\mathfrak{y}}$. 
Collecting this all together we get that the part of the  Loschmidt amplitude pertinent the work distribution in the region $W<4m$. After Fourier transforming we find \begin{widetext}

\begin{eqnarray}\nonumber
P(W)&=&\mathcal{F}\delta\left(W\right)+\sum_{\text{even}\,n}^{\lfloor \frac{\gamma}{\pi-\gamma}\rfloor} m_{b,n}\mathcal{F}L\frac{g_{b,n}}{4}\delta\left(W-m_{b,n}\right)+\frac{\mathcal{F}L}{4\pi}\theta\left(W-2m\right)W\frac{\text{Re}\Big\{\mathcal{K}_{i\frac{\pi}{2}}\left(\frac{\gamma}{\pi}\text{arcosh}\left[\frac{W}{2m}\right]\right)\Big\}}{\sqrt{W^2/4m^2-1}}\\
&&+\sum_{\text{even}\,n}^{\lfloor \frac{\gamma}{\pi-\gamma}\rfloor}\frac{\mathcal{F}L}{4\pi}\theta\left(W-2m-m_{b,n}\right)\frac{\text{Re}\Big\{\alpha_n\left(\frac{\gamma}{\pi}\text{arcosh}\left[\frac{W-m_{b,n}}{2m}\right]\right)\mathcal{K}_{i\frac{\pi}{2}}\left(\frac{\gamma}{\pi}\text{arcosh}\left[\frac{W-m_{b,n}}{2m}\right]\right)\Big\}}{\sqrt{(W-m_{b,n})^2/4m^2-1}}\left(W-m_{b,n}\right)
\end{eqnarray}
\end{widetext}
It is not too difficult to identify the processes involved in each of the terms present here. The first new term is the sum of delta functions at $W=m_{b,n}$ the even parity breather masses. This arises due to transitions to a single breather excitation with zero momentum. Only even parity breathers appear due to the parity of the initial state. The next term, due to the solitons and anti-solitons produces an edge singularity similar to that in the repulsive case with exponent $1/2$. Solitons and anti-solitons appear in the spectrum in both attractive and repulsive regimes and given the negligible effect of the interactions in  this region it should come as no surprise that the same term appear here. The final sum of terms is  also due to the appearance of bound states. They cause new edge singularities to appear within the region $2m<W<4m$ at $W=2m+m_{b.n}$ and correspond to a transition to a state which has two solitons in addition to a zero momentum breather. One can then proceed systematically in this manner and obtain $P(W)$ for larger and larger $W$. 

\section{Conclusion}
\begin{figure}
 \includegraphics[width=.45\textwidth]{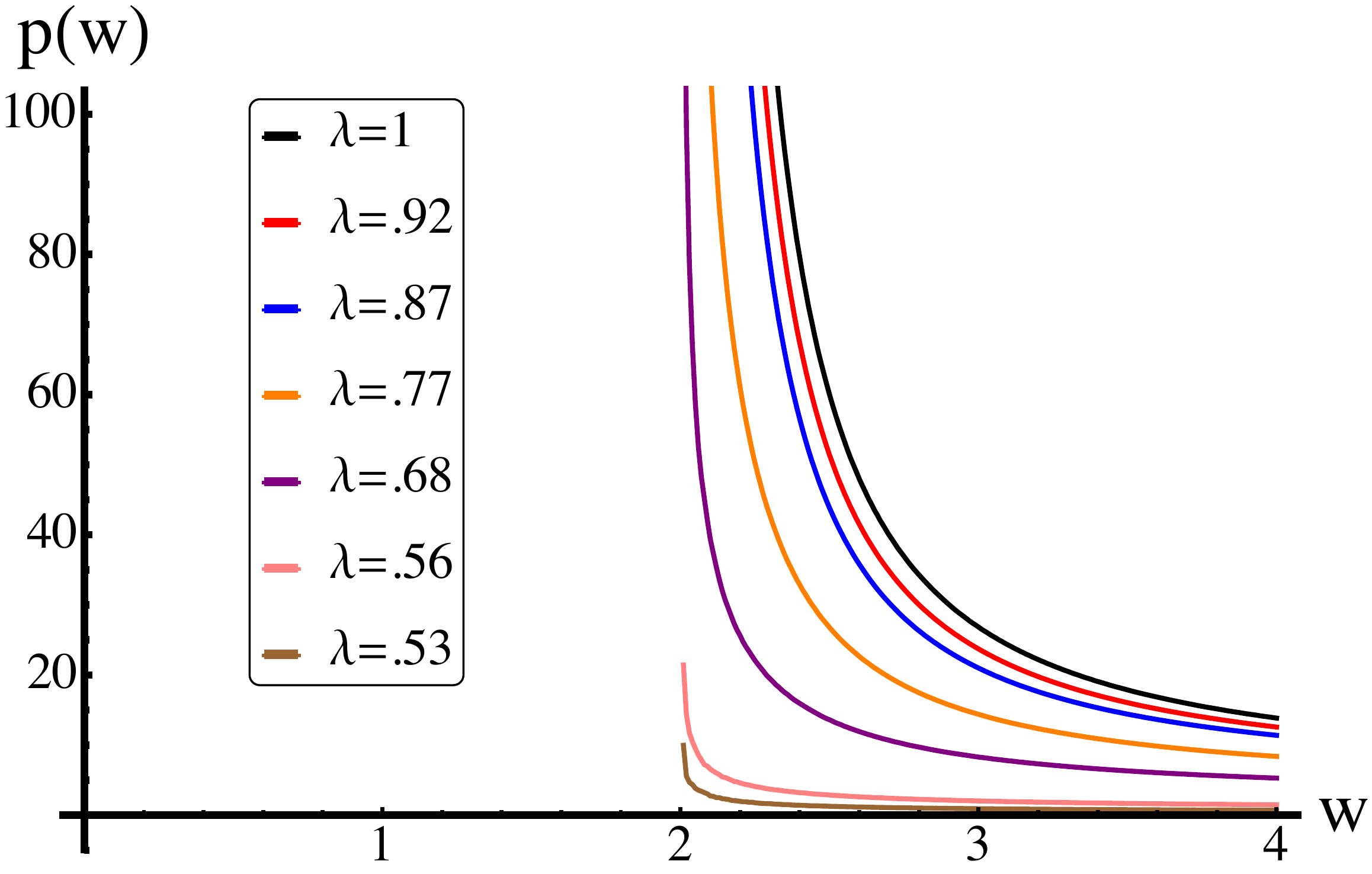}
\caption{Plot of rescaled work distribution function, $p(w)$ at different values of $\lambda=8\pi/\beta^2-1=\gamma/(\pi-\gamma)$ in the region of the threshold for $\xi=0$. The distribution has been rescaled by the fidelity and system size, $p(w)=4\pi P(W)/(mL\mathcal{F})$ and is plotted as a function of work measured from $\delta E$ in units of $m$, $w=(W-\delta E)/m$. We see that the  distribution exhibits an edge singularity with exponent, $-3/2$ and becomes more strongly peaked at the threshold as $\lambda\to1/2$    }
\end{figure}

In this article we have studied the Loschmidt amplitude and work statistics of an interaction quench in the Sine-Gordon/Massive Thirring model. Utilizing a convenient regularizaion of the continuum model we have formulated the amplitude as the partition function of the six vertex model. For a particular class of initial states this can be calculated exactly. The initial states considered includes the ground state and highest excited state of the system with very large initial mass. These quenches can be readily achieved in cold atom systems by varying the optical lattice depth or interaction strength. The solution is given in terms of a single non-linear integral equation which is valid for all parameter regimes and for real time. We perform several non trivial checks on this NLIE including recovering results in the non interacting case and go on to study the work probability distribution.

The work distribution function is studied in the region of $0<W-\delta E<4m$ for the repulsive and then attractive regimes. In the repulsive regime  we see several generic features including the delta function at $W=\delta E$ weighted by the fidelity and an edge singularity at the threshold value of $W=2m+\delta E$ with either exponent of $1/2$ or $-3/2$ depending upon the initial state.  The appearance of the latter exponent is accompanied by an additional delta function peak at $m+\delta E$ signifying the emission of a single particle with zero momentum from the initial state. This changes in the attractive regime due to the appearance of bound states in the spectrum. We find additional delta peaks below the first edge singularity resulting from the transition to a single breather state with zero momentum. In the quench from the ground state only even parity breathers appear. Further edges are also evident within the region considered, these come from transitions to states involving a zero momentum breather in addition to two solitons or anti-solitons. 

We note an interesting duality  relating quenches in the repulsive and attractive  regimes. Under this duality transformation the free point is mapped to itself. The attractive regime of Sine-Gordon model has a much richer structure than the repulsive case and this duality may be useful in calculations in the attractive regime.

We find that dynamical quantum phase transitions, which are defined as the points in time where  the Loschmidt amplitude becomes non-analytic, cannot take place in any parameter regime for the chosen initial states. However,  for other   choices of initial states a DQPT may take place due to  level crossings of $\Lambda_\text{max}$. This would manifest itself in non-analytic points of the NLIE and auxiliary function, $\mathfrak{a}(\mu)$. 
The quantity $\log{[1+\mathfrak{a}(\mu)]}$ has played the role of a non-equilibrium distribution function which contains all information about the quench. Calculation of observables would necessarily involve integrals over this distribution and  any non-analyticities appearing in the NLIE being carried over. The coincidence of non-analytic behavior in the Loschmidt amplitude and observables could therefore both be explained as resulting from the non-equilibrium distribution function. 

Finally we note that in the non-interacting case the Loschmidt amplitude for finite mass  initial states could also be calculated by replacing $\xi$, the parameter describing the initial state, with an appropriate function of the rapidity. A similar strategy was employed in studying such finite mass quenches of the Sinh-Gordon\cite{Sotiriadis, Palmai} model  and it is an interesting question whether it can be adopted for the Sine-Gordon also.

\acknowledgements{CR is supported by the Peter Lindenfeld Fellowship and NA by NSF Grant DMR 1410583. We are grateful to A. Silva for useful discussions}.




\bibliography{mybib}

\begin{thebibliography}{71}%
\makeatletter
\providecommand \@ifxundefined [1]{%
 \@ifx{#1\undefined}
}%
\providecommand \@ifnum [1]{%
 \ifnum #1\expandafter \@firstoftwo
 \else \expandafter \@secondoftwo
 \fi
}%
\providecommand \@ifx [1]{%
 \ifx #1\expandafter \@firstoftwo
 \else \expandafter \@secondoftwo
 \fi
}%
\providecommand \natexlab [1]{#1}%
\providecommand \enquote  [1]{``#1''}%
\providecommand \bibnamefont  [1]{#1}%
\providecommand \bibfnamefont [1]{#1}%
\providecommand \citenamefont [1]{#1}%
\providecommand \href@noop [0]{\@secondoftwo}%
\providecommand \href [0]{\begingroup \@sanitize@url \@href}%
\providecommand \@href[1]{\@@startlink{#1}\@@href}%
\providecommand \@@href[1]{\endgroup#1\@@endlink}%
\providecommand \@sanitize@url [0]{\catcode `\\12\catcode `\$12\catcode
  `\&12\catcode `\#12\catcode `\^12\catcode `\_12\catcode `\%12\relax}%
\providecommand \@@startlink[1]{}%
\providecommand \@@endlink[0]{}%
\providecommand \url  [0]{\begingroup\@sanitize@url \@url }%
\providecommand \@url [1]{\endgroup\@href {#1}{\urlprefix }}%
\providecommand \urlprefix  [0]{URL }%
\providecommand \Eprint [0]{\href }%
\providecommand \doibase [0]{http://dx.doi.org/}%
\providecommand \selectlanguage [0]{\@gobble}%
\providecommand \bibinfo  [0]{\@secondoftwo}%
\providecommand \bibfield  [0]{\@secondoftwo}%
\providecommand \translation [1]{[#1]}%
\providecommand \BibitemOpen [0]{}%
\providecommand \bibitemStop [0]{}%
\providecommand \bibitemNoStop [0]{.\EOS\space}%
\providecommand \EOS [0]{\spacefactor3000\relax}%
\providecommand \BibitemShut  [1]{\csname bibitem#1\endcsname}%
\let\auto@bib@innerbib\@empty
\bibitem [{\citenamefont {Bloch}\ \emph {et~al.}(2008)\citenamefont {Bloch},
  \citenamefont {Dalibard},\ and\ \citenamefont {Zwerger}}]{CA}%
  \BibitemOpen
  \bibfield  {author} {\bibinfo {author} {\bibfnamefont {I.}~\bibnamefont
  {Bloch}}, \bibinfo {author} {\bibfnamefont {J.}~\bibnamefont {Dalibard}}, \
  and\ \bibinfo {author} {\bibfnamefont {W.}~\bibnamefont {Zwerger}},\ }\href
  {\doibase 10.1103/RevModPhys.80.885} {\bibfield  {journal} {\bibinfo
  {journal} {Rev. Mod. Phys.}\ }\textbf {\bibinfo {volume} {80}},\ \bibinfo
  {pages} {885} (\bibinfo {year} {2008})}\BibitemShut {NoStop}%
\bibitem [{\citenamefont {{Polkovnikov}}\ \emph {et~al.}(2011)\citenamefont
  {{Polkovnikov}}, \citenamefont {{Sengupta}}, \citenamefont {{Silva}},\ and\
  \citenamefont {{Vengalattore}}}]{PolRev}%
  \BibitemOpen
  \bibfield  {author} {\bibinfo {author} {\bibfnamefont {A.}~\bibnamefont
  {{Polkovnikov}}}, \bibinfo {author} {\bibfnamefont {K.}~\bibnamefont
  {{Sengupta}}}, \bibinfo {author} {\bibfnamefont {A.}~\bibnamefont {{Silva}}},
  \ and\ \bibinfo {author} {\bibfnamefont {M.}~\bibnamefont {{Vengalattore}}},\
  }\href {\doibase 10.1103/RevModPhys.83.863} {\bibfield  {journal} {\bibinfo
  {journal} {Reviews of Modern Physics}\ }\textbf {\bibinfo {volume} {83}},\
  \bibinfo {pages} {863} (\bibinfo {year} {2011})},\ \Eprint
  {http://arxiv.org/abs/1007.5331} {arXiv:1007.5331 [cond-mat.stat-mech]}
  \BibitemShut {NoStop}%
\bibitem [{\citenamefont {Smacchia}\ and\ \citenamefont
  {Silva}(2013)}]{Smacchia}%
  \BibitemOpen
  \bibfield  {author} {\bibinfo {author} {\bibfnamefont {P.}~\bibnamefont
  {Smacchia}}\ and\ \bibinfo {author} {\bibfnamefont {A.}~\bibnamefont
  {Silva}},\ }\href {\doibase 10.1103/PhysRevE.88.042109} {\bibfield  {journal}
  {\bibinfo  {journal} {Phys. Rev. E}\ }\textbf {\bibinfo {volume} {88}},\
  \bibinfo {pages} {042109} (\bibinfo {year} {2013})}\BibitemShut {NoStop}%
\bibitem [{\citenamefont {{Andrei}}(2016)}]{AndNonEq}%
  \BibitemOpen
  \bibfield  {author} {\bibinfo {author} {\bibfnamefont {N.}~\bibnamefont
  {{Andrei}}},\ }\href@noop {} {\bibfield  {journal} {\bibinfo  {journal}
  {Summer School 2012, Les Houches, France, (Oxford University Press)}\ }
  (\bibinfo {year} {2016})},\ \Eprint {http://arxiv.org/abs/1606.08911}
  {arXiv:1606.08911 [cond-mat.quant-gas]} \BibitemShut {NoStop}%
\bibitem [{\citenamefont {{Mitra}}(2018)}]{MitraRev}%
  \BibitemOpen
  \bibfield  {author} {\bibinfo {author} {\bibfnamefont {A.}~\bibnamefont
  {{Mitra}}},\ }\href {\doibase 10.1146/annurev-conmatphys-031016-025451}
  {\bibfield  {journal} {\bibinfo  {journal} {Annual Review of Condensed Matter
  Physics}\ }\textbf {\bibinfo {volume} {9}},\ \bibinfo {pages} {245} (\bibinfo
  {year} {2018})},\ \Eprint {http://arxiv.org/abs/1703.09740} {arXiv:1703.09740
  [cond-mat.quant-gas]} \BibitemShut {NoStop}%
\bibitem [{\citenamefont {{Caux}}\ and\ \citenamefont
  {{Essler}}(2013)}]{EssCau}%
  \BibitemOpen
  \bibfield  {author} {\bibinfo {author} {\bibfnamefont {J.-S.}\ \bibnamefont
  {{Caux}}}\ and\ \bibinfo {author} {\bibfnamefont {F.~H.~L.}\ \bibnamefont
  {{Essler}}},\ }\href {\doibase 10.1103/PhysRevLett.110.257203} {\bibfield
  {journal} {\bibinfo  {journal} {Physical Review Letters}\ }\textbf {\bibinfo
  {volume} {110}},\ \bibinfo {eid} {257203} (\bibinfo {year} {2013})},\ \Eprint
  {http://arxiv.org/abs/1301.3806} {arXiv:1301.3806 [cond-mat.stat-mech]}
  \BibitemShut {NoStop}%
\bibitem [{\citenamefont {{Chenu}}\ \emph {et~al.}(2018)\citenamefont
  {{Chenu}}, \citenamefont {{Molina-Vilaplana}},\ and\ \citenamefont {{del
  Campo}}}]{delCamp}%
  \BibitemOpen
  \bibfield  {author} {\bibinfo {author} {\bibfnamefont {A.}~\bibnamefont
  {{Chenu}}}, \bibinfo {author} {\bibfnamefont {J.}~\bibnamefont
  {{Molina-Vilaplana}}}, \ and\ \bibinfo {author} {\bibfnamefont
  {A.}~\bibnamefont {{del Campo}}},\ }\href@noop {} {\bibfield  {journal}
  {\bibinfo  {journal} {arXiv e-prints}\ ,\ \bibinfo {eid} {arXiv:1804.09188}}
  (\bibinfo {year} {2018})},\ \Eprint {http://arxiv.org/abs/1804.09188}
  {arXiv:1804.09188 [quant-ph]} \BibitemShut {NoStop}%
\bibitem [{\citenamefont {{Gorin}}\ \emph {et~al.}(2006)\citenamefont
  {{Gorin}}, \citenamefont {{Prosen}}, \citenamefont {{Seligman}},\ and\
  \citenamefont {{{\v Z}nidari{\v c}}}}]{LosChaos}%
  \BibitemOpen
  \bibfield  {author} {\bibinfo {author} {\bibfnamefont {T.}~\bibnamefont
  {{Gorin}}}, \bibinfo {author} {\bibfnamefont {T.}~\bibnamefont {{Prosen}}},
  \bibinfo {author} {\bibfnamefont {T.~H.}\ \bibnamefont {{Seligman}}}, \ and\
  \bibinfo {author} {\bibfnamefont {M.}~\bibnamefont {{{\v Z}nidari{\v c}}}},\
  }\href {\doibase 10.1016/j.physrep.2006.09.003} {\bibfield  {journal}
  {\bibinfo  {journal} {Phys. Rep}\ }\textbf {\bibinfo {volume} {435}},\
  \bibinfo {pages} {33} (\bibinfo {year} {2006})},\ \Eprint
  {http://arxiv.org/abs/quant-ph/0607050} {quant-ph/0607050} \BibitemShut
  {NoStop}%
\bibitem [{\citenamefont {Heyl}\ \emph {et~al.}(2013)\citenamefont {Heyl},
  \citenamefont {Polkovnikov},\ and\ \citenamefont {Kehrein}}]{HeylPolKeh}%
  \BibitemOpen
  \bibfield  {author} {\bibinfo {author} {\bibfnamefont {M.}~\bibnamefont
  {Heyl}}, \bibinfo {author} {\bibfnamefont {A.}~\bibnamefont {Polkovnikov}}, \
  and\ \bibinfo {author} {\bibfnamefont {S.}~\bibnamefont {Kehrein}},\ }\href
  {\doibase 10.1103/PhysRevLett.110.135704} {\bibfield  {journal} {\bibinfo
  {journal} {Phys. Rev. Lett.}\ }\textbf {\bibinfo {volume} {110}},\ \bibinfo
  {pages} {135704} (\bibinfo {year} {2013})}\BibitemShut {NoStop}%
\bibitem [{\citenamefont {{Campisi}}\ \emph {et~al.}(2011)\citenamefont
  {{Campisi}}, \citenamefont {{H{\"a}nggi}},\ and\ \citenamefont
  {{Talkner}}}]{FlucRMP}%
  \BibitemOpen
  \bibfield  {author} {\bibinfo {author} {\bibfnamefont {M.}~\bibnamefont
  {{Campisi}}}, \bibinfo {author} {\bibfnamefont {P.}~\bibnamefont
  {{H{\"a}nggi}}}, \ and\ \bibinfo {author} {\bibfnamefont {P.}~\bibnamefont
  {{Talkner}}},\ }\href {\doibase 10.1103/RevModPhys.83.771} {\bibfield
  {journal} {\bibinfo  {journal} {Reviews of Modern Physics}\ }\textbf
  {\bibinfo {volume} {83}},\ \bibinfo {pages} {771} (\bibinfo {year} {2011})},\
  \Eprint {http://arxiv.org/abs/1012.2268} {arXiv:1012.2268
  [cond-mat.stat-mech]} \BibitemShut {NoStop}%
\bibitem [{\citenamefont {{Goold}}\ \emph {et~al.}(2018)\citenamefont
  {{Goold}}, \citenamefont {{Plastina}}, \citenamefont {{Gambassi}},\ and\
  \citenamefont {{Silva}}}]{Goold}%
  \BibitemOpen
  \bibfield  {author} {\bibinfo {author} {\bibfnamefont {J.}~\bibnamefont
  {{Goold}}}, \bibinfo {author} {\bibfnamefont {F.}~\bibnamefont {{Plastina}}},
  \bibinfo {author} {\bibfnamefont {A.}~\bibnamefont {{Gambassi}}}, \ and\
  \bibinfo {author} {\bibfnamefont {A.}~\bibnamefont {{Silva}}},\ }\href@noop
  {} {\bibfield  {journal} {\bibinfo  {journal} {ArXiv e-prints}\ } (\bibinfo
  {year} {2018})},\ \Eprint {http://arxiv.org/abs/1804.02805} {arXiv:1804.02805
  [quant-ph]} \BibitemShut {NoStop}%
\bibitem [{\citenamefont {{Talkner}}\ \emph {et~al.}(2007)\citenamefont
  {{Talkner}}, \citenamefont {{Lutz}},\ and\ \citenamefont
  {{H{\"a}nggi}}}]{TakLutHan}%
  \BibitemOpen
  \bibfield  {author} {\bibinfo {author} {\bibfnamefont {P.}~\bibnamefont
  {{Talkner}}}, \bibinfo {author} {\bibfnamefont {E.}~\bibnamefont {{Lutz}}}, \
  and\ \bibinfo {author} {\bibfnamefont {P.}~\bibnamefont {{H{\"a}nggi}}},\
  }\href {\doibase 10.1103/PhysRevE.75.050102} {\bibfield  {journal} {\bibinfo
  {journal} {\pre}\ }\textbf {\bibinfo {volume} {75}},\ \bibinfo {eid} {050102}
  (\bibinfo {year} {2007})},\ \Eprint {http://arxiv.org/abs/cond-mat/0703189}
  {cond-mat/0703189} \BibitemShut {NoStop}%
\bibitem [{\citenamefont {Silva}(2008)}]{Silva08}%
  \BibitemOpen
  \bibfield  {author} {\bibinfo {author} {\bibfnamefont {A.}~\bibnamefont
  {Silva}},\ }\href {\doibase 10.1103/PhysRevLett.101.120603} {\bibfield
  {journal} {\bibinfo  {journal} {Phys. Rev. Lett.}\ }\textbf {\bibinfo
  {volume} {101}},\ \bibinfo {pages} {120603} (\bibinfo {year}
  {2008})}\BibitemShut {NoStop}%
\bibitem [{\citenamefont {Mahan}(2000)}]{Mahan}%
  \BibitemOpen
  \bibfield  {author} {\bibinfo {author} {\bibfnamefont {G.}~\bibnamefont
  {Mahan}},\ }\href {https://books.google.com/books?id=xzSgZ4-yyMEC} {\emph
  {\bibinfo {title} {Many-Particle Physics}}},\ Physics of Solids and Liquids\
  (\bibinfo  {publisher} {Springer},\ \bibinfo {year} {2000})\BibitemShut
  {NoStop}%
\bibitem [{\citenamefont {P\'almai}\ and\ \citenamefont
  {Sotiriadis}(2014)}]{Palmai}%
  \BibitemOpen
  \bibfield  {author} {\bibinfo {author} {\bibfnamefont {T.}~\bibnamefont
  {P\'almai}}\ and\ \bibinfo {author} {\bibfnamefont {S.}~\bibnamefont
  {Sotiriadis}},\ }\href {\doibase 10.1103/PhysRevE.90.052102} {\bibfield
  {journal} {\bibinfo  {journal} {Phys. Rev. E}\ }\textbf {\bibinfo {volume}
  {90}},\ \bibinfo {pages} {052102} (\bibinfo {year} {2014})}\BibitemShut
  {NoStop}%
\bibitem [{\citenamefont {{Palmai}}(2015)}]{Palmai2}%
  \BibitemOpen
  \bibfield  {author} {\bibinfo {author} {\bibfnamefont {T.}~\bibnamefont
  {{Palmai}}},\ }\href {\doibase 10.1103/PhysRevB.92.235433} {\bibfield
  {journal} {\bibinfo  {journal} {\prb}\ }\textbf {\bibinfo {volume} {92}},\
  \bibinfo {eid} {235433} (\bibinfo {year} {2015})},\ \Eprint
  {http://arxiv.org/abs/1506.08200} {arXiv:1506.08200 [cond-mat.stat-mech]}
  \BibitemShut {NoStop}%
\bibitem [{\citenamefont {Gritsev}\ \emph
  {et~al.}(2007{\natexlab{a}})\citenamefont {Gritsev}, \citenamefont {Demler},
  \citenamefont {Lukin},\ and\ \citenamefont {Polkovnikov}}]{GritDemPol}%
  \BibitemOpen
  \bibfield  {author} {\bibinfo {author} {\bibfnamefont {V.}~\bibnamefont
  {Gritsev}}, \bibinfo {author} {\bibfnamefont {E.}~\bibnamefont {Demler}},
  \bibinfo {author} {\bibfnamefont {M.}~\bibnamefont {Lukin}}, \ and\ \bibinfo
  {author} {\bibfnamefont {A.}~\bibnamefont {Polkovnikov}},\ }\href {\doibase
  10.1103/PhysRevLett.99.200404} {\bibfield  {journal} {\bibinfo  {journal}
  {Phys. Rev. Lett.}\ }\textbf {\bibinfo {volume} {99}},\ \bibinfo {pages}
  {200404} (\bibinfo {year} {2007}{\natexlab{a}})}\BibitemShut {NoStop}%
\bibitem [{\citenamefont {{Gambassi}}\ and\ \citenamefont
  {{Silva}}(2011)}]{GamSil1}%
  \BibitemOpen
  \bibfield  {author} {\bibinfo {author} {\bibfnamefont {A.}~\bibnamefont
  {{Gambassi}}}\ and\ \bibinfo {author} {\bibfnamefont {A.}~\bibnamefont
  {{Silva}}},\ }\href@noop {} {\bibfield  {journal} {\bibinfo  {journal} {ArXiv
  e-prints}\ } (\bibinfo {year} {2011})},\ \Eprint
  {http://arxiv.org/abs/1106.2671} {arXiv:1106.2671 [cond-mat.stat-mech]}
  \BibitemShut {NoStop}%
\bibitem [{\citenamefont {{Sotiriadis}}\ \emph {et~al.}(2013)\citenamefont
  {{Sotiriadis}}, \citenamefont {{Gambassi}},\ and\ \citenamefont
  {{Silva}}}]{SotGamSil}%
  \BibitemOpen
  \bibfield  {author} {\bibinfo {author} {\bibfnamefont {S.}~\bibnamefont
  {{Sotiriadis}}}, \bibinfo {author} {\bibfnamefont {A.}~\bibnamefont
  {{Gambassi}}}, \ and\ \bibinfo {author} {\bibfnamefont {A.}~\bibnamefont
  {{Silva}}},\ }\href {\doibase 10.1103/PhysRevE.87.052129} {\bibfield
  {journal} {\bibinfo  {journal} {\pre}\ }\textbf {\bibinfo {volume} {87}},\
  \bibinfo {eid} {052129} (\bibinfo {year} {2013})},\ \Eprint
  {http://arxiv.org/abs/1303.0782} {arXiv:1303.0782 [cond-mat.stat-mech]}
  \BibitemShut {NoStop}%
\bibitem [{\citenamefont {{Gambassi}}\ and\ \citenamefont
  {{Silva}}(2012)}]{GamSil2}%
  \BibitemOpen
  \bibfield  {author} {\bibinfo {author} {\bibfnamefont {A.}~\bibnamefont
  {{Gambassi}}}\ and\ \bibinfo {author} {\bibfnamefont {A.}~\bibnamefont
  {{Silva}}},\ }\href {\doibase 10.1103/PhysRevLett.109.250602} {\bibfield
  {journal} {\bibinfo  {journal} {Physical Review Letters}\ }\textbf {\bibinfo
  {volume} {109}},\ \bibinfo {eid} {250602} (\bibinfo {year} {2012})},\ \Eprint
  {http://arxiv.org/abs/1210.3341} {arXiv:1210.3341 [cond-mat.stat-mech]}
  \BibitemShut {NoStop}%
\bibitem [{\citenamefont {Bayat}\ \emph {et~al.}(2016)\citenamefont {Bayat},
  \citenamefont {Apollaro}, \citenamefont {Paganelli}, \citenamefont
  {De~Chiara}, \citenamefont {Johannesson}, \citenamefont {Bose},\ and\
  \citenamefont {Sodano}}]{Chiara1}%
  \BibitemOpen
  \bibfield  {author} {\bibinfo {author} {\bibfnamefont {A.}~\bibnamefont
  {Bayat}}, \bibinfo {author} {\bibfnamefont {T.~J.~G.}\ \bibnamefont
  {Apollaro}}, \bibinfo {author} {\bibfnamefont {S.}~\bibnamefont {Paganelli}},
  \bibinfo {author} {\bibfnamefont {G.}~\bibnamefont {De~Chiara}}, \bibinfo
  {author} {\bibfnamefont {H.}~\bibnamefont {Johannesson}}, \bibinfo {author}
  {\bibfnamefont {S.}~\bibnamefont {Bose}}, \ and\ \bibinfo {author}
  {\bibfnamefont {P.}~\bibnamefont {Sodano}},\ }\href {\doibase
  10.1103/PhysRevB.93.201106} {\bibfield  {journal} {\bibinfo  {journal} {Phys.
  Rev. B}\ }\textbf {\bibinfo {volume} {93}},\ \bibinfo {pages} {201106}
  (\bibinfo {year} {2016})}\BibitemShut {NoStop}%
\bibitem [{\citenamefont {{Lena}}\ \emph {et~al.}(2016)\citenamefont {{Lena}},
  \citenamefont {{Palma}},\ and\ \citenamefont {{De Chiara}}}]{Chiara2}%
  \BibitemOpen
  \bibfield  {author} {\bibinfo {author} {\bibfnamefont {R.~G.}\ \bibnamefont
  {{Lena}}}, \bibinfo {author} {\bibfnamefont {G.~M.}\ \bibnamefont {{Palma}}},
  \ and\ \bibinfo {author} {\bibfnamefont {G.}~\bibnamefont {{De Chiara}}},\
  }\href {\doibase 10.1103/PhysRevA.93.053618} {\bibfield  {journal} {\bibinfo
  {journal} {\pra}\ }\textbf {\bibinfo {volume} {93}},\ \bibinfo {eid} {053618}
  (\bibinfo {year} {2016})},\ \Eprint {http://arxiv.org/abs/1603.05918}
  {arXiv:1603.05918 [quant-ph]} \BibitemShut {NoStop}%
\bibitem [{\citenamefont {Giamarchi}(2003)}]{TG}%
  \BibitemOpen
  \bibfield  {author} {\bibinfo {author} {\bibfnamefont {T.}~\bibnamefont
  {Giamarchi}},\ }\href@noop {} {\emph {\bibinfo {title} {Quantum Physics in
  One Dimension}}},\ International Series of Monographs on Physics\ (\bibinfo
  {publisher} {Clarendon Press},\ \bibinfo {year} {2003})\BibitemShut {NoStop}%
\bibitem [{\citenamefont {Gogolin}\ \emph {et~al.}(2004)\citenamefont
  {Gogolin}, \citenamefont {Nersesyan},\ and\ \citenamefont
  {Tsvelik}}]{gogolin2004bosonization}%
  \BibitemOpen
  \bibfield  {author} {\bibinfo {author} {\bibfnamefont {A.}~\bibnamefont
  {Gogolin}}, \bibinfo {author} {\bibfnamefont {A.}~\bibnamefont {Nersesyan}},
  \ and\ \bibinfo {author} {\bibfnamefont {A.}~\bibnamefont {Tsvelik}},\ }\href
  {https://books.google.com/books?id=BZDfFIpCoaAC} {\emph {\bibinfo {title}
  {Bosonization and Strongly Correlated Systems}}}\ (\bibinfo  {publisher}
  {Cambridge University Press},\ \bibinfo {year} {2004})\BibitemShut {NoStop}%
\bibitem [{\citenamefont {Gritsev}\ \emph
  {et~al.}(2007{\natexlab{b}})\citenamefont {Gritsev}, \citenamefont
  {Polkovnikov},\ and\ \citenamefont {Demler}}]{GritDemPol2}%
  \BibitemOpen
  \bibfield  {author} {\bibinfo {author} {\bibfnamefont {V.}~\bibnamefont
  {Gritsev}}, \bibinfo {author} {\bibfnamefont {A.}~\bibnamefont
  {Polkovnikov}}, \ and\ \bibinfo {author} {\bibfnamefont {E.}~\bibnamefont
  {Demler}},\ }\href {\doibase 10.1103/PhysRevB.75.174511} {\bibfield
  {journal} {\bibinfo  {journal} {Phys. Rev. B}\ }\textbf {\bibinfo {volume}
  {75}},\ \bibinfo {pages} {174511} (\bibinfo {year}
  {2007}{\natexlab{b}})}\BibitemShut {NoStop}%
\bibitem [{\citenamefont {Fisher}\ \emph {et~al.}(1989)\citenamefont {Fisher},
  \citenamefont {Weichman}, \citenamefont {Grinstein},\ and\ \citenamefont
  {Fisher}}]{FishFish}%
  \BibitemOpen
  \bibfield  {author} {\bibinfo {author} {\bibfnamefont {M.~P.~A.}\
  \bibnamefont {Fisher}}, \bibinfo {author} {\bibfnamefont {P.~B.}\
  \bibnamefont {Weichman}}, \bibinfo {author} {\bibfnamefont {G.}~\bibnamefont
  {Grinstein}}, \ and\ \bibinfo {author} {\bibfnamefont {D.~S.}\ \bibnamefont
  {Fisher}},\ }\href {\doibase 10.1103/PhysRevB.40.546} {\bibfield  {journal}
  {\bibinfo  {journal} {Phys. Rev. B}\ }\textbf {\bibinfo {volume} {40}},\
  \bibinfo {pages} {546} (\bibinfo {year} {1989})}\BibitemShut {NoStop}%
\bibitem [{\citenamefont {{Krinner}}\ \emph {et~al.}(2015)\citenamefont
  {{Krinner}}, \citenamefont {{Stadler}}, \citenamefont {{Husmann}},
  \citenamefont {{Brantut}},\ and\ \citenamefont {{Esslinger}}}]{Coldatom}%
  \BibitemOpen
  \bibfield  {author} {\bibinfo {author} {\bibfnamefont {S.}~\bibnamefont
  {{Krinner}}}, \bibinfo {author} {\bibfnamefont {D.}~\bibnamefont
  {{Stadler}}}, \bibinfo {author} {\bibfnamefont {D.}~\bibnamefont
  {{Husmann}}}, \bibinfo {author} {\bibfnamefont {J.-P.}\ \bibnamefont
  {{Brantut}}}, \ and\ \bibinfo {author} {\bibfnamefont {T.}~\bibnamefont
  {{Esslinger}}},\ }\href {\doibase 10.1038/nature14049} {\bibfield  {journal}
  {\bibinfo  {journal} {\nat}\ }\textbf {\bibinfo {volume} {517}},\ \bibinfo
  {pages} {64} (\bibinfo {year} {2015})},\ \Eprint
  {http://arxiv.org/abs/1404.6400} {arXiv:1404.6400 [cond-mat.quant-gas]}
  \BibitemShut {NoStop}%
\bibitem [{\citenamefont {{Bertini}}\ \emph {et~al.}(2014)\citenamefont
  {{Bertini}}, \citenamefont {{Schuricht}},\ and\ \citenamefont
  {{Essler}}}]{Bertini}%
  \BibitemOpen
  \bibfield  {author} {\bibinfo {author} {\bibfnamefont {B.}~\bibnamefont
  {{Bertini}}}, \bibinfo {author} {\bibfnamefont {D.}~\bibnamefont
  {{Schuricht}}}, \ and\ \bibinfo {author} {\bibfnamefont {F.~H.~L.}\
  \bibnamefont {{Essler}}},\ }\href {\doibase 10.1088/1742-5468/2014/10/P10035}
  {\bibfield  {journal} {\bibinfo  {journal} {Journal of Statistical Mechanics:
  Theory and Experiment}\ }\textbf {\bibinfo {volume} {10}},\ \bibinfo {eid}
  {10035} (\bibinfo {year} {2014})},\ \Eprint {http://arxiv.org/abs/1405.4813}
  {arXiv:1405.4813 [cond-mat.stat-mech]} \BibitemShut {NoStop}%
\bibitem [{\citenamefont {Moca}\ \emph {et~al.}(2017)\citenamefont {Moca},
  \citenamefont {Kormos},\ and\ \citenamefont {Zar\'and}}]{Moca}%
  \BibitemOpen
  \bibfield  {author} {\bibinfo {author} {\bibfnamefont {C.~P.}\ \bibnamefont
  {Moca}}, \bibinfo {author} {\bibfnamefont {M.}~\bibnamefont {Kormos}}, \ and\
  \bibinfo {author} {\bibfnamefont {G.}~\bibnamefont {Zar\'and}},\ }\href
  {\doibase 10.1103/PhysRevLett.119.100603} {\bibfield  {journal} {\bibinfo
  {journal} {Phys. Rev. Lett.}\ }\textbf {\bibinfo {volume} {119}},\ \bibinfo
  {pages} {100603} (\bibinfo {year} {2017})}\BibitemShut {NoStop}%
\bibitem [{\citenamefont {Kormos}\ and\ \citenamefont
  {Zar\'and}(2016)}]{Kormos}%
  \BibitemOpen
  \bibfield  {author} {\bibinfo {author} {\bibfnamefont {M.}~\bibnamefont
  {Kormos}}\ and\ \bibinfo {author} {\bibfnamefont {G.}~\bibnamefont
  {Zar\'and}},\ }\href {\doibase 10.1103/PhysRevE.93.062101} {\bibfield
  {journal} {\bibinfo  {journal} {Phys. Rev. E}\ }\textbf {\bibinfo {volume}
  {93}},\ \bibinfo {pages} {062101} (\bibinfo {year} {2016})}\BibitemShut
  {NoStop}%
\bibitem [{\citenamefont {{Cort{\'e}s Cubero}}\ and\ \citenamefont
  {{Schuricht}}(2017)}]{Cubero1}%
  \BibitemOpen
  \bibfield  {author} {\bibinfo {author} {\bibfnamefont {A.}~\bibnamefont
  {{Cort{\'e}s Cubero}}}\ and\ \bibinfo {author} {\bibfnamefont
  {D.}~\bibnamefont {{Schuricht}}},\ }\href {\doibase 10.1088/1742-5468/aa8c2e}
  {\bibfield  {journal} {\bibinfo  {journal} {Journal of Statistical Mechanics:
  Theory and Experiment}\ }\textbf {\bibinfo {volume} {10}},\ \bibinfo {pages}
  {103106} (\bibinfo {year} {2017})},\ \Eprint
  {http://arxiv.org/abs/1707.09218} {arXiv:1707.09218 [cond-mat.stat-mech]}
  \BibitemShut {NoStop}%
\bibitem [{\citenamefont {{Horv{\'a}th}}\ \emph {et~al.}(2018)\citenamefont
  {{Horv{\'a}th}}, \citenamefont {{Kormos}},\ and\ \citenamefont
  {{Tak{\'a}cs}}}]{Horvath}%
  \BibitemOpen
  \bibfield  {author} {\bibinfo {author} {\bibfnamefont {D.~X.}\ \bibnamefont
  {{Horv{\'a}th}}}, \bibinfo {author} {\bibfnamefont {M.}~\bibnamefont
  {{Kormos}}}, \ and\ \bibinfo {author} {\bibfnamefont {G.}~\bibnamefont
  {{Tak{\'a}cs}}},\ }\href@noop {} {\bibfield  {journal} {\bibinfo  {journal}
  {ArXiv e-prints}\ } (\bibinfo {year} {2018})},\ \Eprint
  {http://arxiv.org/abs/1805.08132} {arXiv:1805.08132 [cond-mat.stat-mech]}
  \BibitemShut {NoStop}%
\bibitem [{\citenamefont {Coleman}(1975)}]{sGMTM}%
  \BibitemOpen
  \bibfield  {author} {\bibinfo {author} {\bibfnamefont {S.}~\bibnamefont
  {Coleman}},\ }\href {\doibase 10.1103/PhysRevD.11.2088} {\bibfield  {journal}
  {\bibinfo  {journal} {Phys. Rev. D}\ }\textbf {\bibinfo {volume} {11}},\
  \bibinfo {pages} {2088} (\bibinfo {year} {1975})}\BibitemShut {NoStop}%
\bibitem [{\citenamefont {Zamolodchikov}\ and\ \citenamefont
  {Zamolodchikov}()}]{ZamZam}%
  \BibitemOpen
  \bibfield  {author} {\bibinfo {author} {\bibfnamefont {A.~B.}\ \bibnamefont
  {Zamolodchikov}}\ and\ \bibinfo {author} {\bibfnamefont {A.~B.}\ \bibnamefont
  {Zamolodchikov}},\ }\enquote {\bibinfo {title} {Factorized s-matrices in two
  dimensions as the exact solutions of certain relativistic quantum field
  theory models},}\ in\ \href {\doibase 10.1142/9789812798336_0005} {\emph
  {\bibinfo {booktitle} {Yang-Baxter Equation in Integrable Systems}}},\ pp.\
  \bibinfo {pages} {82--120}\BibitemShut {NoStop}%
\bibitem [{\citenamefont {Bergknoff}\ and\ \citenamefont {Thacker}(1979)}]{BT}%
  \BibitemOpen
  \bibfield  {author} {\bibinfo {author} {\bibfnamefont {H.}~\bibnamefont
  {Bergknoff}}\ and\ \bibinfo {author} {\bibfnamefont {H.~B.}\ \bibnamefont
  {Thacker}},\ }\href {\doibase 10.1103/PhysRevD.19.3666} {\bibfield  {journal}
  {\bibinfo  {journal} {Phys. Rev. D}\ }\textbf {\bibinfo {volume} {19}},\
  \bibinfo {pages} {3666} (\bibinfo {year} {1979})}\BibitemShut {NoStop}%
\bibitem [{\citenamefont {Thacker}(1981)}]{Thacker}%
  \BibitemOpen
  \bibfield  {author} {\bibinfo {author} {\bibfnamefont {H.~B.}\ \bibnamefont
  {Thacker}},\ }\href {\doibase 10.1103/RevModPhys.53.253} {\bibfield
  {journal} {\bibinfo  {journal} {Rev. Mod. Phys.}\ }\textbf {\bibinfo {volume}
  {53}},\ \bibinfo {pages} {253} (\bibinfo {year} {1981})}\BibitemShut
  {NoStop}%
\bibitem [{\citenamefont {Fowler}\ and\ \citenamefont {Zotos}(1982)}]{Fowler}%
  \BibitemOpen
  \bibfield  {author} {\bibinfo {author} {\bibfnamefont {M.}~\bibnamefont
  {Fowler}}\ and\ \bibinfo {author} {\bibfnamefont {X.}~\bibnamefont {Zotos}},\
  }\href {\doibase 10.1103/PhysRevB.25.5806} {\bibfield  {journal} {\bibinfo
  {journal} {Phys. Rev. B}\ }\textbf {\bibinfo {volume} {25}},\ \bibinfo
  {pages} {5806} (\bibinfo {year} {1982})}\BibitemShut {NoStop}%
\bibitem [{\citenamefont {{Korepin}}\ \emph {et~al.}(1993)\citenamefont
  {{Korepin}}, \citenamefont {{Bogoliubov}},\ and\ \citenamefont
  {{Izergin}}}]{Korepin}%
  \BibitemOpen
  \bibfield  {author} {\bibinfo {author} {\bibfnamefont {V.~E.}\ \bibnamefont
  {{Korepin}}}, \bibinfo {author} {\bibfnamefont {N.~M.}\ \bibnamefont
  {{Bogoliubov}}}, \ and\ \bibinfo {author} {\bibfnamefont {A.~G.}\
  \bibnamefont {{Izergin}}},\ }\href@noop {} {\emph {\bibinfo {title} {Quantum
  Inverse Scattering Method and Correlation Functions, by V.~E.~Korepin and
  N.~M.~Bogoliubov and A.~G.~Izergin, pp.~575.~ISBN 0521373204.~Cambridge, UK:
  Cambridge University Press, August 1993.}}}\ (\bibinfo {year} {1993})\ p.\
  \bibinfo {pages} {575}\BibitemShut {NoStop}%
\bibitem [{\citenamefont {Korepin}(1980)}]{Korepinmtm}%
  \BibitemOpen
  \bibfield  {author} {\bibinfo {author} {\bibfnamefont {V.~E.}\ \bibnamefont
  {Korepin}},\ }\href {\doibase 10.1007/BF01212824} {\bibfield  {journal}
  {\bibinfo  {journal} {Communications in Mathematical Physics}\ }\textbf
  {\bibinfo {volume} {76}},\ \bibinfo {pages} {165} (\bibinfo {year}
  {1980})}\BibitemShut {NoStop}%
\bibitem [{\citenamefont {{Destri}}\ and\ \citenamefont {{de
  Vega}}(1987)}]{Lightcone}%
  \BibitemOpen
  \bibfield  {author} {\bibinfo {author} {\bibfnamefont {C.}~\bibnamefont
  {{Destri}}}\ and\ \bibinfo {author} {\bibfnamefont {H.~J.}\ \bibnamefont {{de
  Vega}}},\ }\href@noop {} {\bibfield  {journal} {\bibinfo  {journal} {Nuclear
  Physics B}\ }\textbf {\bibinfo {volume} {290}},\ \bibinfo {pages} {168}
  (\bibinfo {year} {1987})}\BibitemShut {NoStop}%
\bibitem [{\citenamefont {{Pozsgay}}(2013)}]{poz1}%
  \BibitemOpen
  \bibfield  {author} {\bibinfo {author} {\bibfnamefont {B.}~\bibnamefont
  {{Pozsgay}}},\ }\href {\doibase 10.1088/1742-5468/2013/10/P10028} {\bibfield
  {journal} {\bibinfo  {journal} {Journal of Statistical Mechanics: Theory and
  Experiment}\ }\textbf {\bibinfo {volume} {10}},\ \bibinfo {eid} {10028}
  (\bibinfo {year} {2013})},\ \Eprint {http://arxiv.org/abs/1308.3087}
  {arXiv:1308.3087 [cond-mat.stat-mech]} \BibitemShut {NoStop}%
\bibitem [{\citenamefont {{Piroli}}\ \emph {et~al.}(2017)\citenamefont
  {{Piroli}}, \citenamefont {{Pozsgay}},\ and\ \citenamefont
  {{Vernier}}}]{Poz2}%
  \BibitemOpen
  \bibfield  {author} {\bibinfo {author} {\bibfnamefont {L.}~\bibnamefont
  {{Piroli}}}, \bibinfo {author} {\bibfnamefont {B.}~\bibnamefont {{Pozsgay}}},
  \ and\ \bibinfo {author} {\bibfnamefont {E.}~\bibnamefont {{Vernier}}},\
  }\href {\doibase 10.1088/1742-5468/aa5d1e} {\bibfield  {journal} {\bibinfo
  {journal} {Journal of Statistical Mechanics: Theory and Experiment}\ }\textbf
  {\bibinfo {volume} {2}},\ \bibinfo {pages} {023106} (\bibinfo {year}
  {2017})},\ \Eprint {http://arxiv.org/abs/1611.06126} {arXiv:1611.06126
  [cond-mat.stat-mech]} \BibitemShut {NoStop}%
\bibitem [{\citenamefont {{Piroli}}\ \emph {et~al.}(2018)\citenamefont
  {{Piroli}}, \citenamefont {{Pozsgay}},\ and\ \citenamefont
  {{Vernier}}}]{poz3}%
  \BibitemOpen
  \bibfield  {author} {\bibinfo {author} {\bibfnamefont {L.}~\bibnamefont
  {{Piroli}}}, \bibinfo {author} {\bibfnamefont {B.}~\bibnamefont {{Pozsgay}}},
  \ and\ \bibinfo {author} {\bibfnamefont {E.}~\bibnamefont {{Vernier}}},\
  }\href@noop {} {\bibfield  {journal} {\bibinfo  {journal} {ArXiv e-prints}\ }
  (\bibinfo {year} {2018})},\ \Eprint {http://arxiv.org/abs/1803.04380}
  {arXiv:1803.04380 [cond-mat.stat-mech]} \BibitemShut {NoStop}%
\bibitem [{\citenamefont {Destri}\ and\ \citenamefont
  {de~Vega}(1989)}]{Lightcone2}%
  \BibitemOpen
  \bibfield  {author} {\bibinfo {author} {\bibfnamefont {C.}~\bibnamefont
  {Destri}}\ and\ \bibinfo {author} {\bibfnamefont {H.~J.}\ \bibnamefont
  {de~Vega}},\ }\href {\doibase 10.1088/0305-4470/22/9/022} {\bibfield
  {journal} {\bibinfo  {journal} {J. Phys.}\ }\textbf {\bibinfo {volume}
  {A22}},\ \bibinfo {pages} {1329} (\bibinfo {year} {1989})}\BibitemShut
  {NoStop}%
\bibitem [{\citenamefont {Destri}\ and\ \citenamefont
  {de~Vega}(1992{\natexlab{a}})}]{Lightcone3}%
  \BibitemOpen
  \bibfield  {author} {\bibinfo {author} {\bibfnamefont {C.}~\bibnamefont
  {Destri}}\ and\ \bibinfo {author} {\bibfnamefont {H.~J.}\ \bibnamefont
  {de~Vega}},\ }\href {\doibase 10.1016/0550-3213(92)90405-Z} {\bibfield
  {journal} {\bibinfo  {journal} {Nucl. Phys.}\ }\textbf {\bibinfo {volume}
  {B374}},\ \bibinfo {pages} {692} (\bibinfo {year}
  {1992}{\natexlab{a}})}\BibitemShut {NoStop}%
\bibitem [{\citenamefont {{Vernier}}\ and\ \citenamefont {{Cort{\'e}s
  Cubero}}(2017)}]{Vernier}%
  \BibitemOpen
  \bibfield  {author} {\bibinfo {author} {\bibfnamefont {E.}~\bibnamefont
  {{Vernier}}}\ and\ \bibinfo {author} {\bibfnamefont {A.}~\bibnamefont
  {{Cort{\'e}s Cubero}}},\ }\href {\doibase 10.1088/1742-5468/aa5288}
  {\bibfield  {journal} {\bibinfo  {journal} {Journal of Statistical Mechanics:
  Theory and Experiment}\ }\textbf {\bibinfo {volume} {2}},\ \bibinfo {pages}
  {023101} (\bibinfo {year} {2017})},\ \Eprint
  {http://arxiv.org/abs/1609.03220} {arXiv:1609.03220 [cond-mat.stat-mech]}
  \BibitemShut {NoStop}%
\bibitem [{\citenamefont {Baxter}()}]{Baxter}%
  \BibitemOpen
  \bibfield  {author} {\bibinfo {author} {\bibfnamefont {R.~J.}\ \bibnamefont
  {Baxter}},\ }\enquote {\bibinfo {title} {Exactly solved models in statistical
  mechanics},}\ in\ \href {\doibase 10.1142/9789814415255_0002} {\emph
  {\bibinfo {booktitle} {Integrable Systems in Statistical
  Mechanics}}}\BibitemShut {NoStop}%
\bibitem [{\citenamefont {Wang}\ \emph {et~al.}(2015)\citenamefont {Wang},
  \citenamefont {Yang}, \citenamefont {Cao},\ and\ \citenamefont {Shi}}]{ODBA}%
  \BibitemOpen
  \bibfield  {author} {\bibinfo {author} {\bibfnamefont {Y.}~\bibnamefont
  {Wang}}, \bibinfo {author} {\bibfnamefont {W.-L.}\ \bibnamefont {Yang}},
  \bibinfo {author} {\bibfnamefont {J.}~\bibnamefont {Cao}}, \ and\ \bibinfo
  {author} {\bibfnamefont {K.}~\bibnamefont {Shi}},\ }\href@noop {} {\emph
  {\bibinfo {title} {{Off-diagonal Bethe ansatz for exactly solvable
  models}}}}\ (\bibinfo  {publisher} {Springer},\ \bibinfo {address} {Berlin},\
  \bibinfo {year} {2015})\BibitemShut {NoStop}%
\bibitem [{Note1()}]{Note1}%
  \BibitemOpen
  \bibinfo {note} {A relevant figure of merit is the average energy density
  $\epsilon _\protect \text {av}(t)=\protect \ensuremath {\left < \Phi _i(t)
  \right |}\protect \mathaccentV {bar}016{H}\protect \ensuremath {\left | \Phi
  _i(t) \right >}/L$ which should not be too large in order to invoke the
  Sine-Gordon description.}\BibitemShut {Stop}%
\bibitem [{\citenamefont {{Andraschko}}\ and\ \citenamefont
  {{Sirker}}(2014)}]{AndSirk}%
  \BibitemOpen
  \bibfield  {author} {\bibinfo {author} {\bibfnamefont {F.}~\bibnamefont
  {{Andraschko}}}\ and\ \bibinfo {author} {\bibfnamefont {J.}~\bibnamefont
  {{Sirker}}},\ }\href {\doibase 10.1103/PhysRevB.89.125120} {\bibfield
  {journal} {\bibinfo  {journal} {\prb}\ }\textbf {\bibinfo {volume} {89}},\
  \bibinfo {eid} {125120} (\bibinfo {year} {2014})},\ \Eprint
  {http://arxiv.org/abs/1312.4165} {arXiv:1312.4165 [cond-mat.str-el]}
  \BibitemShut {NoStop}%
\bibitem [{Note2()}]{Note2}%
  \BibitemOpen
  \bibinfo {note} {We could have added a chemical potential term $\DOTSI \intop
  \ilimits@ \mu \left [\psi ^\protect \dag _+\psi _++\psi ^\protect \dag _-\psi
  _-\right ]$ to the MTM Hamiltonian and maintain integrability. Such a term
  introduces twisted boundary conditions to the lattice problem and comes at
  the cost on introducing a second coupled NLIE\cite {DDV1}}\BibitemShut
  {NoStop}%
\bibitem [{\citenamefont {{Fendley}}\ and\ \citenamefont
  {{Saleur}}(1994)}]{BdrySaleur}%
  \BibitemOpen
  \bibfield  {author} {\bibinfo {author} {\bibfnamefont {P.}~\bibnamefont
  {{Fendley}}}\ and\ \bibinfo {author} {\bibfnamefont {H.}~\bibnamefont
  {{Saleur}}},\ }\href {\doibase 10.1016/0550-3213(94)90369-7} {\bibfield
  {journal} {\bibinfo  {journal} {Nuclear Physics B}\ }\textbf {\bibinfo
  {volume} {428}},\ \bibinfo {pages} {681} (\bibinfo {year} {1994})},\ \Eprint
  {http://arxiv.org/abs/hep-th/9402045} {hep-th/9402045} \BibitemShut {NoStop}%
\bibitem [{\citenamefont {{Ghoshal}}\ and\ \citenamefont
  {{Zamolodchikov}}(1994)}]{Ghoshal}%
  \BibitemOpen
  \bibfield  {author} {\bibinfo {author} {\bibfnamefont {S.}~\bibnamefont
  {{Ghoshal}}}\ and\ \bibinfo {author} {\bibfnamefont {A.}~\bibnamefont
  {{Zamolodchikov}}},\ }\href {\doibase 10.1142/S0217751X94001552} {\bibfield
  {journal} {\bibinfo  {journal} {International Journal of Modern Physics A}\
  }\textbf {\bibinfo {volume} {9}},\ \bibinfo {pages} {3841} (\bibinfo {year}
  {1994})},\ \Eprint {http://arxiv.org/abs/hep-th/9306002} {hep-th/9306002}
  \BibitemShut {NoStop}%
\bibitem [{\citenamefont {{Bajnok}}\ \emph {et~al.}(2005)\citenamefont
  {{Bajnok}}, \citenamefont {{Palla}},\ and\ \citenamefont
  {{Tak{\'a}cs}}}]{Bajnok}%
  \BibitemOpen
  \bibfield  {author} {\bibinfo {author} {\bibfnamefont {Z.}~\bibnamefont
  {{Bajnok}}}, \bibinfo {author} {\bibfnamefont {L.}~\bibnamefont {{Palla}}}, \
  and\ \bibinfo {author} {\bibfnamefont {G.}~\bibnamefont {{Tak{\'a}cs}}},\
  }\href {\doibase 10.1016/j.nuclphysb.2005.03.021} {\bibfield  {journal}
  {\bibinfo  {journal} {Nuclear Physics B}\ }\textbf {\bibinfo {volume}
  {716}},\ \bibinfo {pages} {519} (\bibinfo {year} {2005})},\ \Eprint
  {http://arxiv.org/abs/hep-th/0412192} {hep-th/0412192} \BibitemShut {NoStop}%
\bibitem [{\citenamefont {{LeClair}}\ \emph {et~al.}(1995)\citenamefont
  {{LeClair}}, \citenamefont {{Mussardo}}, \citenamefont {{Saleur}},\ and\
  \citenamefont {{Skorik}}}]{LMSS}%
  \BibitemOpen
  \bibfield  {author} {\bibinfo {author} {\bibfnamefont {A.}~\bibnamefont
  {{LeClair}}}, \bibinfo {author} {\bibfnamefont {G.}~\bibnamefont
  {{Mussardo}}}, \bibinfo {author} {\bibfnamefont {H.}~\bibnamefont
  {{Saleur}}}, \ and\ \bibinfo {author} {\bibfnamefont {S.}~\bibnamefont
  {{Skorik}}},\ }\href {\doibase 10.1016/0550-3213(95)00435-U} {\bibfield
  {journal} {\bibinfo  {journal} {Nuclear Physics B}\ }\textbf {\bibinfo
  {volume} {453}},\ \bibinfo {pages} {581} (\bibinfo {year} {1995})},\ \Eprint
  {http://arxiv.org/abs/hep-th/9503227} {hep-th/9503227} \BibitemShut {NoStop}%
\bibitem [{\citenamefont {{Schiller}}\ and\ \citenamefont
  {{Andrei}}(2007)}]{schill}%
  \BibitemOpen
  \bibfield  {author} {\bibinfo {author} {\bibfnamefont {A.}~\bibnamefont
  {{Schiller}}}\ and\ \bibinfo {author} {\bibfnamefont {N.}~\bibnamefont
  {{Andrei}}},\ }\href@noop {} {\bibfield  {journal} {\bibinfo  {journal}
  {ArXiv e-prints}\ } (\bibinfo {year} {2007})},\ \Eprint
  {http://arxiv.org/abs/0710.0249} {arXiv:0710.0249 [cond-mat.str-el]}
  \BibitemShut {NoStop}%
\bibitem [{\citenamefont {{Fendley}}\ and\ \citenamefont
  {{Saleur}}(1998)}]{SalFen}%
  \BibitemOpen
  \bibfield  {author} {\bibinfo {author} {\bibfnamefont {P.}~\bibnamefont
  {{Fendley}}}\ and\ \bibinfo {author} {\bibfnamefont {H.}~\bibnamefont
  {{Saleur}}},\ }\href {\doibase 10.1103/PhysRevLett.81.2518} {\bibfield
  {journal} {\bibinfo  {journal} {Physical Review Letters}\ }\textbf {\bibinfo
  {volume} {81}},\ \bibinfo {pages} {2518} (\bibinfo {year} {1998})},\ \Eprint
  {http://arxiv.org/abs/cond-mat/9804173} {cond-mat/9804173} \BibitemShut
  {NoStop}%
\bibitem [{\citenamefont {Calabrese}\ and\ \citenamefont
  {Cardy}(2006)}]{CardCal}%
  \BibitemOpen
  \bibfield  {author} {\bibinfo {author} {\bibfnamefont {P.}~\bibnamefont
  {Calabrese}}\ and\ \bibinfo {author} {\bibfnamefont {J.}~\bibnamefont
  {Cardy}},\ }\href {\doibase 10.1103/PhysRevLett.96.136801} {\bibfield
  {journal} {\bibinfo  {journal} {Phys. Rev. Lett.}\ }\textbf {\bibinfo
  {volume} {96}},\ \bibinfo {pages} {136801} (\bibinfo {year}
  {2006})}\BibitemShut {NoStop}%
\bibitem [{\citenamefont {Jurcevic}\ \emph {et~al.}(2017)\citenamefont
  {Jurcevic}, \citenamefont {Shen}, \citenamefont {Hauke}, \citenamefont
  {Maier}, \citenamefont {Brydges}, \citenamefont {Hempel}, \citenamefont
  {Lanyon}, \citenamefont {Heyl}, \citenamefont {Blatt},\ and\ \citenamefont
  {Roos}}]{dqptexp}%
  \BibitemOpen
  \bibfield  {author} {\bibinfo {author} {\bibfnamefont {P.}~\bibnamefont
  {Jurcevic}}, \bibinfo {author} {\bibfnamefont {H.}~\bibnamefont {Shen}},
  \bibinfo {author} {\bibfnamefont {P.}~\bibnamefont {Hauke}}, \bibinfo
  {author} {\bibfnamefont {C.}~\bibnamefont {Maier}}, \bibinfo {author}
  {\bibfnamefont {T.}~\bibnamefont {Brydges}}, \bibinfo {author} {\bibfnamefont
  {C.}~\bibnamefont {Hempel}}, \bibinfo {author} {\bibfnamefont {B.~P.}\
  \bibnamefont {Lanyon}}, \bibinfo {author} {\bibfnamefont {M.}~\bibnamefont
  {Heyl}}, \bibinfo {author} {\bibfnamefont {R.}~\bibnamefont {Blatt}}, \ and\
  \bibinfo {author} {\bibfnamefont {C.~F.}\ \bibnamefont {Roos}},\ }\href
  {\doibase 10.1103/PhysRevLett.119.080501} {\bibfield  {journal} {\bibinfo
  {journal} {Phys. Rev. Lett.}\ }\textbf {\bibinfo {volume} {119}},\ \bibinfo
  {pages} {080501} (\bibinfo {year} {2017})}\BibitemShut {NoStop}%
\bibitem [{\citenamefont {{Heyl}}(2018)}]{Heylreview}%
  \BibitemOpen
  \bibfield  {author} {\bibinfo {author} {\bibfnamefont {M.}~\bibnamefont
  {{Heyl}}},\ }\href {\doibase 10.1088/1361-6633/aaaf9a} {\bibfield  {journal}
  {\bibinfo  {journal} {Reports on Progress in Physics}\ }\textbf {\bibinfo
  {volume} {81}},\ \bibinfo {eid} {054001} (\bibinfo {year} {2018})},\ \Eprint
  {http://arxiv.org/abs/1709.07461} {arXiv:1709.07461 [cond-mat.stat-mech]}
  \BibitemShut {NoStop}%
\bibitem [{\citenamefont {Karrasch}\ and\ \citenamefont
  {Schuricht}(2013)}]{KarrSchur}%
  \BibitemOpen
  \bibfield  {author} {\bibinfo {author} {\bibfnamefont {C.}~\bibnamefont
  {Karrasch}}\ and\ \bibinfo {author} {\bibfnamefont {D.}~\bibnamefont
  {Schuricht}},\ }\href {\doibase 10.1103/PhysRevB.87.195104} {\bibfield
  {journal} {\bibinfo  {journal} {Phys. Rev. B}\ }\textbf {\bibinfo {volume}
  {87}},\ \bibinfo {pages} {195104} (\bibinfo {year} {2013})}\BibitemShut
  {NoStop}%
\bibitem [{\citenamefont {Kennes}\ \emph {et~al.}(2018)\citenamefont {Kennes},
  \citenamefont {Schuricht},\ and\ \citenamefont {Karrasch}}]{KenSchurKarra}%
  \BibitemOpen
  \bibfield  {author} {\bibinfo {author} {\bibfnamefont {D.~M.}\ \bibnamefont
  {Kennes}}, \bibinfo {author} {\bibfnamefont {D.}~\bibnamefont {Schuricht}}, \
  and\ \bibinfo {author} {\bibfnamefont {C.}~\bibnamefont {Karrasch}},\ }\href
  {\doibase 10.1103/PhysRevB.97.184302} {\bibfield  {journal} {\bibinfo
  {journal} {Phys. Rev. B}\ }\textbf {\bibinfo {volume} {97}},\ \bibinfo
  {pages} {184302} (\bibinfo {year} {2018})}\BibitemShut {NoStop}%
\bibitem [{\citenamefont {{Vajna}}\ and\ \citenamefont
  {{D{\'o}ra}}(2014)}]{VajBal}%
  \BibitemOpen
  \bibfield  {author} {\bibinfo {author} {\bibfnamefont {S.}~\bibnamefont
  {{Vajna}}}\ and\ \bibinfo {author} {\bibfnamefont {B.}~\bibnamefont
  {{D{\'o}ra}}},\ }\href {\doibase 10.1103/PhysRevB.89.161105} {\bibfield
  {journal} {\bibinfo  {journal} {\prb}\ }\textbf {\bibinfo {volume} {89}},\
  \bibinfo {eid} {161105} (\bibinfo {year} {2014})},\ \Eprint
  {http://arxiv.org/abs/1401.2865} {arXiv:1401.2865 [cond-mat.str-el]}
  \BibitemShut {NoStop}%
\bibitem [{\citenamefont {{Kormos}}\ and\ \citenamefont
  {{Pozsgay}}(2010)}]{KorPoz}%
  \BibitemOpen
  \bibfield  {author} {\bibinfo {author} {\bibfnamefont {M.}~\bibnamefont
  {{Kormos}}}\ and\ \bibinfo {author} {\bibfnamefont {B.}~\bibnamefont
  {{Pozsgay}}},\ }\href {\doibase 10.1007/JHEP04(2010)112} {\bibfield
  {journal} {\bibinfo  {journal} {Journal of High Energy Physics}\ }\textbf
  {\bibinfo {volume} {4}},\ \bibinfo {eid} {112} (\bibinfo {year} {2010})},\
  \Eprint {http://arxiv.org/abs/1002.2783} {arXiv:1002.2783 [hep-th]}
  \BibitemShut {NoStop}%
\bibitem [{Note3()}]{Note3}%
  \BibitemOpen
  \bibinfo {note} {We note that the -3/2 exponent would lead to a divergence
  when integrated over $W$. The divergence is however compensated by the
  vanishing of the fidelity (in the infinite volume limit) so that the integral
  over the work distribution yields 1.}\BibitemShut {Stop}%
\bibitem [{\citenamefont {{Sotiriadis}}\ \emph {et~al.}(2014)\citenamefont
  {{Sotiriadis}}, \citenamefont {{Takacs}},\ and\ \citenamefont
  {{Mussardo}}}]{Sotiriadis}%
  \BibitemOpen
  \bibfield  {author} {\bibinfo {author} {\bibfnamefont {S.}~\bibnamefont
  {{Sotiriadis}}}, \bibinfo {author} {\bibfnamefont {G.}~\bibnamefont
  {{Takacs}}}, \ and\ \bibinfo {author} {\bibfnamefont {G.}~\bibnamefont
  {{Mussardo}}},\ }\href {\doibase 10.1016/j.physletb.2014.04.058} {\bibfield
  {journal} {\bibinfo  {journal} {Physics Letters B}\ }\textbf {\bibinfo
  {volume} {734}},\ \bibinfo {pages} {52} (\bibinfo {year} {2014})},\ \Eprint
  {http://arxiv.org/abs/1311.4418} {arXiv:1311.4418 [cond-mat.stat-mech]}
  \BibitemShut {NoStop}%
\bibitem [{\citenamefont {Destri}\ and\ \citenamefont
  {de~Vega}(1992{\natexlab{b}})}]{DDV1}%
  \BibitemOpen
  \bibfield  {author} {\bibinfo {author} {\bibfnamefont {C.}~\bibnamefont
  {Destri}}\ and\ \bibinfo {author} {\bibfnamefont {H.~J.}\ \bibnamefont
  {de~Vega}},\ }\href {\doibase 10.1103/PhysRevLett.69.2313} {\bibfield
  {journal} {\bibinfo  {journal} {Phys. Rev. Lett.}\ }\textbf {\bibinfo
  {volume} {69}},\ \bibinfo {pages} {2313} (\bibinfo {year}
  {1992}{\natexlab{b}})}\BibitemShut {NoStop}%
\bibitem [{\citenamefont {{Bortz}}\ and\ \citenamefont
  {{G{\"o}hmann}}(2005)}]{BorGoh}%
  \BibitemOpen
  \bibfield  {author} {\bibinfo {author} {\bibfnamefont {M.}~\bibnamefont
  {{Bortz}}}\ and\ \bibinfo {author} {\bibfnamefont {F.}~\bibnamefont
  {{G{\"o}hmann}}},\ }\href {\doibase 10.1140/epjb/e2005-00272-6} {\bibfield
  {journal} {\bibinfo  {journal} {European Physical Journal B}\ }\textbf
  {\bibinfo {volume} {46}},\ \bibinfo {pages} {399} (\bibinfo {year} {2005})},\
  \Eprint {http://arxiv.org/abs/cond-mat/0504370} {cond-mat/0504370}
  \BibitemShut {NoStop}%
\bibitem [{\citenamefont {{Ahn}}\ and\ \citenamefont
  {{Nepomechie}}(2004)}]{2bdrySG}%
  \BibitemOpen
  \bibfield  {author} {\bibinfo {author} {\bibfnamefont {C.}~\bibnamefont
  {{Ahn}}}\ and\ \bibinfo {author} {\bibfnamefont {R.~I.}\ \bibnamefont
  {{Nepomechie}}},\ }\href {\doibase 10.1016/j.nuclphysb.2003.11.012}
  {\bibfield  {journal} {\bibinfo  {journal} {Nuclear Physics B}\ }\textbf
  {\bibinfo {volume} {676}},\ \bibinfo {pages} {637} (\bibinfo {year}
  {2004})},\ \Eprint {http://arxiv.org/abs/hep-th/0309261} {hep-th/0309261}
  \BibitemShut {NoStop}%
\bibitem [{\citenamefont {{Doikou}}\ and\ \citenamefont
  {{Nepomechie}}(1999)}]{BreatherBdry}%
  \BibitemOpen
  \bibfield  {author} {\bibinfo {author} {\bibfnamefont {A.}~\bibnamefont
  {{Doikou}}}\ and\ \bibinfo {author} {\bibfnamefont {R.~I.}\ \bibnamefont
  {{Nepomechie}}},\ }\href {\doibase 10.1088/0305-4470/32/20/301} {\bibfield
  {journal} {\bibinfo  {journal} {Journal of Physics A Mathematical General}\
  }\textbf {\bibinfo {volume} {32}},\ \bibinfo {pages} {3663} (\bibinfo {year}
  {1999})},\ \Eprint {http://arxiv.org/abs/hep-th/9903066} {hep-th/9903066}
  \BibitemShut {NoStop}%
\bibitem [{Zam()}]{Zam}%
  \BibitemOpen
  \href@noop {} {\ }\BibitemShut {NoStop}%
\end{thebibliography}%
\begin{widetext}
\appendix
\section{Derivation of the NLIE}

In this appendix we derive the NLIE, \eqref{NLIEa}. Starting from \eqref{NLIE} 
\begin{eqnarray}\nonumber
 \log{\mathfrak{a}(u)}
 =\log{K_\xi(u)}+2M\log\left[\frac{\sinh{(u-\Theta)}\sinh{(u+\Theta-\eta)}}{\sinh{(u+\Theta)}\sinh{(u-\Theta+\eta)}}\right]-\oint_C\frac{\mathrm{d}\mu}{2\pi i} f'(\mu)\log{\left[\frac{1+\mathfrak{a}(\mu)}{1+K(\mu)}\right]}
\end{eqnarray}
where 
\begin{eqnarray}
 f(\mu)=\log\frac{\sinh{(u-\mu+\eta)}}{\sinh{(u-\mu-\eta)}}.
\end{eqnarray}
Before doing this we define the following functions
\begin{eqnarray}\label{phi}
 \phi(u,x)=i\log{\frac{\sinh{(ix+u)}}{\sinh{(ix-u)}}}\\
 \phi'(u,x)=\frac{2\sin{(2x)}}{\cosh{(2u)-\cos(2x)}}
\end{eqnarray}
with Fourier transform
\begin{eqnarray}\label{FT}
\int_{-\infty}^\infty \mathrm{d}u\,e^{-iu\omega}\frac{\phi'(u,x)}{2\pi} =\frac{\sinh{\left[(\pi-2x)\omega/2\right]}}{\sinh{(\pi\omega/2)}}.
\end{eqnarray}
The derivation follows the standard pattern when dealing with such equations\cite{Klumper,DDV1, BorGoh}. We start by splitting the integral into the contributions from the two contours
\begin{eqnarray}\nonumber
\log{\mathfrak{a}(u)}=\log{K_\xi(u)}+2Mi\left[\phi(u-\Theta+\eta/2,\gamma/2)+\phi(u+\Theta-\eta/2,\gamma/2)\right]\\\nonumber
-\int_{C_-}\frac{\mathrm{d}\mu}{2\pi }\phi'(u-\mu,\gamma) \log{\left[\frac{1+\mathfrak{a}(\mu)}{1+K_\xi(\mu)}\right]}\\-\int_{C_+}\frac{\mathrm{d}\mu}{2\pi }\phi'(u-\mu,\gamma) \log{\left[\frac{1+\mathfrak{a}(\mu)}{1+K_\xi(\mu)}\right]}.
\end{eqnarray}
After pulling a factor of $\left[\log{\mathfrak{a}(u)}-\log{K_\xi(u)}\right]$ out of the integral involving the $C_-$ contour we get 
\begin{eqnarray}\nonumber
\log{\left[\frac{\mathfrak{a}(u)}{K_\xi(u)}\right]}+\int_{C_-}\frac{\mathrm{d}\mu}{2\pi }\phi'(u-\mu,\gamma)\log{\left[\frac{\mathfrak{a}(\mu)}{K_\xi(\mu)}\right]} &=&
2Mi\Upsilon(u)
-\int_{C_-}\frac{\mathrm{d}\mu}{2\pi }\phi'(u-\mu,\gamma) \log{\left[\frac{1+\mathfrak{a}^{-1}(\mu)}{1+K_\xi^{-1}(\mu)}\right]}\\&&-\int_{C_+}\frac{\mathrm{d}\mu}{2\pi }\phi'(u-\mu,\gamma) \log{\left[\frac{1+\mathfrak{a}(\mu)}{1+K_\xi(\mu)}\right]}
\end{eqnarray}
where $\Upsilon(u)=\phi(u-\Theta+\eta/2,\gamma/2)+\phi(u+\Theta-\eta/2,\gamma/2)$. For $\gamma\leq \pi/2$ we can choose $\zeta=\gamma/2-\epsilon$ with $\epsilon=0^+$. Then defining $$\mathfrak{b(u)}=\mathfrak{a}(u+i\gamma/2),~~~\bar{\mathfrak{b}}(u)=\mathfrak{a}^{-1}(u-i\gamma/2)= \mathfrak{b}^{-1}(u-i\gamma)$$ and $$\varkappa_\xi(u)=K_\xi(u+i\gamma/2),~~~\bar{\varkappa}_\xi(u)=K_\xi^{-1}(u-i\gamma/2)$$ we get 
\begin{eqnarray}\nonumber
\log{\left[\frac{\mathfrak{b}(u)}{\varkappa_\xi(u)}\right]}+\int_{-\infty}^\infty\frac{\mathrm{d}\mu}{2\pi }\phi'(u-\mu+i\gamma-i\epsilon,\gamma)\log{\left[\frac{\mathfrak{b}(\mu-i\gamma)}{\varkappa_\xi(\mu-i\gamma)}\right]} =
-2Mi\Upsilon(u+i\gamma/2)\\\nonumber
+\int_{-\infty}^\infty\frac{\mathrm{d}\mu}{2\pi }\phi'(u-\mu,\gamma) \log{\left[\frac{1+\mathfrak{b}(\mu)}{1+\varkappa_\xi(\mu)}\right]}\\-\int_{-\infty}^\infty\frac{\mathrm{d}\mu}{2\pi }\phi'(u-\mu+i\gamma-i\epsilon,\gamma) \log{\left[\frac{1+\bar{\mathfrak{b}}(\mu)}{1+\bar{\varkappa}_\xi(\mu)}\right]}
\end{eqnarray}
Rearranging this using Fourier transforms brings us to
\begin{eqnarray}\nonumber
\log{\left[\frac{\mathfrak{b}(u)}{\varkappa_\xi(u)}\right]}
=2MiJ*\Upsilon(u+i\gamma/2)
+\int_{-\infty}^\infty\mathrm{d}\mu\,G(u-\mu,\gamma) \log{\left[\frac{1+\mathfrak{b}(\mu)}{1+\varkappa_\xi(\mu)}\right]}\\-\int_{-\infty}^\infty \mathrm{d}\mu\,G(u-\mu+i\gamma-i\epsilon,\gamma) \log{\left[\frac{1+\bar{\mathfrak{b}}(\mu)}{1+\bar{\varkappa}_\xi(\mu)}\right]}
\end{eqnarray}
where $*$ denotes the convolution $f*g(x)=\int\mathrm{d}y\,f(x-y)g(y)$ and 
\begin{eqnarray}
J(x)=\int_{-\infty}^\infty\frac{\mathrm{d}\omega}{2\pi}e^{i\omega x}\frac{\sinh{(\pi\omega/2)}}{2\cosh{(\gamma\omega/2)}\sinh{\left[(\pi-\gamma)\omega/2\right]}}\\
G(x)\int_{-\infty}^\infty\frac{\mathrm{d}\omega}{2\pi}e^{i\omega x}\frac{\sinh{\left[(\pi-2\gamma)\omega/2\right]}}{2\cosh{(\gamma\omega/2)}\sinh{\left[(\pi-\gamma)\omega/2\right]}}
\end{eqnarray}
The convolution $J*\Upsilon(u+i\gamma/2)$ can be evaluated by Fourier transform which in the large $\Theta$ limit gives
\begin{eqnarray}
J*\Upsilon(u+i\gamma/2)= -4e^{-\frac{\pi}{\gamma}\Theta}\cosh{\left(\frac{\pi}{\gamma}u\right)}.
\end{eqnarray}
Finally, using $m=4L\exp{(-\pi\Theta/\gamma)}/N$ gives us the non linear integral equation for $\gamma\leq\pi/2$
\begin{eqnarray}\nonumber
\log{\left[\frac{\mathfrak{b}(u)}{\varkappa_\xi(u)}\right]}
=-2mit\cosh{\left(\frac{\pi}{\gamma}u\right)}
+\int_{-\infty}^\infty\mathrm{d}\mu\,G(u-\mu,\gamma) \log{\left[\frac{1+\mathfrak{b}(\mu)}{1+\varkappa_\xi(\mu)}\right]}\\\label{NLIEb}-\int_{-\infty}^\infty \mathrm{d}\mu\,G(u-\mu+i\gamma-i\epsilon,\gamma) \log{\left[\frac{1+\bar{\mathfrak{b}}(\mu)}{1+\bar{\varkappa}_\xi(\mu)}\right]}.
\end{eqnarray}

For the regime $\pi>\gamma>\pi/2$ the choice of contour $C$ is different, namely we must choose $\zeta<\pi/2-\gamma/2$ in terms of the original function $\mathfrak{a}(u)$. Shifting the contour back to this value gives us the general case
\begin{eqnarray}\nonumber
\log{\left[\frac{\mathfrak{a}(u)}{K_\xi(u)}\right]}
=-2mt\sinh{\left(\frac{\pi}{\gamma}u\right)}
+\int_{-\infty}^{\infty}\mathrm{d}\mu\,G(u-\mu-i\zeta,\gamma) \log{\left[\frac{1+\mathfrak{a}(\mu+i\zeta)}{1+K_\xi(u+i\zeta)}\right]}\\-\int_{-\infty}^{\infty} \mathrm{d}\mu\,G(u-\mu+i\zeta,\gamma) \log{\left[\frac{1+\mathfrak{a}^{-1}(\mu-i\zeta)}{1+K_\xi^{-1}(\mu-i\zeta)}\right]}.
\end{eqnarray}
Collecting all the $K_\xi$ terms together  on the right hand side gives \eqref{NLIEa}.

\section{Derivation of the Loschmidt amplitude}
In this appendix we derive the amplitude in terms of the auxiliary function given by \eqref{Ga}. Taking the log of $\mathcal{G}(t)$ as it appears in \eqref{Glam} 
\begin{eqnarray}
\log{\mathcal{G}(t)}=N\log\Lambda(-\Theta)-N\log{\left[\braket{v}{v}\right]}-2NM\log\left[\sinh{(2\Theta-\eta)\sinh{(\eta)}}\right]
 \end{eqnarray}
Inserting, the expression for $\Lambda(-\Theta)$,  \eqref{Lambda} into this gives 
 \begin{eqnarray}\nonumber
\log{\mathcal{G}(t)}=N\sum_j^M\left[g(\lambda_j)+g(-\lambda_j)\right]+N\log\left[\frac{\sinh{(2\Theta+\eta)}}{\sinh{(2\Theta)}}\frac{\sinh^2{(\Theta
+(\xi-\eta/2))}}{\braket{v}{v}}\right]\\
+2MN\log{\left[\frac{\sinh{(2\Theta)}\sinh{(\eta)}}{\sinh{(2\Theta-\eta)}\sinh{(\eta)}}\right]}
 \end{eqnarray}
 where we have defined 
 \begin{eqnarray}
 g(u)=\log{\left[\frac{\sinh{(-\Theta+u+\eta)}}{\sinh{(-\Theta+u)}}\right]}=-i\phi(-\Theta+u+\eta/2,\gamma/2)
 \end{eqnarray}
Using the integral form of the sum  \eqref{int} we get
 \begin{eqnarray}\nonumber
\log{\mathcal{G}(t)}=-N\oint_C\frac{\mathrm{d}\mu}{2\pi i} g'(\mu)\log{\left[\frac{1+\mathfrak{a}(\mu)}{1+K_\xi(\mu)}\right]}+ N\log\left[\frac{\sinh{(2\Theta+\eta)}}{\sinh{(2\Theta)}}\frac{\sinh^2{(\Theta+(\xi-\eta/2))}}{\braket{v}{v}}\right]\\
-N\log{\left[\frac{1+\mathfrak{a}(\Theta)}{1+K_\xi(\Theta)}\right]}.~~~~~~~~~
\end{eqnarray}
The last term above comes from the pole at $\mu=\Theta$ in $g'(\mu)$. Using $\mathfrak{a}(\Theta)=0$ and 
taking the large $\Theta$ limit the second and third terms cancel giving 
\begin{eqnarray}
\log{\mathcal{G}(t)}&=&N\oint_C\frac{\mathrm{d}\mu}{2\pi} \phi'(\mu-\Theta+\eta/2,\gamma/2)\log{\left[\frac{1+\mathfrak{a}(\mu)}{1+K_\xi(\mu)}\right]}
\end{eqnarray}
Typically one then proceeds along the lines of the NLIE and extracts a factor of $\log{\mathfrak{a}(u)/K_\xi(u)}$ from the $C_-$ integral. In terms of $\mathfrak{b}(u)$
\begin{eqnarray}\nonumber
\log{\mathcal{G}(t)}&=&N\int_{-\infty}^\infty\frac{\mathrm{d}\mu}{2\pi} \phi'(\mu-\Theta,\gamma/2)\log{\left[\frac{\mathfrak{b}(\mu-i\gamma)}{\varkappa_\xi(\mu-i\gamma)}\right]}+N\int_{-\infty}^\infty\frac{\mathrm{d}\mu}{2\pi} \phi'(\mu-\Theta,\gamma/2)\log{\left[\frac{1+\bar{\mathfrak{b}}(\mu)}{1+\bar{\varkappa}_\xi(\mu)}\right]}\\
&&-N\int_{-\infty}^\infty\frac{\mathrm{d}\mu}{2\pi} \phi'(\mu-\Theta+\eta,\gamma/2)\log{\left[\frac{1+\mathfrak{b}(\mu)}{1+\varkappa_\xi(\mu)}\right]}\\\nonumber
&=&2imtN\int_{-\infty}^\infty\frac{\mathrm{d}\mu}{2\pi} \phi'(\mu-\Theta,\gamma/2)\cosh{\left(\frac{\pi}{\gamma}u\right)}\\
&&+\frac{N}{2\gamma}\int_{-\infty}^\infty\mathrm{d}\mu \,\text{sech}\left(\frac{\pi}{\gamma}(\mu-\Theta)\right)\Bigg\{\log{\left[\frac{1+\bar{\mathfrak{b}}(\mu)}{1+\bar{\varkappa}_\xi(\mu)}\right]}+\log{\left[\frac{1+\mathfrak{b}(\mu)}{1+\varkappa_\xi(\mu)}\right]}\Bigg\}.
\end{eqnarray}
In going to the second line we have used \eqref{NLIEb} and \eqref{Gw}. Then using the definition in \eqref{phi} we can turn the first term into 
\begin{eqnarray}\nonumber
2imtN\int_{-\infty}^\infty\frac{\mathrm{d}\mu}{2\pi} \phi'(\mu-\Theta,\gamma/2)\cosh{\left(\frac{\pi}{\gamma}\mu\right)}&=&imtN\int_{-\infty}^\infty\frac{\mathrm{d}\mu}{2\pi}\left[ \phi'(\Theta+\mu,\gamma/2)+\phi'(\mu-\Theta,\gamma/2)\right]\\
&&\quad\quad~~~~~~~~~\times\cosh{\left(\frac{\pi}{\gamma}u\right)}\\\label{gsenergy}
&=&itL\int_{-\infty}^\infty\frac{\mathrm{d}\mu}{\pi}\,m_0\cosh{(2\mu)}m\cosh{\left(\frac{\pi}{\gamma}\mu\right)}
\end{eqnarray}
which is $-iE_0t$ where $E_0$ is the ground state energy. 
Finally we get  
\begin{eqnarray}
\log{\mathcal{G}(t)}=-iE_0t+\frac{N}{2\gamma}\int_{-\infty}^\infty\mathrm{d}\mu \,\text{sech}\left(\frac{\pi}{\gamma}(\mu-\Theta)\right)\Bigg\{\log{\left[\frac{1+\bar{\mathfrak{b}}(\mu)}{1+\bar{\varkappa}_\xi(\mu)}\right]}+\log{\left[\frac{1+\mathfrak{b}(\mu)}{1+\varkappa_\xi(\mu)}\right]}\Bigg\}
\end{eqnarray}

Again for $\gamma>\pi/2$ we use a different contour with the result
\begin{eqnarray}\nonumber
\log{\mathcal{G}(t)}
&=&-iE_0t+\frac{N}{2i\gamma}\int_{-\infty}^\infty\mathrm{d}\mu\, \text{csch}\left(\frac{\pi}{\gamma}(\mu-\Theta-i\zeta)\right)\log{\left[\frac{1+\mathfrak{a}^{-1}(\mu-i\zeta)}{1+K_\xi^{-1}(\mu-i\zeta)}\right]}\\
&&-\frac{N}{2i\gamma}\int_{-\infty}^\infty\mathrm{d}\mu\,\text{csch}\left(\frac{\pi}{\gamma}(\mu-\Theta+i\zeta)\right)\log{\left[\frac{1+\mathfrak{a}(\mu+i\zeta)}{1+K_\xi(\mu+i\zeta)}\right]}
\end{eqnarray}
Taking the continuum limit and collecting all the $K_\xi$ terms together we arrive at \eqref{Gaa}.


\section{Boundary Phase shift}
The renormalized boundary S matrix, $\mathbb{K}_\xi(u)$ plays an important role. In this section we review its explicit  calculation mostly following the work in \citenum{BdrySaleur, 2bdrySG,BreatherBdry}. It is defined by 
\begin{eqnarray}\label{Kb}
\log{\mathbb{K}_\xi(u)}=J*\left[\log{K_\xi(u)}+i\phi(u,\gamma)\right]
\end{eqnarray} 
with $K_\xi$ given in \eqref{k}. More explicitly this is 
\begin{eqnarray}\nonumber
\mathbb{K}_\xi(u)=-\exp{\Bigg\{\int_{-\infty}^\infty\mathrm{d}\omega e^{i\omega u}}\frac{\sinh{(\pi\omega/2)}}{2\omega\cosh{(\gamma\omega/2)}\sinh{\left[(\pi-\gamma)\omega/2\right]}}\left[\frac{\sinh{\left[(\pi-2\gamma)\omega/2\right]}}{\sinh{(\pi\omega/2)}}\right.\\\left.
+\frac{2\cosh{(\pi\omega/4)}\sinh{\left[(\pi-2\gamma)\omega/4\right]}}{\sinh{(\pi\omega/2)}}-2\frac{\sinh{(\chi\omega/2)}}{\sinh{(\pi\omega/2)}}\right]\Bigg\}
\end{eqnarray}
where  $\chi=\pi-\gamma+2i\xi$ which means we have either $\chi=\pi-\gamma$, for $\xi=0$ or $\chi=-\gamma$ for $\xi=i\pi/2$.
It is standard to use the renormalized rapidity $\theta=\pi u/\gamma$ as well as the coupling constant $\lambda=\nu-1$ frequently used in studies of the Sine-Gordon model
\begin{eqnarray}
\lambda=\frac{8\pi}{\beta^2}-1=\frac{\gamma}{\pi-\gamma}.
\end{eqnarray}
After taking $\omega\to\frac{\pi}{\pi-\gamma}\omega$ and performing  some algebra  we get
\begin{eqnarray}\nonumber
\mathbb{K}_\xi(\frac{\gamma}{\pi}\theta)&=&-\exp\Bigg\{i\int_{-\infty}^\infty\frac{\mathrm{d}\omega}{\omega}\sin{\left(\omega \frac{2\lambda}{\pi}\theta\right)}\left[\frac{\sinh{\left[(\lambda-1)\omega/2\right]}}{2\cosh{\left[\lambda\omega/2\right]}\sinh{(\omega/2)}}
\right.\\\label{Kt}&&\left.-2\frac{\sinh{\left[3\lambda\omega/2\right]}\sinh{\left[(\lambda-1)\omega/2\right]}}{\sinh{(\omega/2)}\sinh{\left[2\lambda\omega\right]}}-\frac{\sinh{\left(\bar{\chi}\omega\right)}}{\cosh{\left[\lambda\omega\right]}\sinh{(\omega)}}\right]\Bigg\}~~~~~~~~~
\end{eqnarray}
with $\bar{\chi}=1,-\lambda$. The first term can be rewritten in a convenirent form by sending $\omega\to2\omega$
\begin{eqnarray}
\exp\Bigg\{i\int_{-\infty}^\infty\frac{\mathrm{d}\omega}{\omega}\sin{\left(\omega \frac{4\lambda}{\pi}\theta\right)}\left[\frac{\sinh{\left[(\lambda-1)\omega\right]}}{2\cosh{\left[\lambda\omega\right]}\sinh{(\omega)}}
\right]
=S^{-1}_{++}\left(2\theta\right)
\end{eqnarray}
where $S_{++}(x)$ is the physical soliton-soliton phase shift. Using the results from the appendix in [\!\!\citenum{Zam, BdrySaleur}]
\begin{eqnarray}\label{Spp}
S^{-1}_{++}\left(2\theta\right)&=&-\frac{\Gamma\left(\lambda+\frac{2i\lambda}{\pi}\theta\right)}{\Gamma(\lambda)\Gamma(1+\frac{2i\lambda}{\pi}\theta)}\prod_{l=1}^{\infty}\frac{F_l(0)F_l(i\pi)}{F_l(2\theta)F_l(i\pi-2\theta)}\\
F_l(\theta)&=&\frac{\Gamma\left(2l\lambda+\frac{i\lambda}{\pi}\theta\right)\Gamma\left(1+2l\lambda+\frac{i\lambda}{\pi}\theta\right)}{\Gamma\left((2l+1)\lambda+\frac{i\lambda}{\pi}\theta\right)\Gamma\left(1+(2l-1)\lambda+\frac{i\lambda}{\pi}\theta\right)}
\end{eqnarray}
Note that in both the attractive and repulsive regimes there are no poles of this function within the physical strip $0<\text{Im}(\theta)<i\pi/2$. The remaining term in $\mathbb{K}_\xi(\frac{\gamma}{\pi}\theta)$ has also be calculated and appears in [\!\!\citenum{BreatherBdry}], 
\begin{eqnarray}\nonumber
\left[S_0\left(\theta\right)S_1\left(\theta\right)\right]^2&=&\exp\Bigg\{-2i\int_{-\infty}^\infty\frac{\mathrm{d}\omega}{\omega}\sin{\left(\omega \frac{2\lambda}{\pi}\theta\right)}\left[\frac{\sinh{\left[3\lambda\omega/2\right]}\sinh{\left[(\lambda-1)\omega/2\right]}}{\sinh{(\omega/2)}\sinh{\left[2\lambda\omega\right]}}\right.\\\label{R2}&&\left.~~~~~~~~~~~~~~~~~~~\quad\quad\quad\quad+\frac{\sinh{\left(\bar{\chi}\omega\right)}}{2\cosh{\left[\lambda\omega\right]}\sinh{(\omega)}}\right]\Bigg\}.
\end{eqnarray}
with 
\begin{eqnarray}\nonumber
S_0\left(\theta\right)=\prod_{n=0}^\infty\frac{\Gamma\left(\frac{2i\lambda}{\pi}\theta+4n \lambda+1\right)\Gamma\left(\frac{2i\lambda}{\pi}\theta+(4n +4)\lambda\right)}{\Gamma\left(\frac{2i\lambda}{\pi}\theta+(4n+3) \lambda+1\right)\Gamma\left(\frac{2i\lambda}{\pi}\theta+(4n +1)\lambda\right)}\\
\times\frac{\Gamma\left(-\frac{2i\lambda}{\pi}\theta+(4n+3) \lambda+1\right)\Gamma\left(-\frac{2i\lambda}{\pi}\theta+(4n +1)\lambda\right)}{\Gamma\left(-\frac{2i\lambda}{\pi}\theta+4n \lambda+1\right)\Gamma\left(-\frac{2i\lambda}{\pi}\theta+(4n +4)\lambda\right)}
\end{eqnarray}
The final factor is 
\begin{eqnarray}\nonumber
S_1(\theta)&=&\frac{\cosh{\left[\lambda(\theta-i\frac{\pi}{2})-i\frac{\pi}{2}\bar{\chi}\right]}}{\pi}\\\nonumber&&\times\prod_{n=0}^\infty\frac{\Gamma\left(-i\frac{\lambda}{\pi}\theta +2n\lambda+\frac{1}{2}(\bar{\chi}+\lambda+1)\right)\Gamma\left(-i\frac{\lambda}{\pi}\theta +2n\lambda-\frac{1}{2}(\bar{\chi}+\lambda-1)\right)}{\Gamma\left(i\frac{\lambda}{\pi}\theta +\lambda(2n+2)+\frac{1}{2}(\bar{\chi}+\lambda+1)\right)\Gamma\left(i\frac{\lambda}{\pi}\theta +\lambda(2n+2)-\frac{1}{2}(\bar{\chi}+\lambda-1)\right)}\\
&&\times\frac{\Gamma\left(i\frac{\lambda}{\pi}\theta +\lambda(2n+1)+\frac{1}{2}(\bar{\chi}+\lambda+1)\right)\Gamma\left(i\frac{\lambda}{\pi}\theta +\lambda(2n+1)-\frac{1}{2}(\bar{\chi}+\lambda-1)\right)}{\Gamma\left(-i\frac{\lambda}{\pi}\theta +\lambda(2n+1)+\frac{1}{2}(\bar{\chi}+\lambda+1)\right)\Gamma\left(-i\frac{\lambda}{\pi}\theta +\lambda(2n+1)-\frac{1}{2}(\bar{\chi}+\lambda-1)\right)}\quad\quad
\end{eqnarray}
which is in agreement\cite{BdrySaleur} with the bootstrap calculations of [\!\!\citenum{Ghoshal}]. 

We are interested in the behaviour of $\mathbb{K}(u)$ in the physical strip $0<\text{Im}(u)<\gamma/2$ or $0<\text{Im}(\theta)<\pi/2$. With this in mind we can extract out the $n=0$ terms 
\begin{eqnarray}\nonumber
\mathbb{K}_\xi(\frac{\gamma}{\pi}\theta)&=&-\left[\frac{\cosh{\left[\lambda(\theta-i\frac{\pi}{2})-i\frac{\pi}{2}\bar{\chi}\right]}}{\pi}\frac{\Gamma\left(-i\frac{\lambda}{\pi}\theta +\frac{1}{2}(\bar{\chi}+\lambda+1)\right)\Gamma\left(-i\frac{\lambda}{\pi}\theta -\frac{1}{2}(\bar{\chi}+\lambda-1)\right)}{\Gamma\left(i\frac{\lambda}{\pi}\theta +2\lambda+\frac{1}{2}(\bar{\chi}+\lambda+1)\right)\Gamma\left(i\frac{\lambda}{\pi}\theta +2\lambda-\frac{1}{2}(\bar{\chi}+\lambda-1)\right)}\right.\\
&&\left.\times\frac{\Gamma\left(i\frac{\lambda}{\pi}\theta +\lambda+\frac{1}{2}(\bar{\chi}+\lambda+1)\right)\Gamma\left(i\frac{\lambda}{\pi}\theta +\lambda-\frac{1}{2}(\bar{\chi}+\lambda-1)\right)}{\Gamma\left(-i\frac{\lambda}{\pi}\theta +\lambda+\frac{1}{2}(\bar{\chi}+\lambda+1)\right)\Gamma\left(-i\frac{\lambda}{\pi}\theta +\lambda-\frac{1}{2}(\bar{\chi}+\lambda-1)\right)}\right]^2 \frac{\Gamma\left(\frac{2i\lambda}{\pi}\theta+1\right)}{\Gamma\left(\frac{2i\lambda}{\pi}\theta+\lambda\right)}\frac{\bar{S}^2_0(\theta)\bar{S}^2_1(\theta)}{\bar{S}_{++}(2\theta)}
\end{eqnarray}
where the barred functions $\bar{S}_{0,1,++}$ contain no zeros or poles within the physical strip. We now specialize to the two cases for $\xi=i\pi/2$, $\bar{\chi}=-\lambda$ this is
\begin{eqnarray}\nonumber
\mathbb{K}_{i\frac{\pi}{2}}(\frac{\gamma}{\pi}\theta)&=&-\left[\frac{\cosh{\left(\lambda\theta\right)}}{\pi}\right]^2\left[\frac{\Gamma\left(-i\frac{\lambda}{\pi}\theta +\frac{1}{2}\right)}{\Gamma\left(i\frac{\lambda}{\pi}\theta +2\lambda+\frac{1}{2}\right)}\frac{\Gamma\left(i\frac{\lambda}{\pi}\theta +\lambda+\frac{1}{2}\right)}{\Gamma\left(-i\frac{\lambda}{\pi}\theta +\lambda+\frac{1}{2}\right)}\right]^4 \frac{\Gamma\left(\frac{2i\lambda}{\pi}\theta+1\right)}{\Gamma\left(\frac{2i\lambda}{\pi}\theta+\lambda\right)}\frac{\bar{S}^2_0(\theta)\bar{S}^2_1(\theta)}{\bar{S}_{++}(2\theta)}
\end{eqnarray}
examining this one can see that it has poles
\begin{eqnarray}
\theta=i\frac{\pi}{\lambda}m,~~~2m<\lambda
\end{eqnarray}
while additionally it vanishes linearly for $\theta=i\pi/2$. These poles lie within the physical strip provided $\lambda>1$ which is the attractive regime, $\gamma>\pi/2$. For $\xi=0$ or $\bar{\chi}=1$ 
\begin{eqnarray}\nonumber
\mathbb{K}_0(\frac{\gamma}{\pi}\theta)&=&-\left[i\frac{\sinh{\left[\lambda(\theta-i\frac{\pi}{2})\right]}}{\pi}\frac{\Gamma\left(-i\frac{\lambda}{\pi}\theta +\frac{1}{2}\lambda+1\right)\Gamma\left(-i\frac{\lambda}{\pi}\theta -\frac{1}{2}\lambda\right)}{\Gamma\left(i\frac{\lambda}{\pi}\theta +\frac{5}{2}\lambda+1\right)\Gamma\left(i\frac{\lambda}{\pi}\theta +\frac{3}{2}\lambda\right)}\right.\\
&&\left.\times\frac{\Gamma\left(i\frac{\lambda}{\pi}\theta +\frac{3}{2}\lambda+1\right)\Gamma\left(i\frac{\lambda}{\pi}\theta +\frac{1}{2}\lambda\right)}{\Gamma\left(-i\frac{\lambda}{\pi}\theta +\frac{3}{2}\lambda+1\right)\Gamma\left(-i\frac{\lambda}{\pi}\theta +\frac{1}{2}\lambda\right)}\right]^2 \frac{\Gamma\left(\frac{2i\lambda}{\pi}\theta+1\right)}{\Gamma\left(\frac{2i\lambda}{\pi}\theta+\lambda\right)}\frac{\bar{S}^2_0(\theta)\bar{S}^2_1(\theta)}{\bar{S}_{++}(2\theta)}
\end{eqnarray}
Therefore we get poles at 
\begin{eqnarray}
\theta=i\frac{\pi}{2\lambda}m,~~~m<\lambda
\end{eqnarray}
in addition to a pole at $\theta=i\pi/2$. 

\section{Non interacting  Quenches}
In this section we consider the following Hamiltonian
\begin{eqnarray}
H=\int-i\left[\psi^\dag_+\partial_x\psi_+-\psi^\dag_-\partial_x\psi_-\right]+m_0\left[\psi^\dag_+\psi_-+\psi^\dag_-\psi_+\right]
\end{eqnarray}
of massive Dirac fermions. We  study quenches of $m_i\to m_0$ where the inital mass  $m_i$ is arbitrary.  In particular we want to to calculate the Loschmidt amplitude
\begin{eqnarray}
\mathcal{G}(t)=\bra{\Phi_i}e^{-iHt}\ket{\Phi_i}
\end{eqnarray}
and confirm the non interacting limit reached in the text. We will proceed by rewriting the Hamiltonian in momentum space such that
\begin{eqnarray}
\bar{H}=\sum_p\omega_i(p)\left[\psi^\dag_a(p)\psi_a(p)-\psi^\dag_b(p)\psi_b(p)\right]\\
H=\sum_p\omega(p)\left[\psi^\dag_\alpha(p)\psi_\alpha(p)-\psi^\dag_\beta(p)\psi_\beta(p)\right]
\end{eqnarray}
where $\omega_i(p)=\sqrt{p^2+m_i^2}$ and $\omega(p)=\sqrt{p^2+m_0^2}$.
Since both Hamiltonians  are quadratic they are simply related by a rotation to  each other by a Bogoliubov transformation
\begin{eqnarray}
\begin{pmatrix}
\psi_\alpha(p)\\
\psi_\beta(p)
\end{pmatrix}=\begin{pmatrix}\bar{c}_p &\bar{s}_p\\
-\bar{s}_p& \bar{c}_p
\end{pmatrix}\begin{pmatrix}
\psi_a(p)\\
\psi_b(p)
\end{pmatrix}.
\end{eqnarray}
with the ratio of the coefficients  
\begin{eqnarray}
\frac{\bar{s}_p}{\bar{c}_p}=\frac{\sqrt{(\omega(p)-p)(\omega_i(p)+p)}-\sqrt{(\omega_i(p)-p)(\omega(p)+p)}}{\sqrt{(\omega(p)+p)(\omega_i(p)+p)}+\sqrt{(\omega_i(p)-p)(\omega(p)-p)}}.
\end{eqnarray}
If the initial state is the ground state of the theory with $m_i$ it is written as
\begin{eqnarray}
\ket{\Phi_i}&=&\prod_{p}\psi^\dag_b(p)\ket{0}\\
&=&\prod_{p}\{\bar{s}_p\psi^\dag_\alpha(p)+\bar{c}_p\psi^\dag_\beta(p)\}\ket{0}
\end{eqnarray}
Time evolving this initial state we get 
\begin{eqnarray}
e^{-iHt}\ket{\Phi_i}&=&\prod_{p}\{e^{-i\omega(p)t}\bar{s}_p\psi^\dag_\alpha(p)+e^{i\omega(p)t}\bar{c}_p\psi^\dag_\beta(p)\}\ket{0}\\
&=&\prod_{p}\{e^{-i\omega(p)t}\bar{s}^{2}_p+e^{i\omega(p)t}\bar{c}_p^2\}\psi^\dag_b(p)\ket{0}+\dots
\end{eqnarray}
where the dots contain $\psi^\dag_a(p)$ terms. Taking the overlap with the initial state gives
\begin{eqnarray}
\bra{\Phi_i}e^{-iHt}\ket{\Phi_i}&=&\prod_{p}\left[e^{-i\omega(p)t}\bar{s}_p^2+e^{i\omega(p)t}\bar{c}_p^2\right]\\
&=&e^{it\sum_p\omega(p)}\prod_{p}\bar{c}^{2}_p\prod_{p}\left[1+\left[\frac{\bar{s}_{p}}{\bar{c}_p}\right]^2e^{-2i\omega(p)t}\right]
\end{eqnarray}
In the thermodynamic limit the log of the amplitude is found to be
\begin{eqnarray}\nonumber
\log{\mathcal{G}(t)}&=&itL\int_{-\infty}^\infty\frac{\mathrm{d}p}{2\pi}\omega(p)+2L\int_{-\infty}^\infty\frac{\mathrm{d}p}{2\pi}\log{\left[\bar{c}_p\right]}\\
&&+L\int_{-\infty}^\infty\frac{\mathrm{d}p}{2\pi}\log{\left[1+\left(\frac{\bar{s}_{p}}{\bar{c}_p}\right)^2e^{-2i\omega(p)t}\right]}
\end{eqnarray}
which  is the form for any quench where the initial and final Hamiltonians are quadratic. 
In the limit of $m_i\to\infty$ we get 
\begin{eqnarray}\label{rotations1}
\bar{s}_p&=&\frac{\sqrt{(\omega(p)-p)}-\sqrt{(\omega(p)+p)}}{\sqrt{2\omega(p)}}\\\label{rotations2}
\bar{c}_p&=&\frac{\sqrt{(\omega(p)-p)}+\sqrt{(\omega(p)+p)}}{\sqrt{2\omega(p)}}
\end{eqnarray}
We can write this in terms of rapidities $p=m_0\sinh{2\mu}$ to get 
\begin{eqnarray}\nonumber
\log{\mathcal{G}(t)}&=&itL\int_{-\infty}^\infty\frac{\mathrm{d}\mu}{\pi}m_0^2\cosh^2{(2\mu)}-m_0L\int_{-\infty}^\infty\frac{\mathrm{d}\mu}{2\pi}\cosh{(2\mu)}\log{\left[1+\tanh^2(\mu)\right]}\\
&&+m_0\int_{-\infty}^\infty\frac{\mathrm{d}\mu}{\pi}\cosh{(2\mu)}\log{\left[1+\tanh^2{(\mu)}e^{-2im\cosh{(\lambda)}t}\right]}
\end{eqnarray}
in agreement with the expression in the  text. The quench from the highest excited state of $\bar{H}$ can be obtained by taking $m_i\to-\infty$ in \eqref{rotations1} and \eqref{rotations2} which gives $\bar{s}_p/\bar{c}_p=\coth{\mu}$ instead which is again what is found in the text. For a finite initial mass, $m_i$ it is not too difficult to check that that 
\begin{eqnarray}
\left(\frac{\bar{s}_p}{\bar{c}_p}\right)^2=\left[\frac{\sinh{\left(\mu-\frac{1}{2}\sinh^{-1}\left[\frac{m_0}{m_i}\sinh{(2\mu)}\right]\right)}}{\cosh{\left(\mu-\frac{1}{2}\sinh^{-1}\left[\frac{m_0}{m_i}\sinh{(2\mu)}\right]\right)}}\right]^2.
\end{eqnarray}
\end{widetext}
\end{document}